\documentclass[prx,reprint,floatfix,letterpaper,twocolumn,nofootinbib,showpacs,shortbibliography]{revtex4-1}
\usepackage[caption=false]{subfig}
\usepackage{graphicx} 
\usepackage{epstopdf}
\usepackage{array}
\usepackage{verbatim}
\usepackage{amsmath,amsfonts,amssymb,amscd}
\usepackage{amsthm}
\usepackage{tabularx}
\usepackage{stmaryrd}
\usepackage{enumerate}
\usepackage[ruled]{algorithm2e}
\usepackage{enumitem}
\usepackage{wasysym}
\usepackage[dvipsnames]{xcolor}
\usepackage[normalem]{ulem}
\usepackage{bbold}
\input{Qcircuit}

\usepackage{hyperref}
\hypersetup{
    bookmarksnumbered=true, 
    unicode=false, 
    pdfstartview={FitH}, 
    pdftitle={}, 
    pdfauthor={}, 
    pdfsubject={}, 
    pdfcreator={}, 
    pdfproducer={}, 
    pdfkeywords={}, 
    pdfnewwindow=true, 
    colorlinks=true, 
    linkcolor=blue, 
    citecolor=blue, 
    filecolor=blue, 
    urlcolor=blue 
}

\theoremstyle{plain}
\newtheorem{thm}{Theorem}

\newtheorem{lem}[thm]{Lemma}

\theoremstyle{definition}

\newcommand{\eq}[1]{(\hyperref[eq:#1]{\ref*{eq:#1}})}

\renewcommand{\sec}[1]{\hyperref[sec:#1]{Section~\ref*{sec:#1}}}
\newcommand{\thrm}[1]{\hyperref[thm:#1]{Theorem~\ref*{thm:#1}}}
\newcommand{\lemm}[1]{\hyperref[lemm:#1]{Lemma~\ref*{lemm:#1}}}
\newcommand{\prop}[1]{\hyperref[prop:#1]{Proposition~\ref*{prop:#1}}}
\newcommand{\corr}[1]{\hyperref[corr:#1]{Corollary~\ref*{corr:#1}}}
\newcommand{\fig}[1]{\hyperref[fig:#1]{Figure~\ref*{fig:#1}}}

\DeclareMathAlphabet{\matheu}{U}{eus}{m}{n}

\newcolumntype{L}[1]{>{\raggedright}p{#1}}
\newcolumntype{C}[1]{>{\centering}p{#1}}
\newcolumntype{R}[1]{>{\raggedleft}p{#1}}
\newcolumntype{D}{>{\centering\arraybackslash}X}

\definecolor{darkgreen}{rgb}{0,0.5,0}
\definecolor{darkblue}{rgb}{0,0,0.5}

\newcommand{\schlafli}{Schl\"afli }

\begin{document}
\title{A four-dimensional toric code with non-Clifford transversal gates}
\author{Tomas Jochym-O'Connor}
\email{tjoc@ibm.com}
\affiliation{\small IBM Quantum, IBM T.J. Watson Research Center, Yorktown Heights, NY 10598}

\author{Theodore J. Yoder}
\email{ted.yoder@ibm.com}
\affiliation{\small IBM Quantum, IBM T.J. Watson Research Center, Yorktown Heights, NY 10598}

\begin{abstract}
The design of a four-dimensional toric code is explored with the goal of finding a lattice capable of implementing a logical~$\mathsf{CCCZ}$ gate transversally. The established lattice is the octaplex tessellation, which is a regular tessellation of four-dimensional~Euclidean space whose underlying 4-cell is the octaplex, or hyper-diamond. This differs from the conventional 4D~toric code lattice, based on the hypercubic tessellation, which is symmetric with respect to logical~$X$ and $Z$ and only allows for the implementation of a transversal Clifford gate. This work further develops the established connection between topological dimension and transversal gates in the Clifford hierarchy, generalizing the known designs for the implementation of transversal~$\mathsf{CZ}$ and~$\mathsf{CCZ}$ in two and three dimensions, respectively.
\end{abstract}

\maketitle

\section{Motivation}

Quantum error correction is expected to play an essential role in the development of large-scale quantum computers. Namely, identifying, controlling, and correcting physical errors will be necessary in running long quantum computations and to that end the theory of quantum fault tolerance has been developed~\cite{AB97}. One of the essential primitives in fault tolerance theory is the notion of a transversal gate, that is a logical gate that can be implemented by addressing each qubit within a codeblock, individually, in parallel. The primary benefit of such gates is they prevent the propagation of noise between different qubits in the code, which is typically problematic.

Topological error correcting codes are among the most well-studied forms of quantum error correcting codes. These codes are defined by their spatially-local stabilizer checks in $D$-dimensions, while embedding their logical information in macroscopic non-local degrees of freedom. Such codes provide numerous computing advantages including a pathway for experimental qubit layout~\cite{FMMC12, YK17, litinski19, chamberland20a, chamberland20b}, efficient decoding algorithms~\cite{dennis02, delfosse14, DN17, KD19, KP19, VBK20}, and provable target threshold error rates for numerous local noise models~\cite{KBM09, bombin+12, FMMC12}. In this work we explore the set of transversal gates in one such code, the 4D~toric code embedded in the octaplex tessellation. Unlike the symmetric 4D~toric code on the hypercubic lattice, where there is a symmetry between the $X$ and $Z$~stabilizers for the purposes of quantum self-correction~\cite{dennis02}, the version presented here is asymmetric with respect to $X$ and~$Z$ as required by disjointness~\cite{JoKY18} in order to have the ability to implement a non-Clifford logical gate transversally.

There is a rich set of literature relating topological codes to logical gates in quantum error correction. Namely, the set of transversal gates accessible to a topological code is related to the spatial dimension in which it is embedded. In the color code family, all Clifford gates are transversal for 2D color codes~\cite{BM06}, the 3D color code has a transversal~$T$ gate ($e^{i\pi Z/8})$, and more generally $D$-dimensional color codes have transversal~$e^{i \pi Z/2^D}$ gates~\cite{Bombin15a, KB15}. It should be noted, while these higher-dimensional codes have more exotic gates, they cannot have both a transversal~$e^{i \pi Z/2^D}$ gate with $D>2$ while also having a transversal Hadamard as this would violate no-go theorems for transversal, universal gate sets~\cite{EK09, ZCC11}.

In order to better define the relationship between spatial dimension and the transversal gates that are accessible, consider the Clifford hierarchy. The $n$-qubit Clifford hierarchy is defined recursively, where the first level is the $n$-qubit Pauli group~$\mathcal{P}_n$:
\begin{align}
\mathcal{C}_n^1 = \mathcal{P}_n, \ \mathcal{C}_n^k = \{ U \in \mathcal{U}(n) :  UPU^{\dagger} \in C_{n}^{k-1} \ \forall P \in P_n\}
\end{align}
In general, topological codes in $D$~dimensions will be limited to having transversal gates that are in the $D$-th level of the Clifford hierarchy~\cite{BK13}. The 2D toric code has a transversal controlled-$Z$ ($\mathsf{CZ}$) gate, which is in the second level of the Clifford hierarchy. Recently, Vasmer and Browne showed that the 3D~toric code has a transversal controlled-$\mathsf{CZ}$ $(\mathsf{CCZ})$~gate~\cite{VB19}, a gate in the third level of the Clifford hierarchy, provided the codes are chosen appropriately. Rather surprisingly, the different codeblocks are not identical.

Given the aforementioned prior work, it is natural to ask whether some choice of 4D~toric code has a transversal multi-qubit gate in the fourth level of the Clifford hierarchy. In this work, we explore 4D~toric codes and show a particular choice of 4D~toric code that has an underlying~$\mathsf{CCCZ}$ gate, a gate in the fourth level of the Clifford hierarchy. The underlying lattice of the code is not a simple generalization of the cubic lattice, such has a hypercubic tessellation, but a rather more exotic lattice that allows for the appropriate overlap conditions of the underlying stabilizers of the four codeblocks.

The paper is organized as follows: Sec.~\ref{sec:Background} is reserved for a review of prior results. In \ref{sec:TransversalConditions} we review the necessary algebraic conditions for CSS codes to have transversal multi-controlled-$Z$ gates. In Sec.~\ref{sec:Schlafli} we review \schlafli symbols and summarize their relationship to constructing 2D and 3D toric codes in Secs.~\ref{sec:2DTC} and~\ref{sec:3DTC}, respectively. In Sec.~\ref{sec:DualCond} we present the conditions we will require for searching for a 4D~toric code with a transversal~$\mathsf{CCCZ}$. Sec.~\ref{sec:4DTC} provides all the details of the 4D~lattice containing a transversal~$\mathsf{CCCZ}$, the octaplex tessellation, while Sec.~\ref{sec:Boundaries} discusses placing boundaries on such a lattice. Sec.~\ref{sec:Metachecks} discusses metachecks and single-shot error correction in the context of this code. Finally, Sec.~\ref{sec:Discussion} provides a discussion on the difficulty in finding higher-dimensional generalizations and other open questions.

\section{Background}
\label{sec:Background}

\subsection{Transversal (multi-)controlled-$Z$ gates in CSS codes}
\label{sec:TransversalConditions}

Suppose we have multiple codeblocks, each composed of $n$~qubits and containing equal numbers of logical qubits. We denote a Pauli operator $P$ on qubit~$i$ of codeblock~$c$ as $P^{(c)}_i$. We denote an $X$~stabilizer generator of codeblock~$c$ as $\mathcal{X}^{(c)}_i$, where $i$ is a label for the generator. Similarly, the $Z$~stabilizer generators are labeled as~$\mathcal{Z}^{(c)}_i$. Logical operators of codeblock $c$ of $X$-type and $Z$-type are represented by $\overline{\mathcal{X}}^{(c)}_j$ and $\overline{\mathcal{Z}}^{(c)}_j$, respectively. When contextually appropriate, we use $\mathcal{X}^{(c)}_i$, $\overline{\mathcal{X}}^{(c)}_j$, etc.~to represent the supports of these operators as well.

The controlled-$Z$ ($\mathsf{CZ}$) is a two-qubit Clifford gate that under conjugation maps Pauli~$X$ on one qubit to itself times Pauli~$Z$ on the other qubit. Mathematically, this relation is expressed as $X_i^{(a)} \rightarrow X_i^{(a)} Z_i^{(b)}$, where the $\mathsf{CZ}$~gate is acting on qubits~$i$ of codeblocks~$a, b$. We label this a $\mathsf{CZ}_i^{(a,b)}$ gate. Additionally, $\mathsf{CZ}$ leaves any Pauli~$Z$ operator unchanged under conjugation. Therefore, if we have two CSS codeblocks~\cite{CS96, Steane96b} of $n$~qubits and we perform a transversal~$\mathsf{CZ}$, that is $\prod_{i=1}^n\mathsf{CZ}_i^{(a,b)}$, then the stabilizer generators are transformed as follows:
\begin{align*}
\mathcal{X}_i^{(a)} & = \prod_{j \in \mathcal{X}_i^{(a)}} X_j^{(a)}  \rightarrow  \mathcal{X}_i^{(a)} \cdot \prod_{k \in \mathcal{X}_i^{(a)}} Z_k^{(b)} \\
\mathcal{Z}_i^{(a)} &\rightarrow \mathcal{Z}_i^{(a)} .
\end{align*}
As such, in order for the codespace to be preserved we require that the $X$~stabilizers from one codeblock map onto $Z$~stabilizers in the other codeblock. Equivalently, as long as the $X$~stabilizers of a given codeblock overlap an even number of times with the $X$~stabilizers and $X$~logical operators of the other codeblock, the codespace will be preserved. That is: $\forall \ i,j$,
\begin{subequations}
\label{eq:IdentityReq2}
\begin{align}
| \mathcal{X}_i^{(a)} \cap \mathcal{X}_j^{(b)} | &= 0, \\ 
| \mathcal{X}_i^{(a)} \cap \overline{\mathcal{X}}_j^{(b)} | &= 0.
\end{align}
\end{subequations}
The expression~$\mathcal{M} \cap \mathcal{N}$ indicates the overlap in support of two Pauli operators~$\mathcal{M}, \ \mathcal{N}$ while $| \mathcal{O} |$ is the total weight of a given operator~$\mathcal{O}$ modulo 2. Given the above equations are satisfied, transversal~$\mathsf{CZ}$ will be a logical operator. Yet, in order for it to implement a logical~$\mathsf{CZ}$ gate there are additional conditions the logical operators must satisfy: $\forall \ i,j$,
\begin{align} \tag{2c}
| \overline{\mathcal{X}}_i^{(a)} \cap \overline{\mathcal{X}}_j^{(b)} | = \delta_{ij}.
\end{align}

We can generalize these constraints to the controlled-controlled-$Z$~$(\mathsf{CCZ})$ gate as well. Note that under conjugation, the $\mathsf{CCZ}$ gate maps Pauli~$X$ on one codeblock onto $\mathsf{CZ}$ on the other two codeblocks, that is $X_i^{(a)} \rightarrow X_i^{(a)} \mathsf{CZ}_i^{(b,c)}$. Additionally, given it is again a diagonal operator $\mathsf{CCZ}$ leaves and Pauli~$Z$ operator unchanged. The $X$~stabilizer of one codeblock will therefore undergo the following transformation under the action of transversal $\mathsf{CCZ}$:
\begin{align*}
\mathcal{X}_i^{(a)} = \prod_{j \in \mathcal{X}_i^{(a)}} X_j^{(a)}   &\rightarrow  \mathcal{X}_i^{(a)} \cdot \prod_{k \in \mathcal{X}_i^{(a)}} \mathsf{CZ}_k^{(b,c)}.
\end{align*}
Therefore, in order to preserve the stabilizer group, we require that $\prod_{k \in \mathcal{X}_i^{(a)}} \mathsf{CZ}_k^{(b,c)}$ be logical identity, which is equivalent to imposing Eqs.~\ref{eq:IdentityReq2}, but restricted to the support of the $X$~stabilizer on codeblock~$a$. Therefore, the generalized requirement is: $\forall \ i,j,k$,
\begin{subequations}
\label{eq:IdentityReq3}
\begin{align}
| \mathcal{X}_i^{(a)} \cap \mathcal{X}_j^{(b)}  \cap \mathcal{X}_k^{(c)} | &= 0, \label{eq:IdentityReq3a}\\ 
| \mathcal{X}_i^{(a)} \cap \mathcal{X}_j^{(b)} \cap \overline{\mathcal{X}}_k^{(c)} | &= 0 \\
| \mathcal{X}_i^{(a)} \cap \overline{\mathcal{X}}_j^{(b)} \cap \overline{\mathcal{X}}_k^{(c)} | &= 0.
\end{align}
\end{subequations}
Given the above requirements are satisfied, the transversal~$\mathsf{CCZ}$ maps Pauli~$X$ onto a transversal~$\mathsf{CZ}$, limited by the support of the Pauli~$X$ of the original code. This $\mathsf{CZ}$ must in turn perform a logical~$\mathsf{CZ}$ operation on the other two codes, which impose the additional set of constraints: $\forall \ i,j,k$,
\begin{align} \tag{3d}
| \overline{\mathcal{X}}_i^{(a)} \cap \overline{\mathcal{X}}_j^{(b)} \cap \overline{\mathcal{X}}_k^{(c)} | = [i=j=k],
\end{align}
where we have used the Iverson bracket~\cite{Iverson1962} to denote the overlap being odd when all logical indices are matching, and 0 otherwise.

Finally, one can straightforwardly generalize this process to multi-controlled-$Z$ operations. The set of requirements for the transversal $\mathsf{CCCZ}$~gate to implement a logical~$\mathsf{CCCZ}$ on four codeblocks is: $\forall \ i,j,k,l$,
\begin{subequations}
\label{eq:IdentityReq4}
\begin{align}
|\mathcal{X}_i^{(a)} \cap \mathcal{X}_j^{(b)}  \cap \mathcal{X}_k^{(c)} \cap \mathcal{X}_l^{(d)}| &= 0,\label{eq:Cond1} \\ 
|\mathcal{X}_i^{(a)} \cap \mathcal{X}_j^{(b)} \cap \mathcal{X}_k^{(c)} \cap \overline{\mathcal{X}}_l^{(d)}| &= 0, \label{eq:Cond2} \\
|\mathcal{X}_i^{(a)} \cap \mathcal{X}_j^{(b)} \cap \overline{\mathcal{X}}_k^{(c)} \cap \overline{\mathcal{X}}_l^{(d)}| &= 0, \label{eq:Cond3} \\
|\mathcal{X}_i^{(a)} \cap \overline{\mathcal{X}}_j^{(b)} \cap \overline{\mathcal{X}}_k^{(c)} \cap \overline{\mathcal{X}}_l^{(d)}| &= 0, \label{eq:Cond4} \\
|\overline{\mathcal{X}}_i^{(a)} \cap \overline{\mathcal{X}}_j^{(b)} \cap \overline{\mathcal{X}}_k^{(c)} \cap \overline{\mathcal{X}}_l^{(d)}| &= [i=j=k=l]. \label{eq:Cond5}
\end{align}
\end{subequations}

\subsection{Pauli sandwich trick}
\label{sec:PauliSandwich}

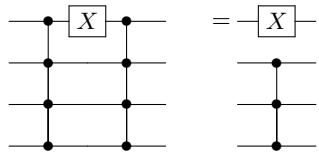
\begin{figure}
\begin{align*}
\Qcircuit @C=.1em @R=.2em @! {
& \ctrl{3} & \gate{X} & \ctrl{3} & \qw & \\
& \ctrl{2} & \qw & \ctrl{2} & \qw & \\
& \ctrl{1} & \qw & \ctrl{1} & \qw & \\
& \ctrl{-1} & \qw & \ctrl{-1} & \qw & \\
}
=
\Qcircuit @C=.1em @R=0.2em @! {
& \gate{X} & \qw  \\
& \ctrl{2} & \qw  \\
& \ctrl{1} & \qw  \\
& \ctrl{-1} & \qw \\
}
\end{align*}
\caption{Pauli sandwich trick: inserting a Pauli~$X$ operator between two $\mathsf{CCCZ}$ gates results in a $\mathsf{CCZ}$ gate on the other three qubits along with the implemented Pauli~$X$. This can be generalized for any multi-controlled-$\mathsf{Z}$ operation.}
\label{fig:TargetedCCZ}
\end{figure}

One advantage of implementing a $\mathsf{CCCZ}$~gate transversally is that if the resulting logical gate couples multiple logical qubits, such as the 3D~toric code with periodic boundaries or in instances of Pin~codes~\cite{VuBr19}, one may still be able to achieve a targeted $\mathsf{CCZ}$ gate by repeated uses of the transversal gate. The idea is to insert a logical $X$~gate (which can always be implemented transversally in a stabilizer code) between two instances of the transversal gate implementing logical~$\mathsf{CCCZ}$, which results in implementing a~$\mathsf{CCZ}$~gate on the other three logical codeblocks, see Fig.~\ref{fig:TargetedCCZ}.

\begin{figure}
\begin{align*}
\Qcircuit @C=.5em @R=.5em  {
& \ctrl{6} & \qw & \qw & \qw & \gate{\overline{\mathcal{X}}} & \ctrl{6} & \qw & \qw & \qw & \qw\\
& \qw & \ctrl{4} & \qw & \qw & \qw & \qw & \ctrl{4} & \qw & \qw & \qw \\
& \qw & \qw & \ctrl{6} & \qw & \qw & \qw & \qw & \ctrl{6} & \qw & \qw \\
& \qw & \qw & \qw & \ctrl{4} & \qw & \qw & \qw & \qw & \ctrl{4} & \qw \\
&&&&&& &  &  &  & \\
& \qw & \ctrl{8} & \qw & \qw & \qw & \qw & \ctrl{8} & \qw & \qw & \qw \\
& \ctrl{6} & \qw & \qw & \qw & \qw & \ctrl{6} & \qw & \qw & \qw & \qw \\
& \qw & \qw & \qw & \ctrl{4} & \qw & \qw & \qw & \qw & \ctrl{4} & \qw \\
& \qw & \qw & \ctrl{2} & \qw & \qw & \qw & \qw & \ctrl{2} & \qw & \qw \\
&&&&&\\
& \qw & \qw & \ctrl{6} & \qw & \qw & \qw & \qw & \ctrl{6} & \qw & \qw \\
& \qw & \qw & \qw & \ctrl{4} & \qw & \qw & \qw & \qw & \ctrl{4} & \qw \\
& \ctrl{6} & \qw & \qw & \qw & \qw & \ctrl{6} & \qw & \qw & \qw & \qw \\
& \qw & \ctrl{4} & \qw & \qw & \qw & \qw & \ctrl{4} & \qw & \qw & \qw \\
&&&&&\\
& \qw & \qw & \qw & \ctrl{-4} & \qw & \qw & \qw & \qw & \ctrl{-4} & \qw \\
& \qw & \qw & \ctrl{-6} & \qw & \qw & \qw & \qw & \ctrl{-6} & \qw & \qw \\
& \qw & \ctrl{-4} & \qw & \qw & \qw & \qw & \ctrl{-4} & \qw & \qw & \qw \\
& \ctrl{-6} & \qw & \qw & \qw & \qw & \ctrl{-6} & \qw & \qw & \qw & \qw \\
}
=
\Qcircuit @C=.5em @R=.65em  {
& \ctrl{6} & \gate{\overline{\mathcal{X}}} & \ctrl{6} & \qw \\
& \qw & \qw & \qw & \qw \\
& \qw & \qw & \qw & \qw \\
& \qw & \qw & \qw & \qw \\
&&&&&& &  &  &  & \\
& \qw & \qw & \qw & \qw \\
& \ctrl{6} & \qw & \ctrl{6} & \qw \\
& \qw & \qw & \qw & \qw \\
& \qw & \qw & \qw & \qw \\
&&&&&\\
& \qw & \qw & \qw & \qw \\
& \qw & \qw & \qw & \qw \\
& \ctrl{6} & \qw & \ctrl{6} & \qw \\
& \qw & \qw & \qw & \qw \\
&&&&&\\
& \qw & \qw & \qw & \qw \\
& \qw & \qw & \qw & \qw \\
& \qw & \qw & \qw & \qw \\
& \ctrl{-6} & \qw & \ctrl{-6} & \qw  \\
}
=
\Qcircuit @C=.5em @R=.65em  {
& \qw & \gate{\overline{\mathcal{X}}} & \qw & \qw \\
& \qw & \qw & \qw & \qw \\
& \qw & \qw & \qw & \qw \\
& \qw & \qw & \qw & \qw \\
&&&&&& &  &  &  & \\
& \qw & \qw & \qw & \qw \\
& \qw & \ctrl{6} & \qw & \qw \\
& \qw & \qw & \qw & \qw \\
& \qw & \qw & \qw & \qw \\
&&&&&\\
& \qw & \qw & \qw & \qw \\
& \qw & \qw & \qw & \qw \\
& \qw & \ctrl{6} & \qw & \qw \\
& \qw & \qw & \qw & \qw \\
&&&&&\\
& \qw & \qw & \qw & \qw \\
& \qw & \qw & \qw & \qw \\
& \qw & \qw & \qw & \qw \\
& \qw & \ctrl{-6} & \qw & \qw  \\
}
\end{align*}
\caption{Assume transversal~$\mathsf{CCCZ}$ implements the logical gate as shown containing a set of four logical~$\mathsf{CCCZ}$ gates. Then, by sandwiching a single logical~$X$ operator on a given logical qubit between rounds of transversal~$\mathsf{CCCZ}$ the resulting logical action is a targeted~$\mathsf{CCZ}$. Given any Pauli gate can be implemented transversally in a stabilizer code, the global action also remains transversal.}
\label{fig:4CCCZ}
\end{figure}
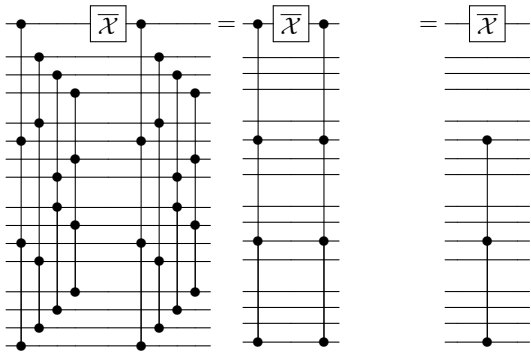

For example, as will be shown in this work, there will be a version of a 4D~toric code with periodic boundaries, encoding four logical qubits, that exhibits a transversal~$\mathsf{CCCZ}$. The resulting action of the gate will couple the four logical qubits by implementing four separate logical~$\mathsf{CCCZ}$ gates across the four codeblocks, see Fig.~\ref{fig:4CCCZ}. By sandwiching the appropriate logical~$X$ gate we can achieve a transversal implementation of a targeted~$\mathsf{CCZ}$ logical gate.

\subsection{Schl\"afli symbols}
\label{sec:Schlafli}

\begin{figure}[t]
\centering
\subfloat[$\{3\}$]{
\includegraphics[trim={15.25cm 10.75cm 15.25cm 10.75cm},clip,width = 0.2\linewidth]{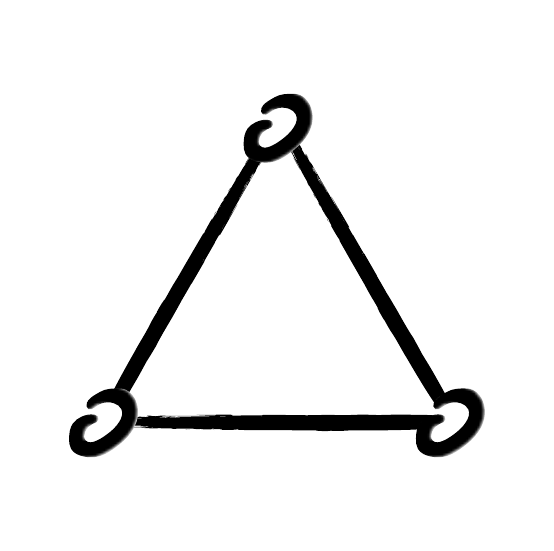}
\label{fig:Schlafli_3}}
\hfill
\subfloat[$\{4\}$]{
\includegraphics[trim={15.25cm 10.75cm 15.25cm 10.75cm},clip,width = 0.2\linewidth]{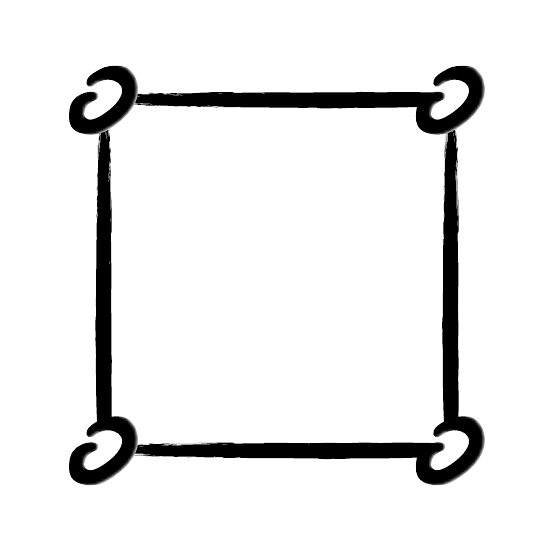}
\label{fig:Schlafli_4}}
\hfill
\subfloat[$\{6\}$]{
\includegraphics[trim={15.25cm 10.75cm 15.25cm 10.75cm},clip,width = 0.2\linewidth]{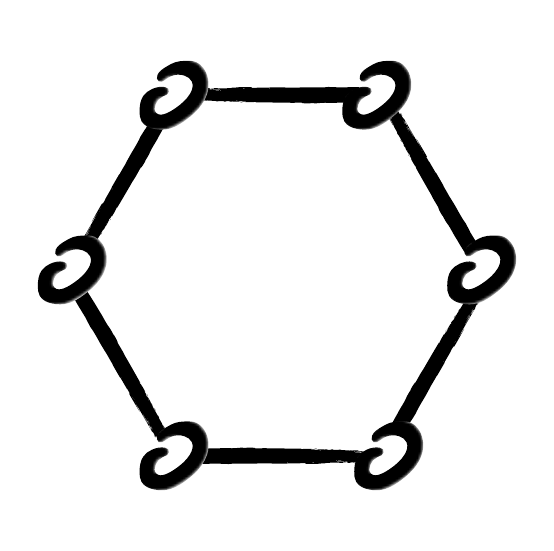}
\label{fig:Schlafli_6}}
\newline
\subfloat[$\{3,3\}$]{
\includegraphics[trim={15.25cm 10.75cm 15.25cm 10.75cm},clip,width = 0.2\linewidth]{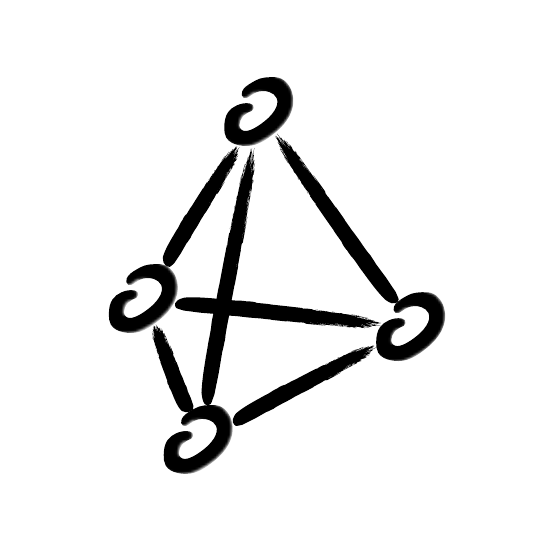}
\label{fig:Schlafli_33}}
\hfill
\subfloat[$\{3,4\}$]{
\includegraphics[trim={15.25cm 10.75cm 15.25cm 10.75cm},clip,width = 0.2\linewidth]{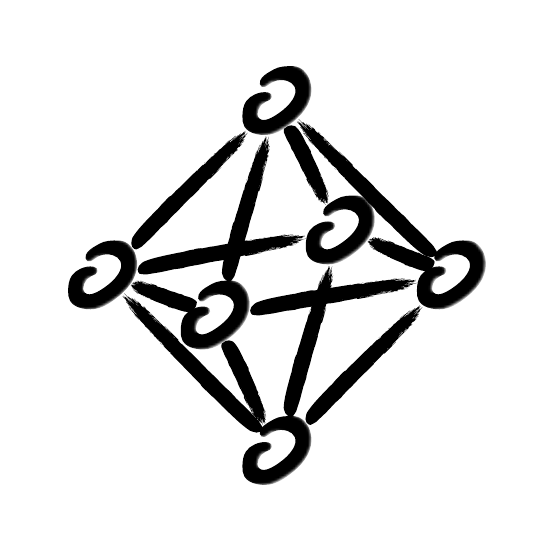}
\label{fig:Schlafli_34}}
\hfill
\subfloat[$\{4,3\}$]{
\includegraphics[trim={15.25cm 10.75cm 15.25cm 10.75cm},clip,width = 0.2\linewidth]{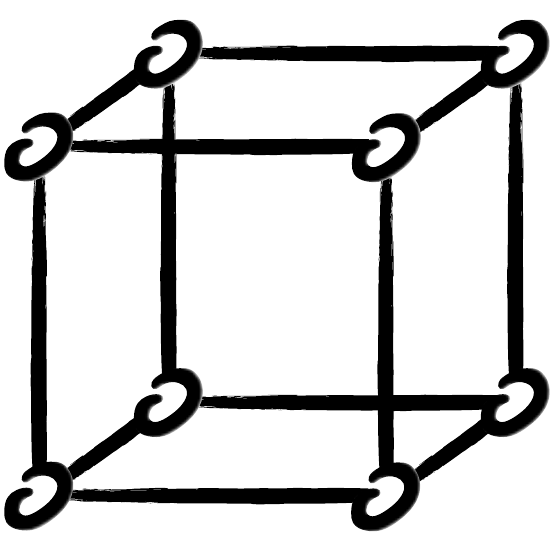}
\label{fig:Schlafli_43}}
\caption{Various polytopes and their associated \schlafli symbols. A \schlafli symbol~$\{ n \}$ represents an $n$-sided polygon, while polytopes with \schlafli symbols~$\{n,m\}$ have $m$ $n$-sided polygons around each of their vertices. }
\label{fig:Schlafli}
\end{figure}

In order to clarify the upcoming constructions to 4D~lattices, we will briefly describe Schl\"afli symbols and their relationship to regular tessellations. A Schl\"afli symbol is a succinct description of regular polytopes and tessellations that is defined recursively. Given an integer~$n$, the Schl\"afli symbol~$\{ n \}$ represents an $n$-sided regular convex polygon. A Schl\"afli symbol with two integer entries~$\{ n, m\}$ represents a geometric object with~$m$ symmetrically distributed objects~$\{n\}$ around each vertex. For example, a $\{3,3\}$ represents a tetrahedron as each vertex has 3~adjacent faces corresponding to equilateral triangles~$\{3\}$, see Fig.~\ref{fig:Schlafli} for further examples. 

\begin{figure}[t]
\centering
\subfloat[$\{4,4\}$]{
\includegraphics[trim={11.5cm 7cm 11.5cm 7cm},clip,width = 0.4\linewidth]{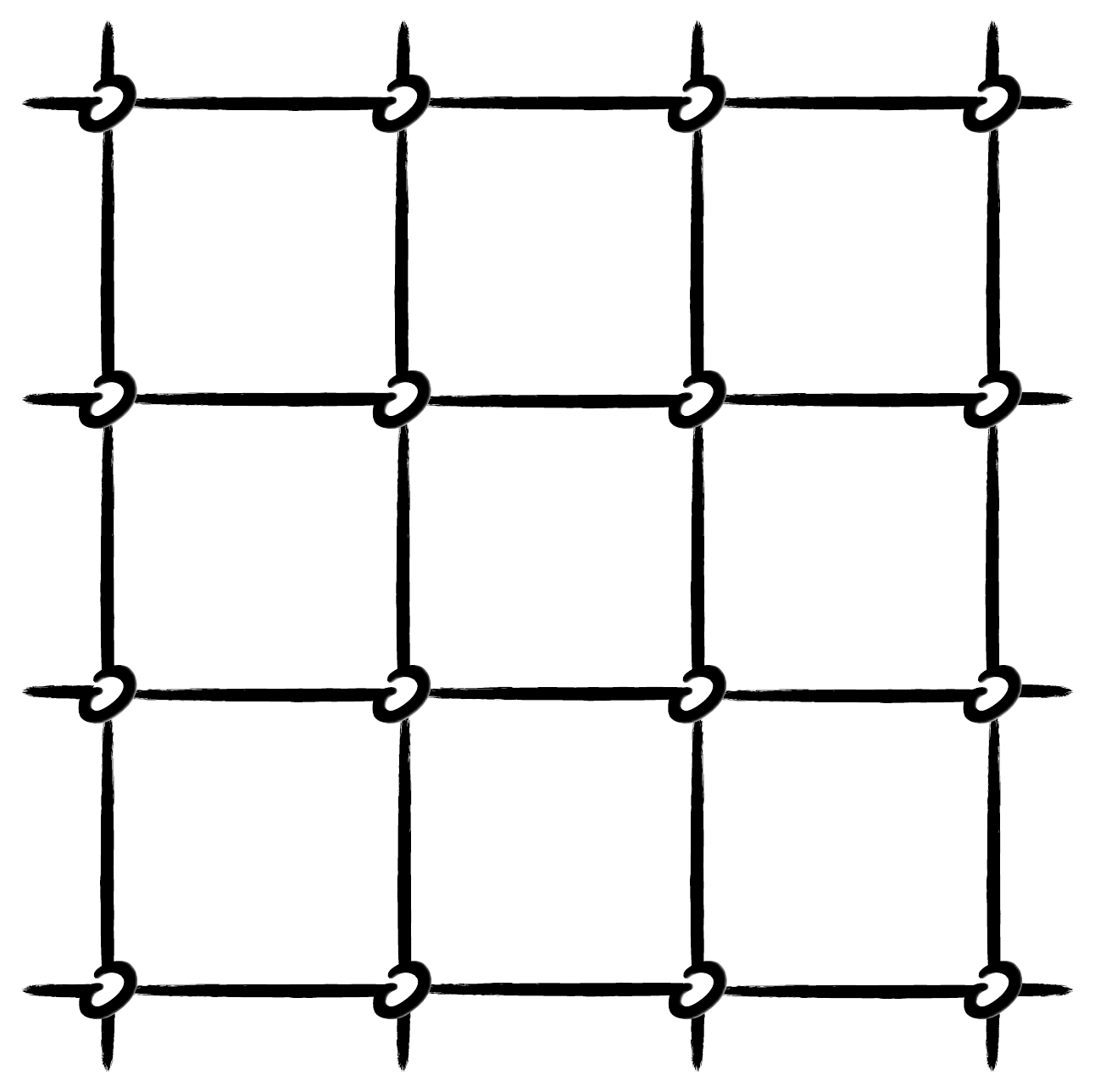}
\label{fig:Schlafli_44}}
\hfill
\subfloat[$\{6,3\}$]{
\includegraphics[trim={11.5cm 7cm 11.5cm 7cm},clip,width = 0.4\linewidth]{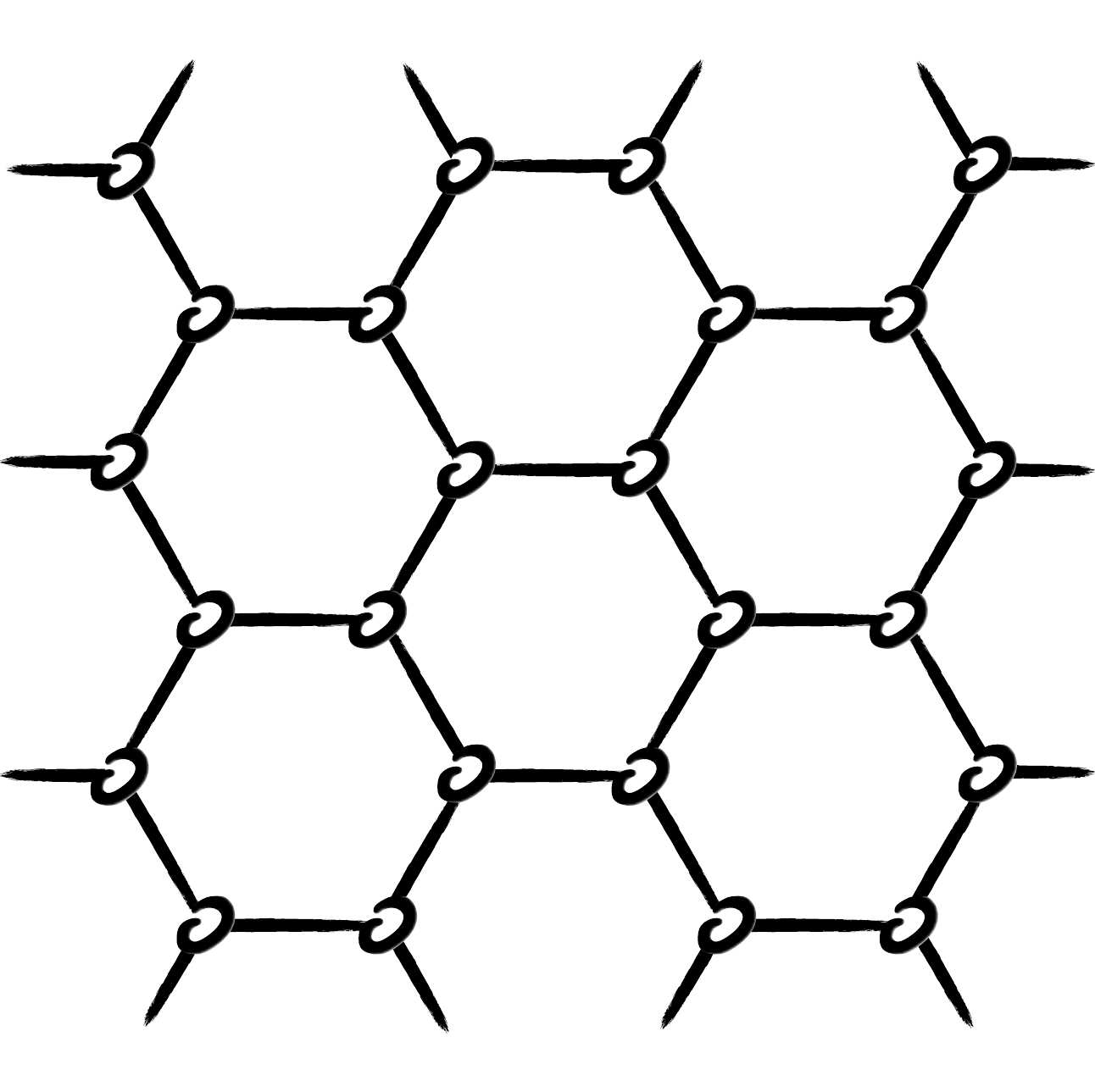}
\label{fig:Schlafli_63}}
\caption{Regular tessellations of two-dimensional Euclidean space given by \schlafli symbols~$\{n,m\}$. Each vertex is surrounded by $m$~$n$-sided polygons.}
\label{fig:Schlafli_2Dlat}
\end{figure}

While it is most straightforward to view Schl\"afli symbols~$\{ n, m\}$ as regular convex polytopes in three dimensions, they may also represent tessellations in two-dimensional Euclidean or hyperbolic space. For example, the Schl\"afli symbol~$\{4,4\}$ represents a square tessellation of 2D~space as every vertex is adjacent to 4~squares. The distinction between these two interpretations can either be made based on context or by use of the words cell/polytope and lattice/tessellation.

In a similar manner, a Schl\"afli symbol~$\{ n,m,l\}$ can represent a regular 4-cell or a tessellation of 3D~space, where adjacent to every edge (1-cell) are $l$~objects $\{n,m\}$. Importantly for our discussions, the cubic lattice in 3D is described by Schl\"afli symbol~$\{4,3,4\}$ as adjacent to every edge are 4~cubes ($\{4,3\}$). 

Finally, the full generalization of the recursive definition of the \schlafli symbol is as follows: an object with \schlafli symbol~$\{ r_1,\cdots, r_d\}$ has $r_d$~objects with \schlafli symbol~$\{r_1,\cdots,r_{d-1}\}$ adjacent to every $(d-2)$-cell.

We define the \emph{vertex operator} of a regular polytope/tessellation to be an object centered at a vertex whose $x$-cells are placed along $(x+1)$-cells of the original polytope tessellation. For example, the vertex operator of an octohedron is a square, as adjacent to every vertex are four edges
whom each share a face with two of the other aforementioned edges. A shorthand method for determining the vertex operator is again through the \schlafli symbol, as the vertex operator of an object $\{ r_1,\cdots, r_d\}$ is an object whose \schlafli symbol comes from removing the first entry:~$\{ r_2, \cdots, r_d \}$. We shall occasionally also refer to an \emph{edge operator} which is just the vertex operator of a vertex operator whose \schlafli symbol is determined by removing the first two entries.

\subsection{2D toric code}
\label{sec:2DTC}

Consider a square lattice $\{4,4\}$ with periodic boundary conditions in both spatial dimensions. The 2D toric code is defined by placing physical qubits on the edges of the graph and defining the $Z$~stabilizers at the vertices and the $X$~stabilizers on the plaquettes\footnote{The labelling of the $X$ and $Z$~stabilizers are reversed from the traditional definition of the 2D~toric code, as will become evident later in this work. The choice however is equivalent.}. 

\begin{figure}[t]
\centering
\subfloat[$X$~stabilizers are defined on plaquettes (blue) while $Z$~stabilizers are defined on vertices (yellow).]{
\includegraphics[trim={8cm 5cm 8cm 5cm},clip,width = 0.45\linewidth]{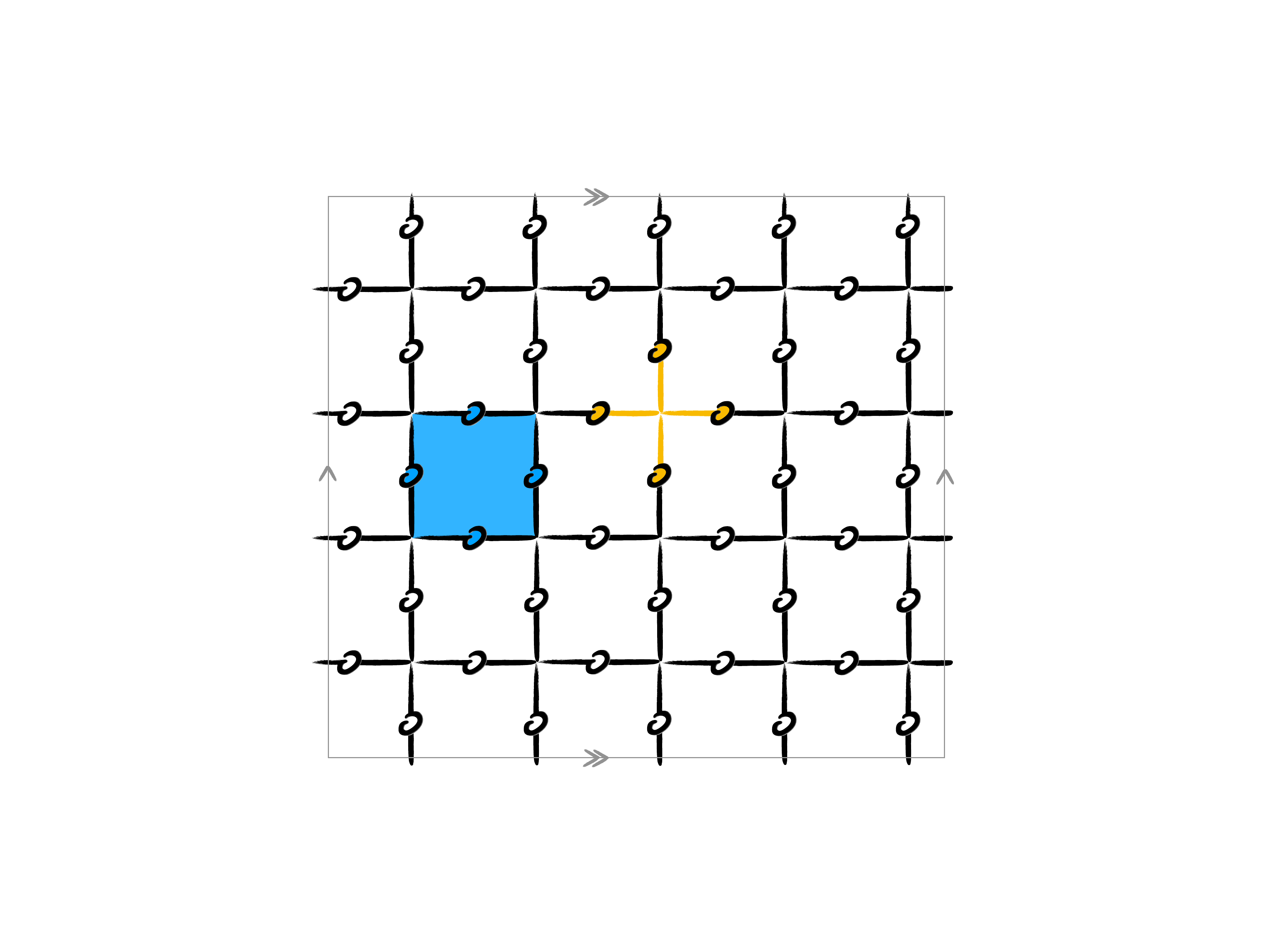}
\label{fig:2DTC}}
\hfill
\subfloat[Equivalent description of the 2D~toric code where plaquettes and vertex operators have mirroring descriptions.]{
\includegraphics[trim={8cm 5cm 8cm 5cm},clip,width = 0.45\linewidth]{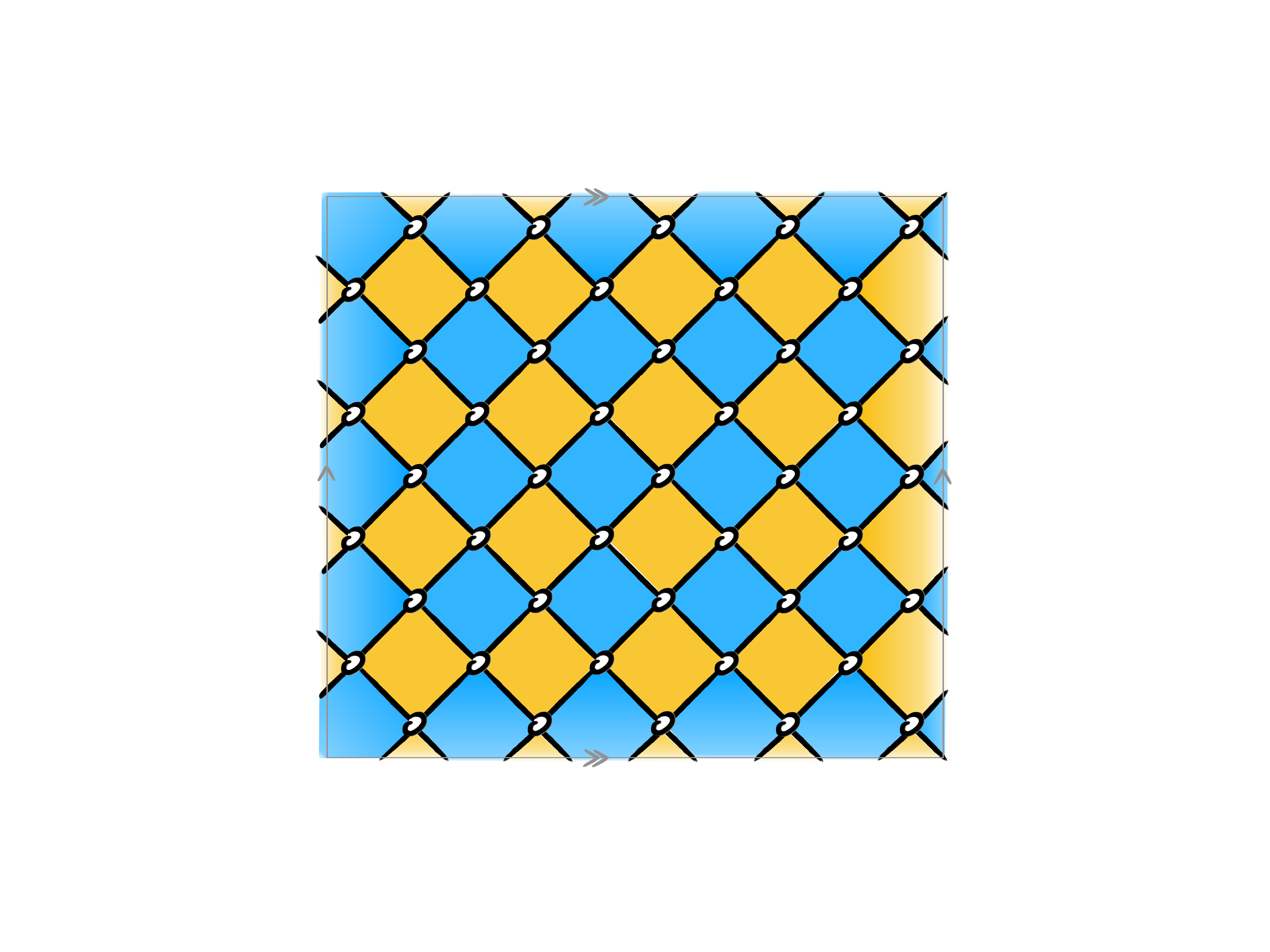}
\label{fig:2DTC_vertexfigs}}
\caption{2D~toric code with periodic boundary conditions.}
\label{fig:3D_2coloring}
\end{figure}

Given that the stabilizer code is CSS, we can treat any set of errors by considering separately the individual $X$ and $Z$ components of the error (as the measurement of the stabilizers will project the error onto distinct sets of $X$ and $Z$~errors). Any Pauli string of $X/Z$ errors that does not form a closed loop will violate the stabilizers at the end points of the error string. As such, we say that the errors result in point-like excitations in the 2D toric code, and the fact they come in pairs as a consequence of the underlying $\mathbb{Z}_2$~symmetry. Any closed loop of $X/Z$ errors will commute with the stabilizers and thus return a state in the codespace. If the closed loop is contractible, then it is the product of all stabilizers within the loop, while if it is non-contractible it forms a logical operator.

We can obtain a transversal controlled-$Z$~($\mathsf{CZ}$) gate between two layers of the 2D toric code if the second copy has swapped the locations of the $X$ and $Z$~stabilizers. It is then straightforward to verify the conditions of section~\ref{sec:TransversalConditions} as the $X$~stabilizers will clearly overlap an even number of times by construction. Moreover, given the switching of the $X$ and $Z$~logical operators between the two code copies, it is clear that logical~$X$ from one codeblock will be mapped onto logical~$Z$ on the other codeblock, implementing a logical~$\mathsf{CZ}$.\footnote{Note, that for \emph{any} CSS code we can take a second copy and reverse the roles of $X/Z$ to obtain a transversal $\mathsf{CZ}$.}

\subsection{3D toric code}
\label{sec:3DTC}

\begin{figure}[t]
\centering
\subfloat[$X$~stabilizers of two codeblocks]{
\includegraphics[trim={11cm 6.5cm 11cm 6.5cm},clip,width = 0.45\linewidth]{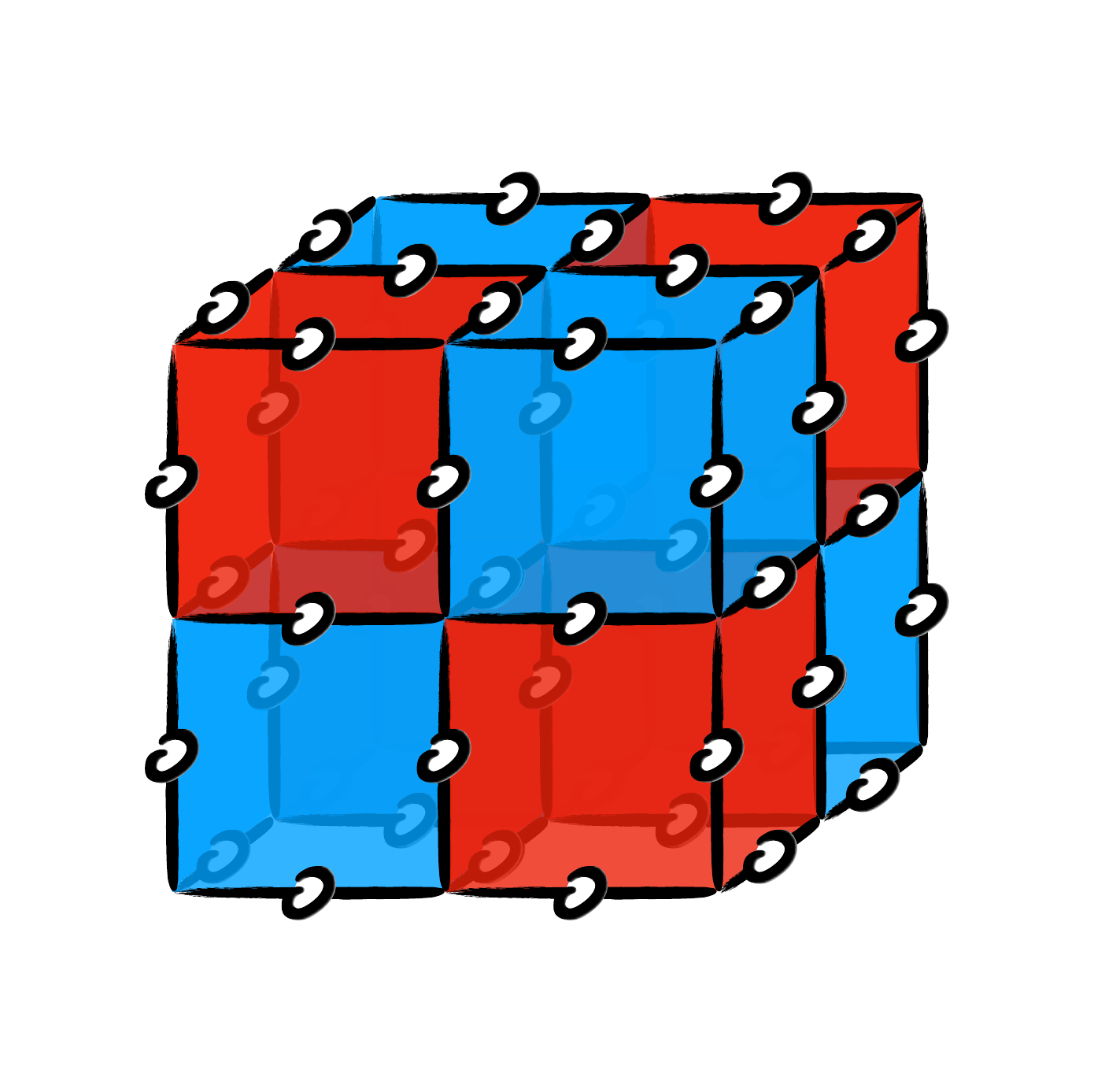}
\label{fig:3DTC}}
\hfill
\subfloat[$X$~stabilizers of all three codeblocks]{
\includegraphics[trim={11cm 6.5cm 11cm 6.5cm},clip,width = 0.45\linewidth]{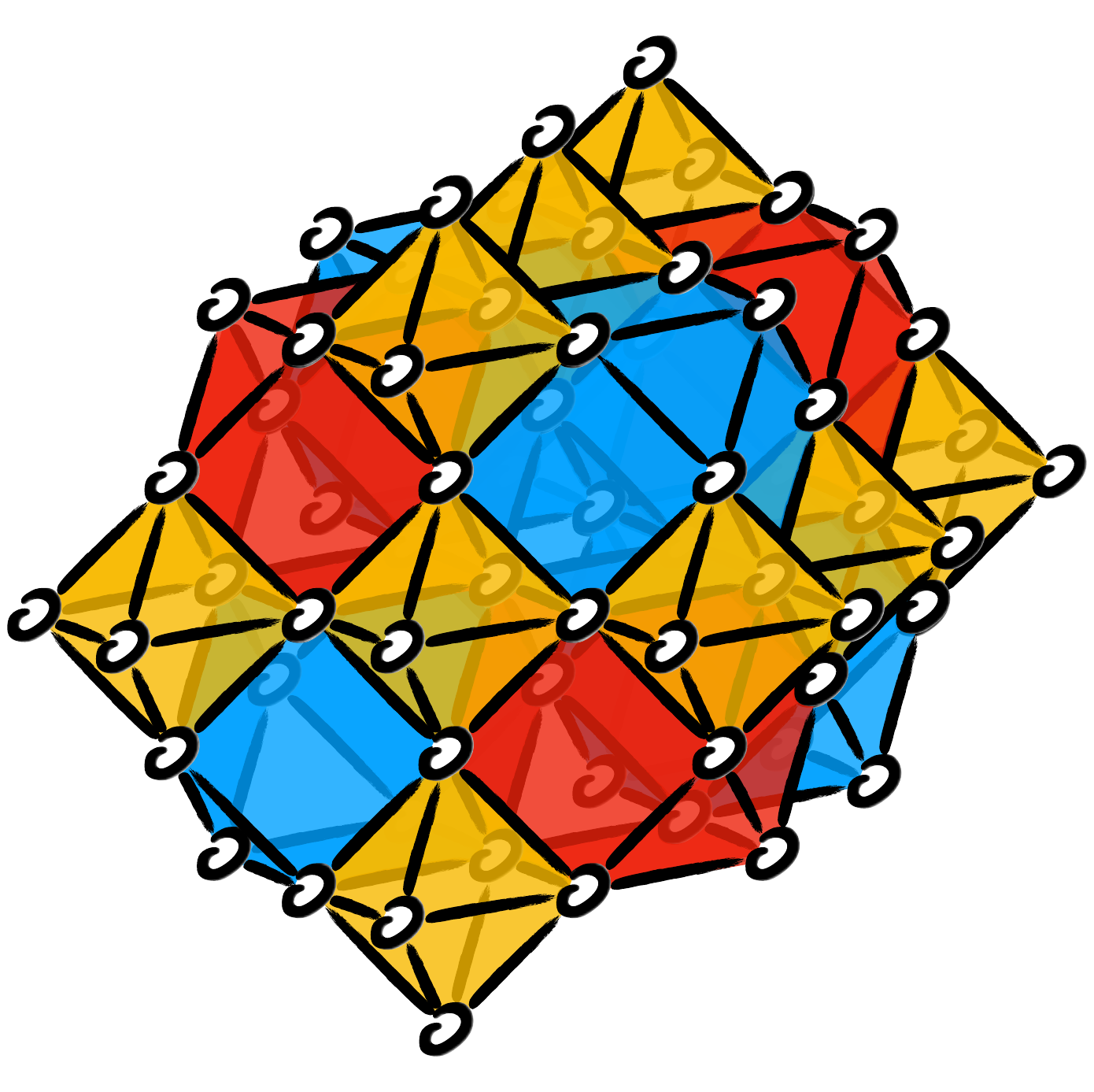}
\label{fig:3DTC_vertexfigs}}
\hfill
\subfloat[Intersection of two planar logical operators]{
\includegraphics[trim={11cm 5.65cm 11cm 5.7cm},clip,width = 0.45\linewidth]{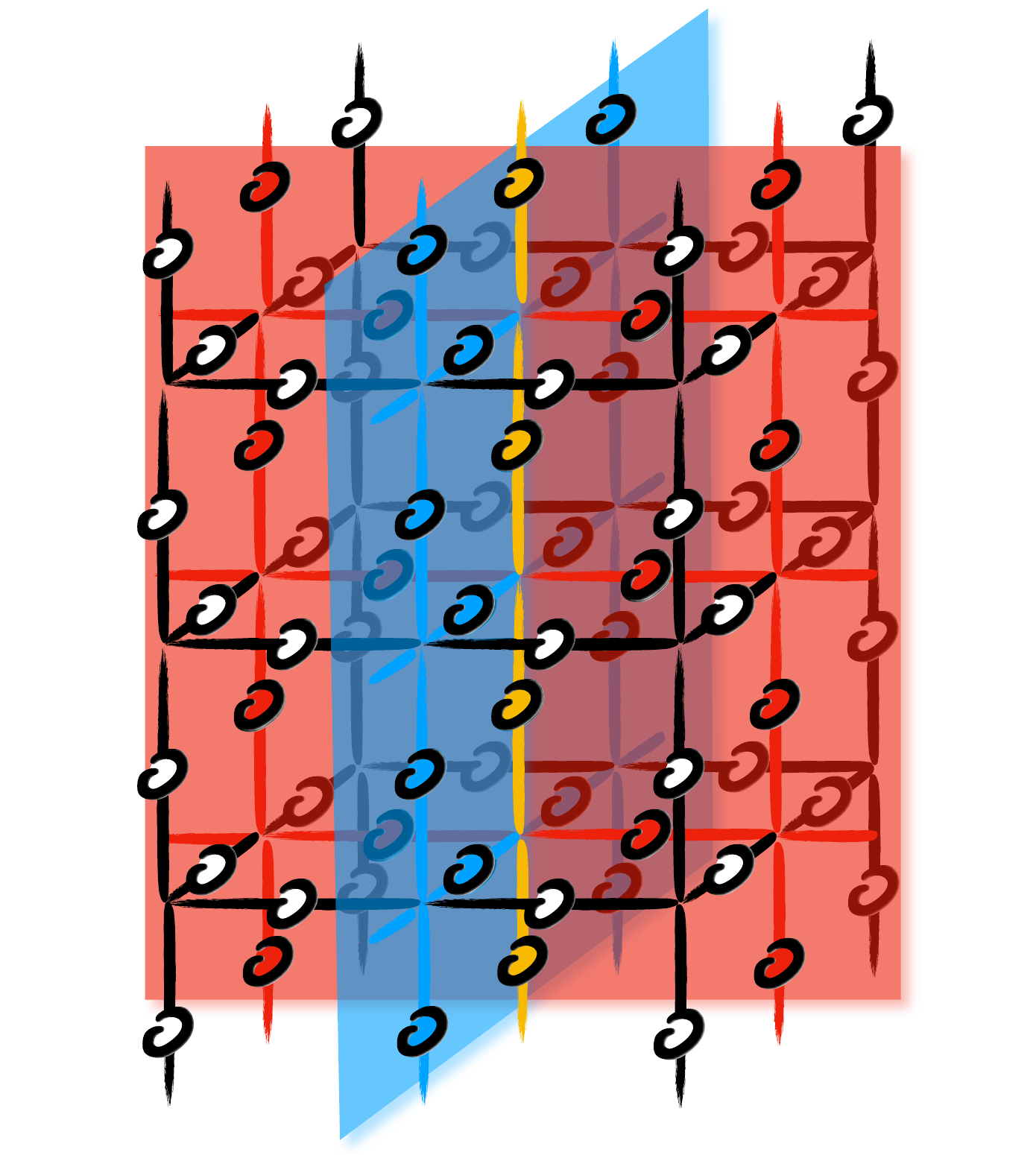}
\label{fig:3DTC_logOps}}
\hfill
\subfloat[Dual lattice where $X$~stabilizers from red and blue codeblocks are vertices]{
\includegraphics[trim={11cm 6.5cm 11cm 6.5cm},clip,width = 0.45\linewidth]{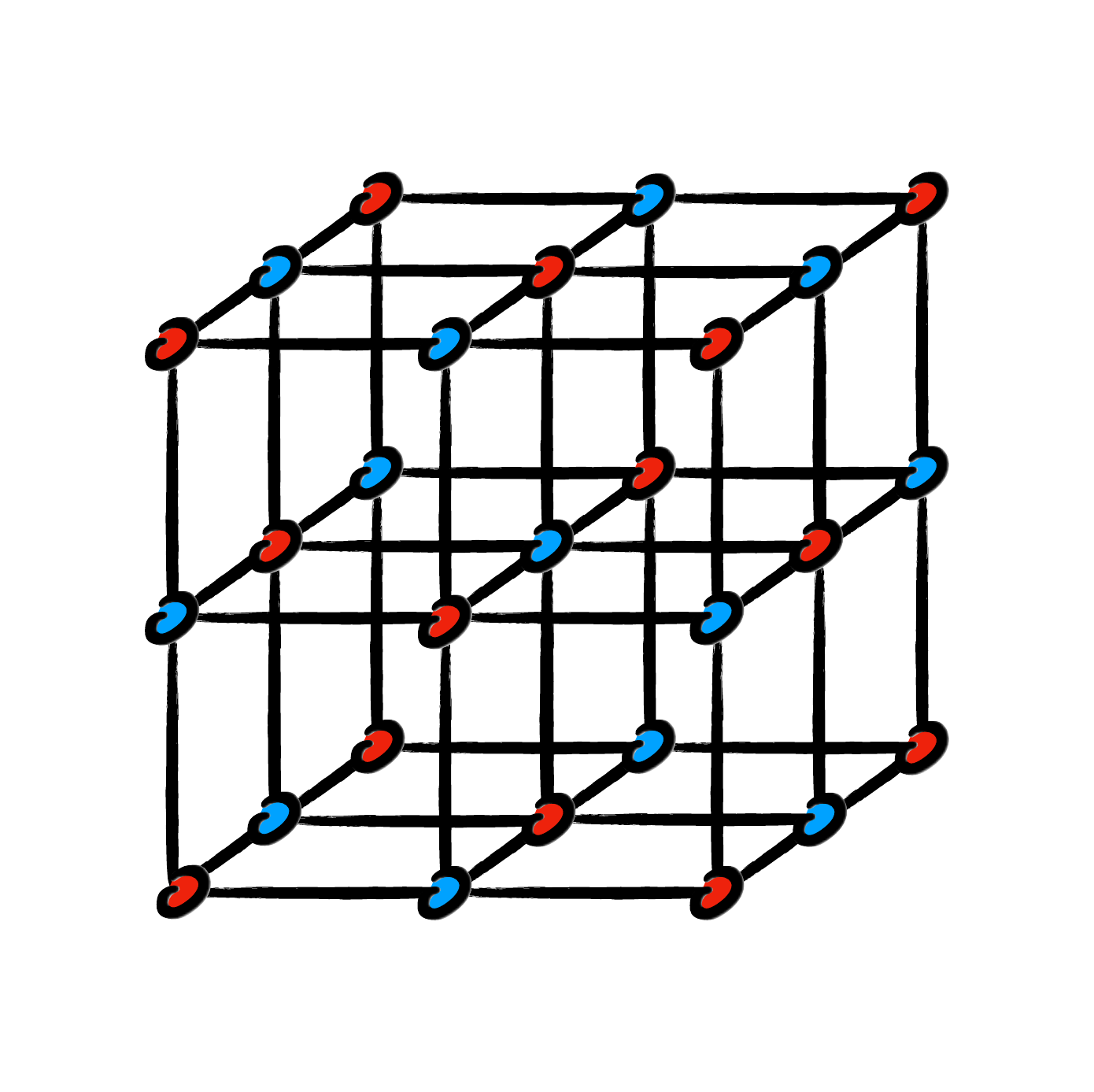}
\label{fig:434colored}}
\caption{Different descriptions of the 3D~toric code. (a)~2-coloring of the 3-cells where qubits reside on edges. $X$~stabilizers of the red (blue) code correspond to qubits supported on individual red (blue) 3-cells. (b)~Addition of $X$~stabilizers from third codeblock, labelled codeblock 0, which corresponds to edges sharing the same vertex. (c)~$X$~logical operators for first two codeblocks whose intersection forms a 1D~closed loop corresponding to the $Z$~logical operator of codeblock~0. (d)~Dual lattice, where physical qubits will reside on 2-cells (faces) and $X$~stabilizers from first two codeblocks will be represented by vertices. Support of the $X$~stabilizers is given by all faces sharing a given vertex.}
\label{fig:3D_2coloring}
\end{figure}

In this subsection we review the transversal~$\mathsf{CCZ}$ gate for the 3D~toric code due to Vasmer and Browne~\cite{VB19}. Consider the cubic lattice $\{4,3,4\}$ with qubits residing on edges of the lattice. There are three qubits per edge, one for each codeblock. We can then color the cubes in the lattice in two colors, say red and blue, such that cubes of the same color do not share a face (but can share an edge), see Fig.~\ref{fig:3DTC}. Again, we will focus on the case of periodic boundary conditions in all three spatial dimensions, this condition can be relaxed with an appropriate choice of boundaries.

The first codeblock is defined to have $X$~stabilizers supported on the red cubes, while the second codeblock is defined to have $X$~stabilizers supported on the blue cubes. These are weight-12 stabilizers because there are 12~edges to a cube. The third codeblock has weight-6 $X$~stabilizers defined by the vertices of the lattice, with the stabilizer supported on neighboring edges, as shown in Fig.~\ref{fig:3DTC_vertexfigs}. 

Any $Z$~error will give rise to a pair of violated $X$ stabilizers, as in the 2D~toric code. As such, a non-contractible loop spanning the lattice forms a logical~$Z$ operator. There are three independent logical operators for the 3D~toric code defined on a periodic lattice, one for each dimension. The $X$~logical operator is formed from a 2D~plane orthogonal to its conjugate $Z$~logical string pair. See Fig.~\ref{fig:3DTC_logOps} for a pictorial representation of planar $X$~logical operators for the red and blue codeblocks, whose intersection forms a non-contractible loop whose support is that of a $Z$~logical operator for the yellow codeblock.

\begin{figure}[t]
\centering
\subfloat[Codeblock 1]{
\includegraphics[trim={11cm 6.5cm 11cm 6.5cm},clip,width = 0.45\linewidth]{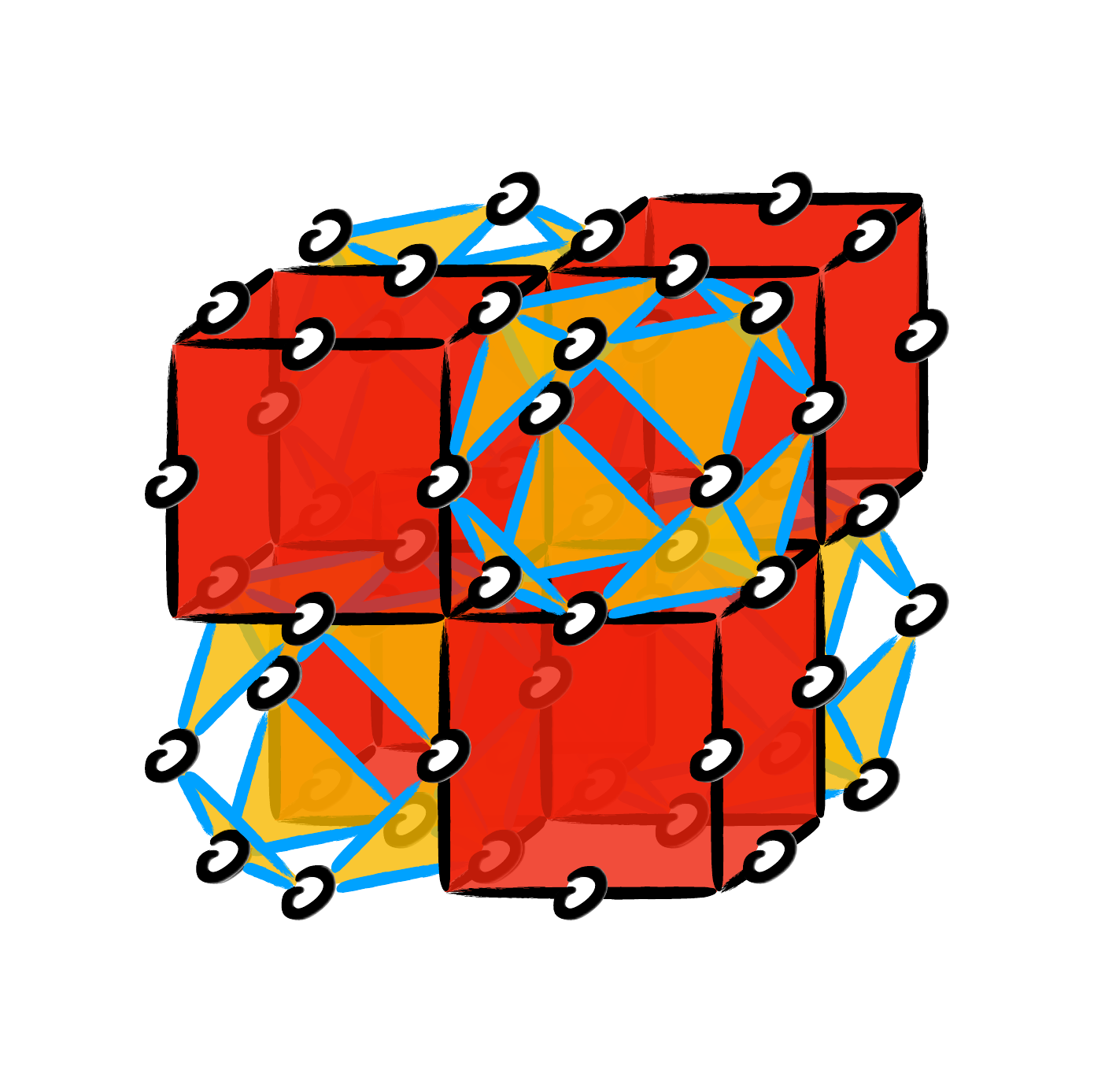}
\label{fig:3DTC_Zstab1}}
\hfill
\subfloat[Codeblock 0]{
\includegraphics[trim={11cm 6.5cm 11cm 6.5cm},clip,width = 0.45\linewidth]{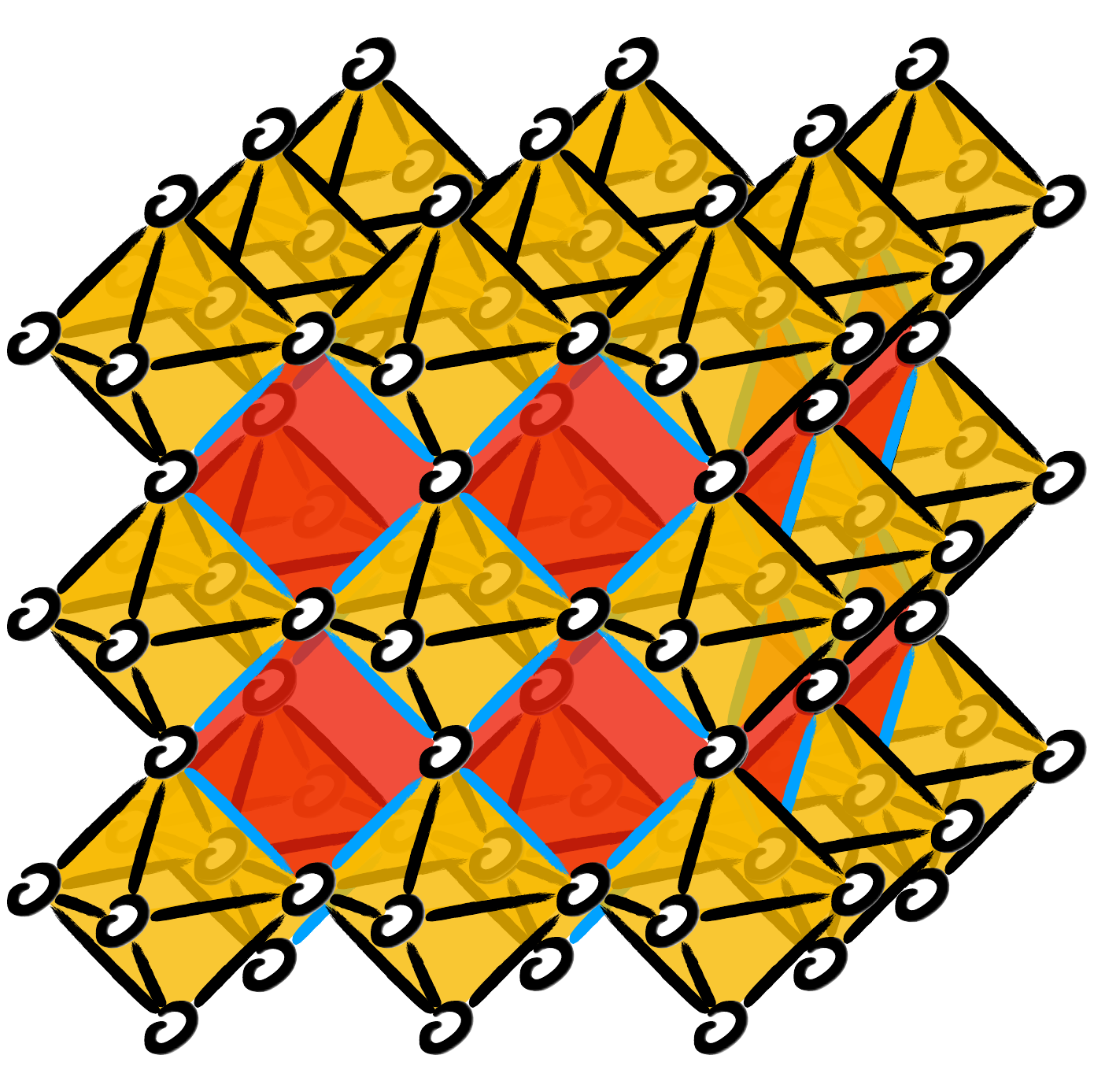}
\label{fig:3DTC_Zstab3}}
\caption{$X$~stabilizers and $Z$~stabilizers of the 3D toric code. (a)~Codeblock where the $X$~stabilizers are given by red 3-cells, while the $Z$~stabilizers are in intersection of the compliment set of 3-cells with the vertex operators, given by yellow faces with blue edges. (b)~Codeblock where the $X$~stabilizers are given by yellow vertex operators. The $Z$~stabilizers are faces of the intersection of two complimentary 3-cells, shown as red faces with blue edges.}
\label{fig:3DTC_Zstab}
\end{figure}

We can then straightforwardly verify the orthogonality conditions, presented in Sec.~\ref{sec:TransversalConditions}, for the existence of a transversal $\mathsf{CCZ}$~gate. The intersection of $X$~stabilizers from the first two codeblocks corresponds to a weight-4 face belonging to the cubes, such a face will intersect any neighboring vertex operator at the two corresponding adjacent edges of that vertex belonging to the face. Therefore, the intersection of $X$~stabilizers from the three different codeblocks will either be trivial or weight~2. The $Z$~stabilizers of a codeblock can then be given by all possible intersections of pairs of $X$~stabilizers from the other two codeblocks, see Fig.~\ref{fig:3DTC_Zstab}.

The transversal~$\mathsf{CCZ}$ gate is a logical~$\mathsf{CCZ}$ gate as described above the intersection of pairs of logical~$X$ operators from different codeblocks corresponds to the support of the~$Z$ logical operator of the third codeblock, as required.

\subsection{The dual picture}
\label{sec:DualCond}

The dual of a lattice of dimension~$D$ is again a $D$-dimensional lattice, where every $D$-cell is replaced by a vertex (0-cell), every $(D-1)$-cell is replaced by an edge (1-cell) connecting two vertices when the corresponding original $D$-cells share the $(D-1)$-cell, and so forth. Conveniently, the dual to any object with Schl\"afli symbol~$\{ r_1,\cdots, r_d\}$ is the object with \schlafli symbol~$\{r_d, \cdots, r_1\}$. That is, one reverses the ordering of the integers in the symbol. For example, the dual of a cube~$\{4,3\}$ is an octahedron~$\{3,4\}$ and the cubic lattice~$\{4,3,4\}$ is self-dual. 

In the original 2D~toric code picture, the $X$~stabilizers are associated to faces, while the $Z$~stabilizers are associated to vertices. The dual of the square lattice is also a square lattice, with faces and vertices interchanged. Therefore, to describe the 2D-toric code in the dual picture, qubits are still placed on edges, but $X$~stabilizers are instead associated to vertices and $Z$~stabilizers to faces.

The 3D-toric codes on the cubic lattice can also be described in the dual picture. The dual of the cubic lattice is again the cubic lattice, but qubits now reside on faces rather than edges. Moreover, $X$ stabilizers of the first two codeblocks are associated with vertices in the dual lattice. The choice of $X$~stabilizers for the two codeblocks is equivalent to a 2-coloring of the vertices in the dual lattice, where vertices of the same color share only faces (never an edge), as shown in Fig.~\ref{fig:434colored}. The $X$~stabilizers of the third codeblock (originally the vertex operators) are associated with cubes in the dual lattice. The condition, see Eq.~\ref{eq:IdentityReq3a}, that $X$~stabilizers from the three codeblocks only ever intersect at an even number of qubits can be viewed in the dual picture as any edge (intersection between two vertices corresponding to $X$~stabilizers of codeblocks 1 \& 2) sharing only 2~faces with a given cube.

An additional important property from the 2D and 3D~toric codes is that the $Z$~logical operator is a non-contractible 1D~loop. From the view point of excitations, this arises due to excitations coming in pairs. When a single or connected string of $Z$~errors occur, the only violated syndromes are those at the endpoints of the string, and thus closing the string annihilates the excitations and forms a logical operator. The logical operator is non-trivial (not the product of stabilizers, therefore not the identity operator) if it forms a non-contractible loop. Essential to this reasoning is that $Z$~errors lead to a pair of violated $X$~syndromes, in all codeblocks. Therefore, in the dual picture, any qubit (edge in 2D, face in 3D) includes exactly two vertices of the any one color. In 2D this condition is satisfied trivially, however in 3D it demands that faces must be squares with opposite corners colored the same.

Therefore, in our quest for regular tessellations in $D$-dimensions that yield interesting multi-controlled-$Z$ gates, we propose the following criteria:
\begin{enumerate}
\item Qubits are placed on edges of the underlying graph. Conversely, qubits are placed on $(D-1)$-cells in the dual lattice.
\item The vertices of the dual lattice are $(D-1)$-colorable, such that any two vertices of the same color share at most a $(D-1)$-cell
\item Every $(D-1)$-cell in the dual lattice has two vertices of each of the $(D-1)$ colors.
\end{enumerate}

In this dual-lattice description, qubits correspond to $(D-1)$-cells. The $X$~stabilizers of codeblocks~$1, \hdots, D-1$ correspond to vertices of the corresponding colors, $1,\hdots, D-1$. That is, given a vertex, the corresponding $X$-stabilizer is supported on qubits at all $(D-1)$-cells that contain the vertex. Codeblock~0 is different from the rest in that its $X$~stabilizers are defined by the $D$-cells in the dual lattice, where each $X$~stabilizer is supported on the set of $(D-1)$-cells belonging to a given $D$-cell.

\section{4D code with transversal $\mathsf{CCCZ}$}
\label{sec:4DTC}

\begin{figure}[t]
\centering
\subfloat[3-coloring of an octahedron]{
\includegraphics[trim={15.25cm 10.75cm 15.25cm 10.75cm},clip,width = 0.45\linewidth]{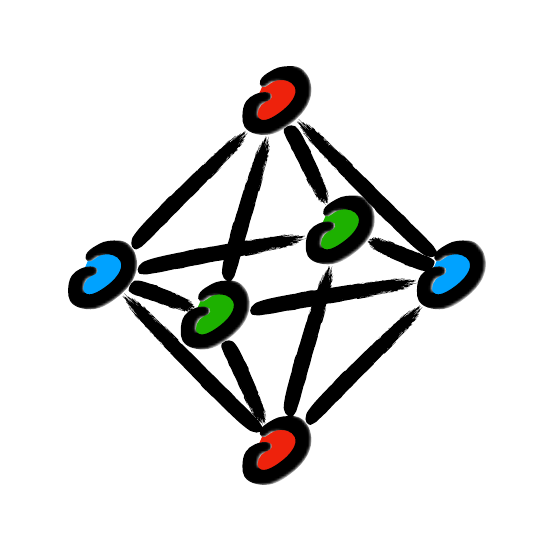}
\label{fig:34coloring}}
\hfill
\subfloat[3-coloring of an octaplex]{
\includegraphics[trim={10.5cm 6cm 10.5cm 6cm},clip,width = 0.45\linewidth]{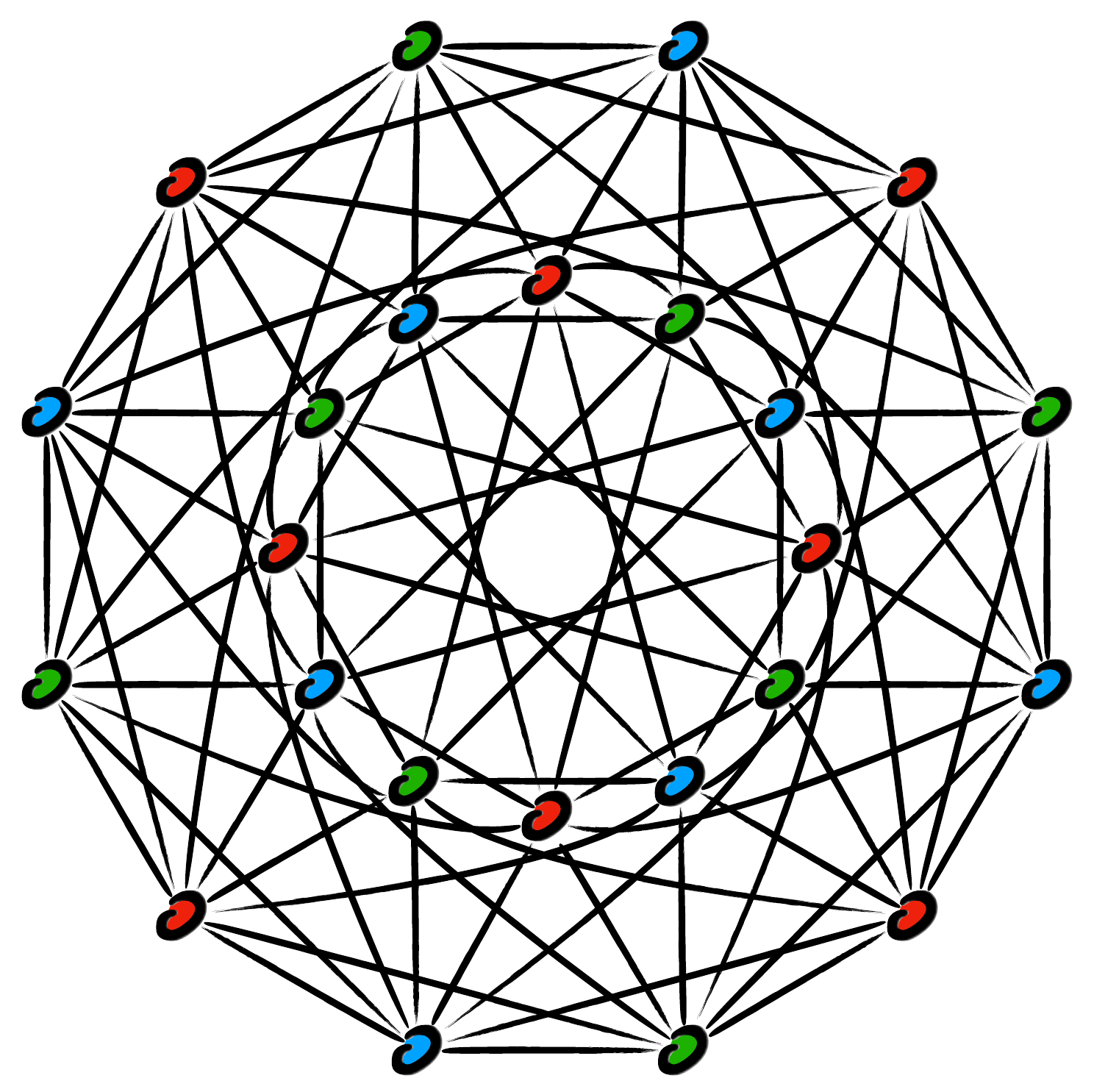}
\label{fig:24cell}}
\caption{The octaplex, \schlafli symbol~$\{3,4,3\}$, is composed of 24~vertices which can be 3-colored. The vertices in turn compose the 24~octahedra (3-cells), where each edge is surrounded by 3~octahedra by definition.}
\label{fig:3coloring_octaplex}
\end{figure}

We seek a tessellation of 4D~space with the required desiderata laid out in the previous section. We consider the dual lattice, where $X$~stabilizers from the first three codeblocks are given by vertices. Since the lattice should be 3-colorable, such that the 3-cells have two vertices of each color, we require the underlying lattice to be composed of octahedra. Therefore, the \schlafli symbol of the tessellation should read:~$\{3,4,a,b\}$, where $a$ and $b$ are integer degrees of freedom. There is only one such regular tessellation in Euclidean 4-space, the octaplex tessellation\footnote{Other equivalent names include 24-cell honeycomb and icositetrachoric honeycomb.}: $\{ 3,4,3,3 \}$. In fact, the only other regular tessellations in Euclidean 4-space are its dual, the hexadecachoron tessellation\footnote{Also can be called the 16-cell honeycomb.}:~$\{3,3,4,3\}$ and the tesseractic tessellation~$\{4,3,3,4\}$. In Fig.~\ref{fig:3coloring_octaplex} we present the 3-coloring of the octahedron and the octaplex~$\{3,4,3\}$ which form the unit cell for the octaplex tessellation. It should be noted that this lattice was theorized as a potential candidate for the implementation of 4D~$\mathsf{CCCZ}$ using the theory of Coxeter diagrams~\cite{VasmerThesis}, which share many of the features of \schlafli symbols.

\subsection{Stabilizer weights}
\label{sec:SchlafliWeights}

The $X$~stabilizers of codeblock 0 are represented by 4-cells in the 4D~tessellation (recall we are working with the dual lattice). That is, each $X$~stabilizer is supported on all 3-cells contained within a 4-cell. By definition, the 4D tessellation is composed of octaplexes $\{3,4,3\}$. Now, since each octaplex is self-dual, the number of 3-cells within a given octaplex, and the weight of any $X$ stabilizer, is equal to its number of vertices:~24. In order to determine the weight of the $Z$~stabilzers, we must consider the common intersection of $X$~stabilizers from each of the other codeblocks. Each $X$~stabilizer from the other codeblocks is represented by a different colored vertex and they will only commonly intersect if they form a triangular face in the tessellation. The weight of the $Z$~stabilizer will thus be the number of 3-cells which contain a given face. This can be determined by recursively taking the vertex figure three times.\footnote{Each time one takes a vertex figure, the objects drop in dimension by one. For example, when taking the vertex figure, edges become vertices, faces become edges, etc.. Thus, applying it a second time, faces become vertices, 3-cells become edges, etc.} The resulting geometrical object will have \schlafli symbol~$\{3\}$, a triangle, and as such each face is adjacent to three 3-cells (vertices in the triangle). Therefore, the $Z$~stabilizers will have weight~3.

By construction, codeblocks~1--3 have $X$~stabilizers that are represented by each of the three colors of vertices in the octaplex tessellation. The support of each $X$~stabilizer is the set of 3-cells containing the given vertex. As such, in order to calculate the weight of the $X$~stabilizers we can consider the vertex~figure:~$\{4,3,3 \}$, a tesseract. Given the tesseract has 24~faces and each face in the vertex figure represents a 3-cell containing a given vertex, each of the $X$~stabilizers from codeblocks~1--3 are each of weight~24. The $Z$~stabilizers will be the intersection of an edge (containing two colored vertices) and a 4-cell. That is, we must determine how many 3-cells support a given edge within a given 4-cell. Again this can simply be read off from the \schlafli symbol, as by definition a~$\{3,4,3\}$ is the geometric object such that 3~octahedra ($\{3,4\}$) surround each edge. Therefore, the $Z$~stabilizers for codeblocks 1--3 also have weight~3.

By construction, the fact that codeblock~0 and codeblocks 1--3 each are composed of $X$~stabilizers of weight 24 and $Z$~stabilizer of weight 3 would appear to be coincidental, yet as will become evident in the description in the following subsection and subsection~\ref{sec:Equivalent}, there is indeed an additional symmetry in the octaplex tessellation that implies that all four codeblocks are equivalent codes.

\subsection{Coordinate system for the octaplex tessellation}
\label{sec:Coordinates}

We now give an explicit construction of the octaplex tessellation. First, take the tesseractic tessellation~$\{ 4,3,3,4\}$ on a periodic lattice of size~$L$ with integer vertices~$(x,y,z,w) \in \mathbb{Z}_L^4$ and tesseracts centered at half-integer coordinates~$(x+\tfrac{1}{2},y+\tfrac{1}{2},z+\tfrac{1}{2},w+\tfrac{1}{2}) \in (\mathbb{Z}_L+\tfrac{1}{2})^4$. The vertices of the \emph{octaplex tessellation} can be identified with the faces of the tesseractic tessellation, that is coordinates~$(x,y,z,w)$ such that two are integers, and two are half-integers\footnote{Such faces correspond to the intersection of 4 tesseracts.}. Vertices share edges if and only if they are distance~$1/\sqrt{2}$ apart in 2-norm. We can then label all of the vertices of the octaplex tessellation as one of three colors:
\begin{subequations}
\begin{align}
\mathcal{V}_r &= \{ (x,y,z+\tfrac{1}{2},w+\tfrac{1}{2}) | \ x,y,z,w \in \mathbb{Z}_L^4\} \nonumber  \\
\label{eq:Vertices1}
&\hspace{1em} \cup \{ (x+\tfrac{1}{2},y+\tfrac{1}{2},z,w) | \ x,y, z,w \in \mathbb{Z}_L^4 \} \\
\mathcal{V}_g &= \{ (x,y+\tfrac{1}{2},z,w+\tfrac{1}{2}) | \ x,y,z,w \in \mathbb{Z}_L^4\} \nonumber \\
&\hspace{1em} \cup \{ (x+\tfrac{1}{2},y,z+\tfrac{1}{2},w) | \ x,y, z,w \in \mathbb{Z}_L^4 \} \\
\mathcal{V}_b&= \{ (x,y+\tfrac{1}{2},z+\tfrac{1}{2},w) | \ x,y,z,w \in \mathbb{Z}_L^4\}  \nonumber \\
\label{eq:Vertices3}
&\hspace{1em} \cup \{ (x+\tfrac{1}{2},y,z,w+\tfrac{1}{2}) | \ x,y, z,w \in \mathbb{Z}_L^4 \} 
\end{align}
\end{subequations}
It is straightforward to verify that no two vertices of the same color will share an edge as they will be at least distance~$1$ away from one another in 2-norm.

The 4-cells of the octaplex tessellation are given by coordinates $(x,y,z,w) \in \mathbb{Z}_L^4$ or $(x+\frac12,y+\frac12,z+\frac12,w+\frac12) \in (\mathbb{Z}_L+\tfrac{1}{2})^4$ (that is the same coordinates as the vertices and centers of the hypercubes of the original tesseractic tessellation) and form 24-body objects called octaplexes. The vertices belonging to a octaplex centered at coordinates~$(x,y,z,w)$ will be all vertices distance~$\tfrac{1}{\sqrt{2}}$ in 2-norm from the corresponding center of the octaplex. For example, for the octaplex centered at the origin, all neighboring vertices will be the set of points~$(\pm1/2, \pm 1/2, 0, 0)$ with each $\pm$ taken independently and their permutations, thus totaling a set of 24 vertices. We label the set of 4-cells by~$\mathcal{O}$.

The 3-cells correspond to intersections of two neighboring 4-cells, which come in three different types: (3i) the intersection of two 4-cells with integer coordinates differing by $\pm 1$ in a single coordinate (the center of this type of 3-cell has three integer coordinates and one half-integer), (3ii) the intersection of two 4-cells with half-integer coordinates again differing by~$\pm 1$ in a single coordinate (centered at points with three half-integer and one integer coordinates), or (3iii) the intersection of 4-cells, one with integer and one with half-integer coordinates, differing by~$\pm \tfrac{1}{2}$ in each coordinate (centered at points with four quarter-integer coordinates, that is, points whose entries are odd multiples of $\tfrac{1}{4}$). 
Each 3-cell contains all vertices that are distance~$\tfrac{1}{2}$ away in 2-norm from its center. Given a 3-cell characterized of type (3i), centered say at $(x,y,z,w+\tfrac{1}{2})$ for integers $x,y,z,w$, there are 6 vertices belonging to the 3-cell: $(x\pm \tfrac{1}{2},y,z,w+\tfrac{1}{2})$, $(x,y\pm \tfrac{1}{2},z,w+\tfrac{1}{2})$, and $(x,y,z\pm \tfrac{1}{2},w+\tfrac{1}{2})$. Thus, the 3-cell is an octahedron with 3-coloring as required by Fig.~\ref{fig:34coloring}. A symmetric argument holds for 3-cells of type (3ii). For the 3-cells of type (3iii), the associated vertices are all~$\pm \tfrac{1}{4}$ in each coordinate such that two are integer and half-integer and thus there are $6=\binom{4}{2}$ such coordinates, again forming an octahedron with appropriate coloring as required. We label the set of 3-cells by~$\mathcal{Q}$. 

It will also be useful to characterize the 2-cells (faces) as they will defined the $Z$~stabilizers for codeblock~0. As described in Sec.~\ref{sec:SchlafliWeights}, the 2-faces are formed from the intersection of three neighboring 3-cells. In fact, each 2-cell will have one neighboring 3-cell of type (3i) or (3ii) and two of type~(3iii). Without loss of generality, consider the following 3-cell of type~(3i):~$(x+\tfrac{1}{2},y+\tfrac{1}{2},z+\tfrac{1}{2},w)$ and in particular the face whose vertices are:~$\{(x+\tfrac{1}{2},y+\tfrac{1}{2},z,w),(x+\tfrac{1}{2},y,z+\tfrac{1}{2},w),(x,y+\tfrac{1}{2},z+\tfrac{1}{2},w)\}$. Such a face cannot belong to a 3-cell of type~(3ii) as any face belonging to such a 3-cell will have to have one of the coordinates being fixed as a half-integer (rather than integer $w$ above). The neighboring 3-cells of type~(3iii) will be of the following form:~$(x+\tfrac{1}{4},y+\tfrac{1}{4},z+\tfrac{1}{4},w\pm\tfrac{1}{4})$. Therefore, we label such a 2-cell to be given by the set of coordinates:~$(x+\tfrac{1}{4},y+\tfrac{1}{4},z+\tfrac{1}{4},w)$\footnote{We note that the chosen labeling does not correspond to the geometric mean of the three vertices belonging to the face. Yet, such a labelling was chosen to simplify notation and can be self-consistent as explicitly discussed at the end of this section.}. Any 2-cell whose coordinates are composed of 3 quarter-integer and one integer coordinate will be denoted type (2i). Symmetrically, any 2-cell whose coordinates are composed of 3~quarter integer and 1 half-integer coordinate will be labelled as type~(2ii).

Finally, it is rather straightforward to verify that 1-cells will always have one integer and half-integer fixed among the vertices at their endpoints, and will alter between an integer and half-integer in the other two coordinates. As such, the 1-cells are all labelled by a set of coordinates composed of one integer, one half-integer and two quarter-integers.

To summarize, the geometric objects of the octaplex tessellation will be specified by the following forms of cartesian coordinates:
\begin{itemize}
\item 0-cells (vertices, $\mathcal{V}_r, \ \mathcal{V}_g, \ \mathcal{V}_b$): Two integer and two half-integer coordinates.
\item 1-cells: One integer, one half-integer, two quarter-integers.
\item 2-cells: (2i)~One integer and three quarter-integer coordinates. (2ii)~One half-integer and three quarter-integer coordinates.
\item 3-cells (physical qubits, $\mathcal{Q}$): (3i)~One integer and three half-integer coordinates. (3ii)~Three integer and one half-integer coordinates. (3iii)~Four quarter-integer coordinates. 
\item 4-cells~$\mathcal{O}$: (4i)~Four integer coordinates. (4ii)~Four half-integer coordinates.
\end{itemize}

An object with dimension $D$ is composed of several objects of dimension $D-1$. In this algebraic construction of the lattice, a $D$-dimensional located at point $P$ is composed of all $(D-1)$-dimensional objects nearest to $P$ in 2-norm. We provide more detail in each of the cases.
\begin{itemize}
\item A 1-cell at coordinate $(x,y+\frac12,z+\frac14,w+\frac14)$ is composed of two 0-cells at $(x,y+\frac12,z+\frac12,w)$ and at $(x,y+\frac12,z,w+\frac12)$.
\item (2i) A 2-cell at coordinate $(x,y+\frac14,z+\frac14,w+\frac14)$ is composed of three 1-cells at $(x,y+\frac12,z+\frac14,w+\frac14)$, $(x,y+\frac14,z+\frac12,w+\frac14)$, and $(x,y+\frac14,z+\frac14,w+\frac12)$. (2ii) A 2-cell at coordinate $(x+\frac12,y+\frac14,z+\frac14,w+\frac14)$ is composed of three 1-cells at $(x+\frac12,y,z+\frac14,w+\frac14)$, $(x+\frac12,y+\frac14,z,w+\frac14)$, and $(x+\frac12,y+\frac14,z+\frac14,w)$.
\item (3i) A 3-cell at $(x,y+\frac12,z+\frac12,w+\frac12)$ is composed of 8 2-cells of type 2i, $(x,y+\frac12\pm\frac14,z+\frac12\pm\frac14,w+\frac12\pm\frac14)$ where each $\pm$ sign can be chosen independently. (3ii) Likewise, a 3-cell at $(x+\frac12,y,z,w)$ is composed of 8 2-cells of type 2ii, $(x+\frac12,y\pm\frac14,z\pm\frac14,w\pm\frac14)$. (3iii) A 3-cell at $P=(x+\frac14,y+\frac14,z+\frac14,w+\frac14)$ is composed of 4 2-cells of each type, located at $P$ plus or minus permutations of the vector $(\frac14,0,0,0)$.
\item (4i) A 4-cell at $(x,y,z,w)$ is composed of 24 3-cells, $(x\pm\frac12,y,z,w)$, $(x,y\pm\frac12,z,w)$, $(x,y,z\pm\frac12,w)$, $(x,y,z,w\pm\frac12)$, and $(x\pm\frac14,y\pm\frac14,z\pm\frac14,w\pm\frac14)$, where again all~$\pm$ can be chosen idependently. (4ii) Replace $x,y,z$, and $w$ with $x+\frac12,y+\frac12,z+\frac12$, and $w+\frac12$ in the (4i) case.
\end{itemize}

In the language of chain complexes, we have provided a simple description of the \emph{boundary operators}.

\begin{figure}[htbp]
\centering
\subfloat[Red faces $\mathcal{V}_r$.]{
\includegraphics[trim={11.5cm 7cm 11.5cm 7cm},clip,width = 0.4\linewidth]{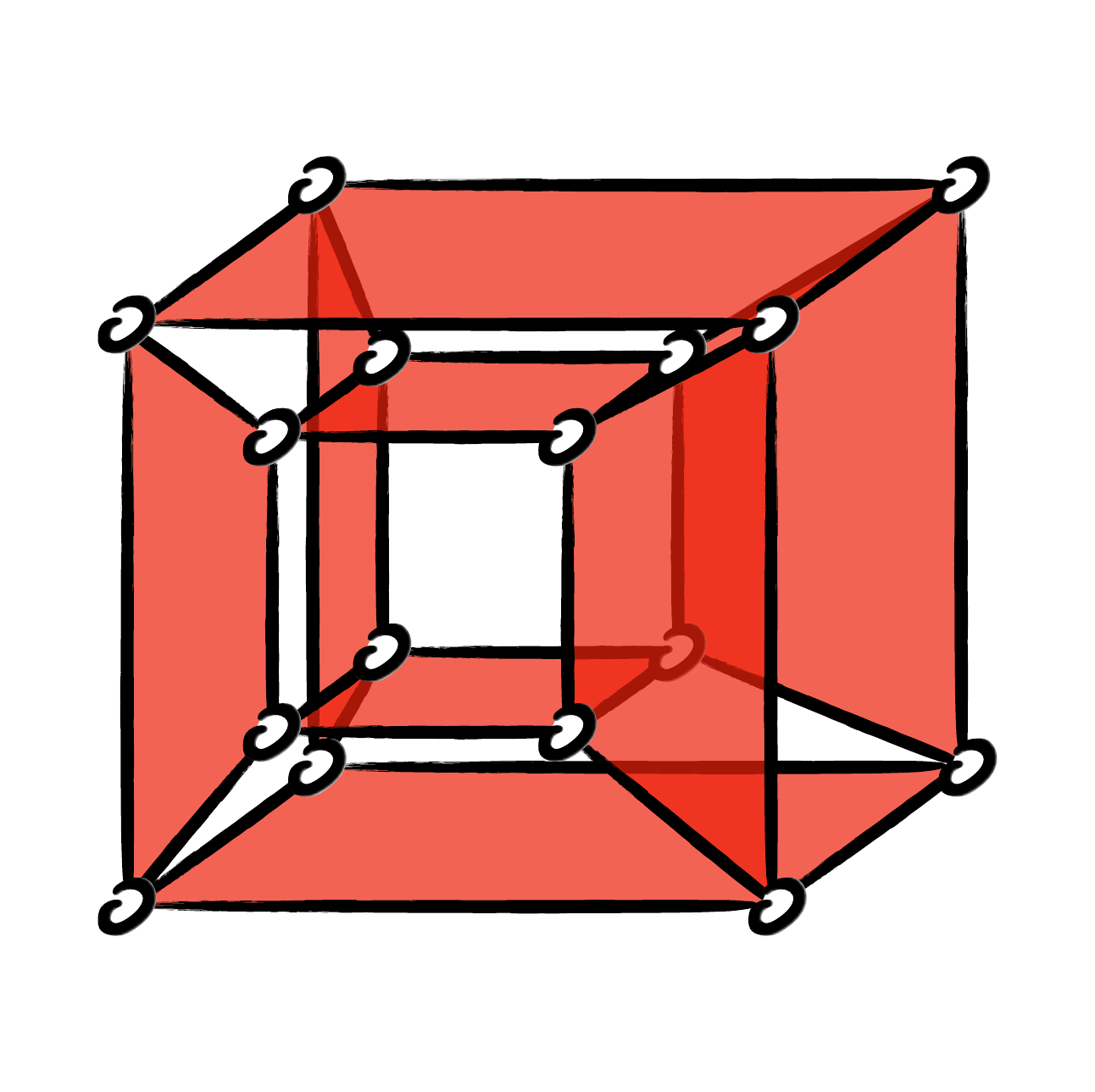}
\label{fig:4DTC_red}}
\hfill
\subfloat[Green faces $\mathcal{V}_g$, front and back faces omitted for clarity.]{
\includegraphics[trim={11.5cm 7cm 11.5cm 7cm},clip,width = 0.4\linewidth]{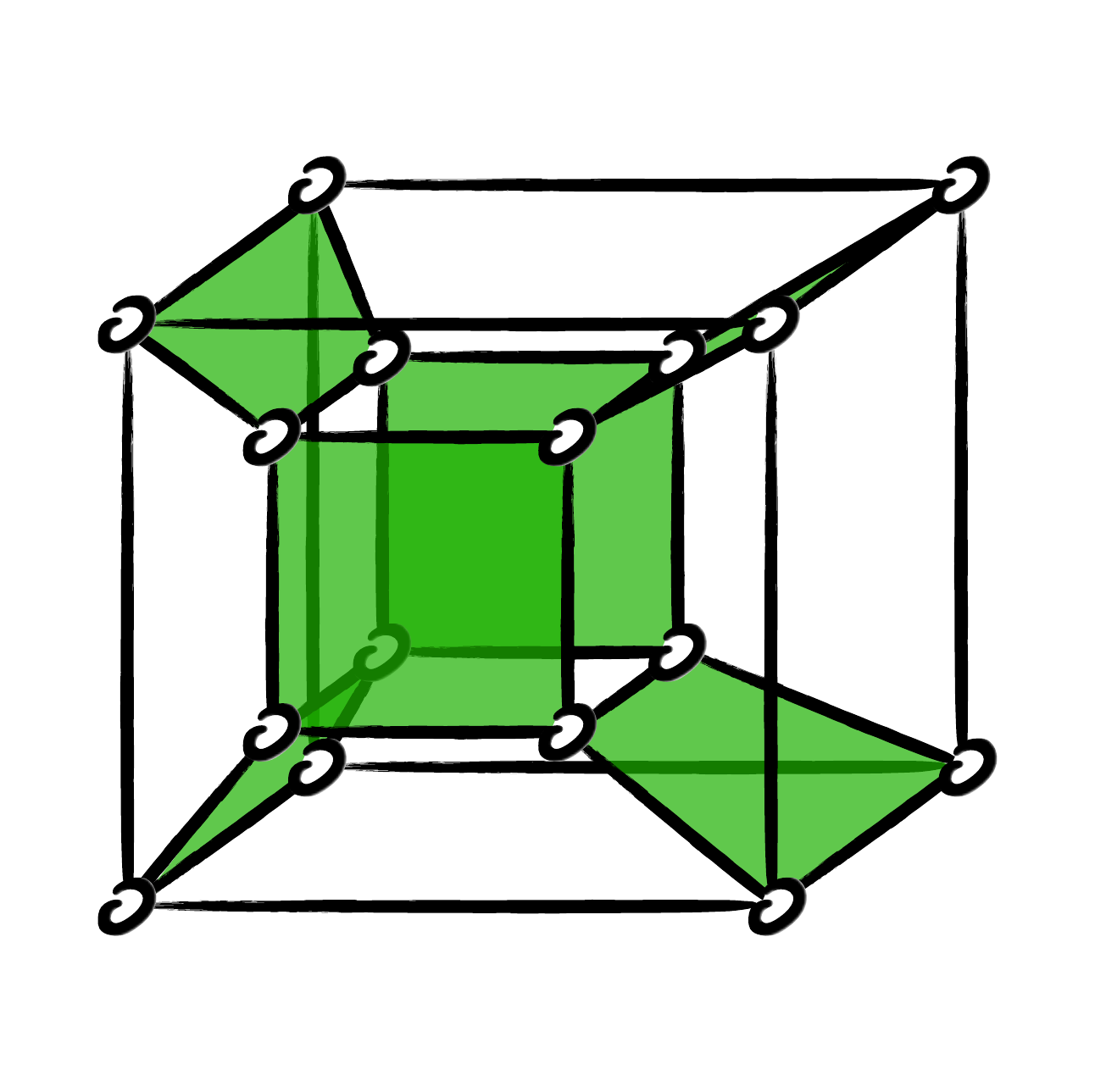}
\label{fig:4DTC_green}}
\newline
\subfloat[Blue faces $\mathcal{V}_b$.]{
\includegraphics[trim={11.5cm 7cm 11.5cm 7cm},clip,width = 0.4\linewidth]{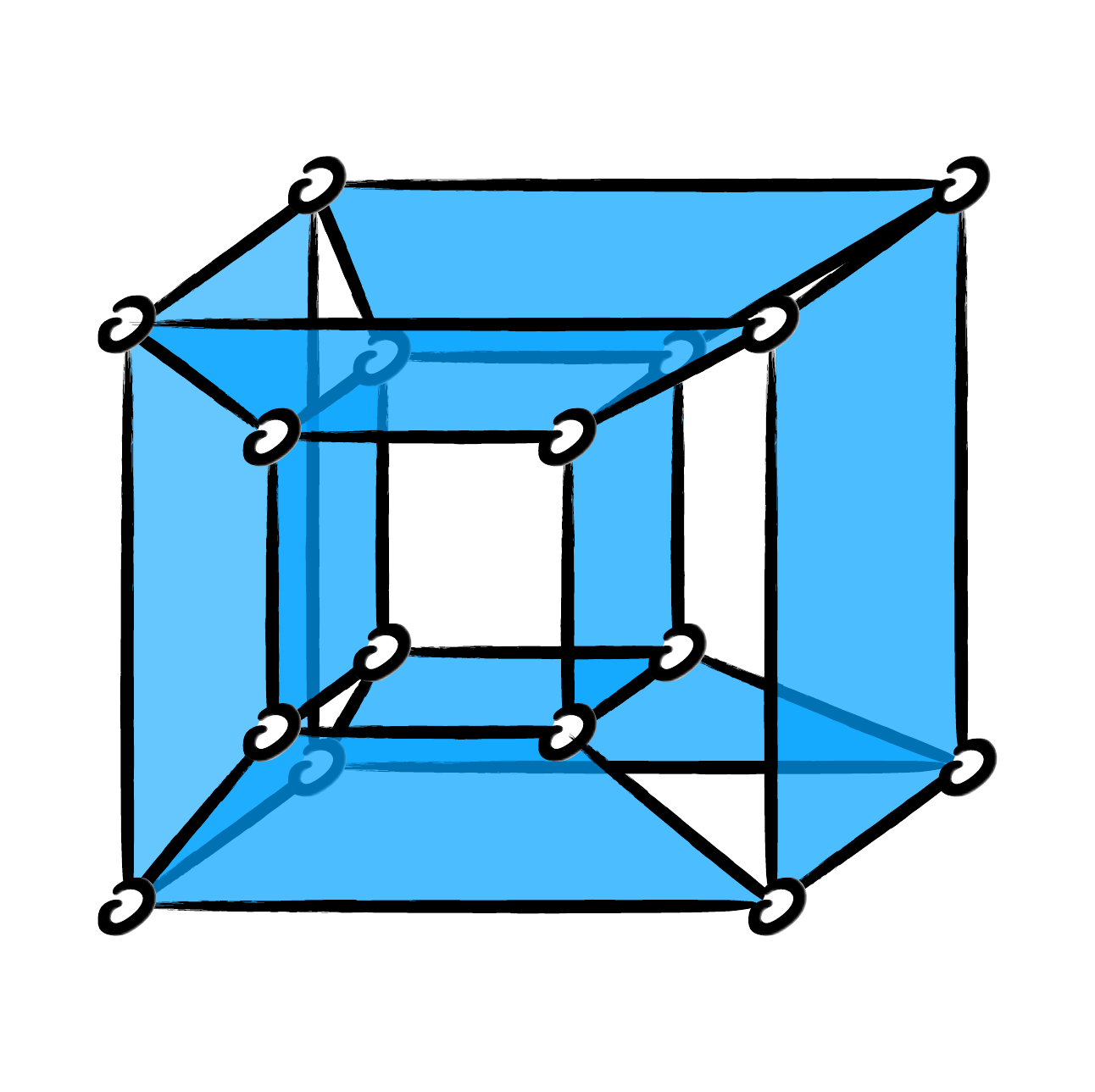}
\label{fig:4DTC_blue}}
\hfill
\subfloat[Subset of vertices of all three colors.]{
\includegraphics[trim={11.5cm 7cm 11.5cm 7cm},clip,width = 0.4\linewidth]{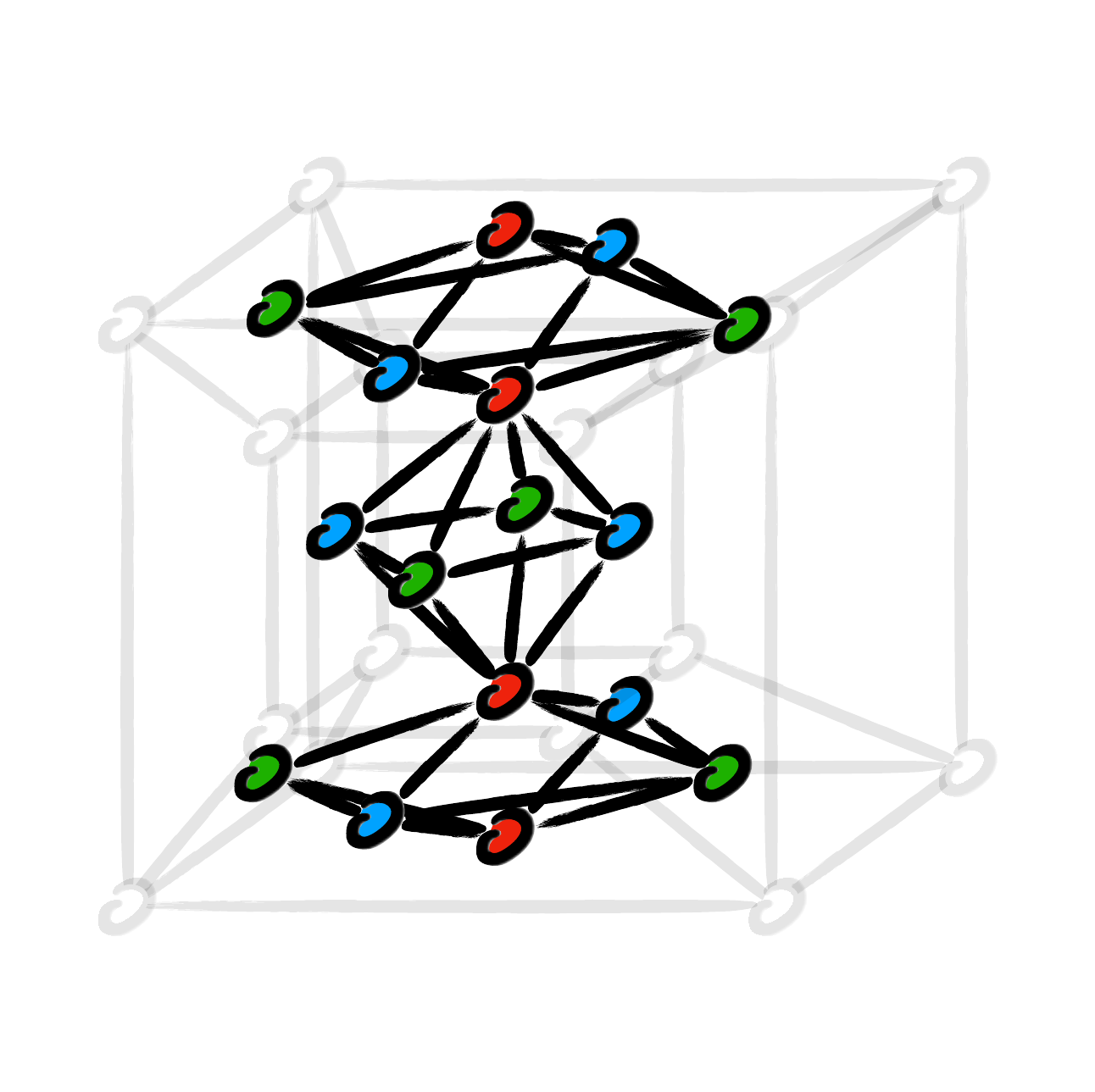}
\label{fig:4DTC_vertices}}
\newline
\subfloat[Alternative subset of vertices of all three colors]{
\includegraphics[trim={11.5cm 7cm 11.5cm 7cm},clip,width = 0.4\linewidth]{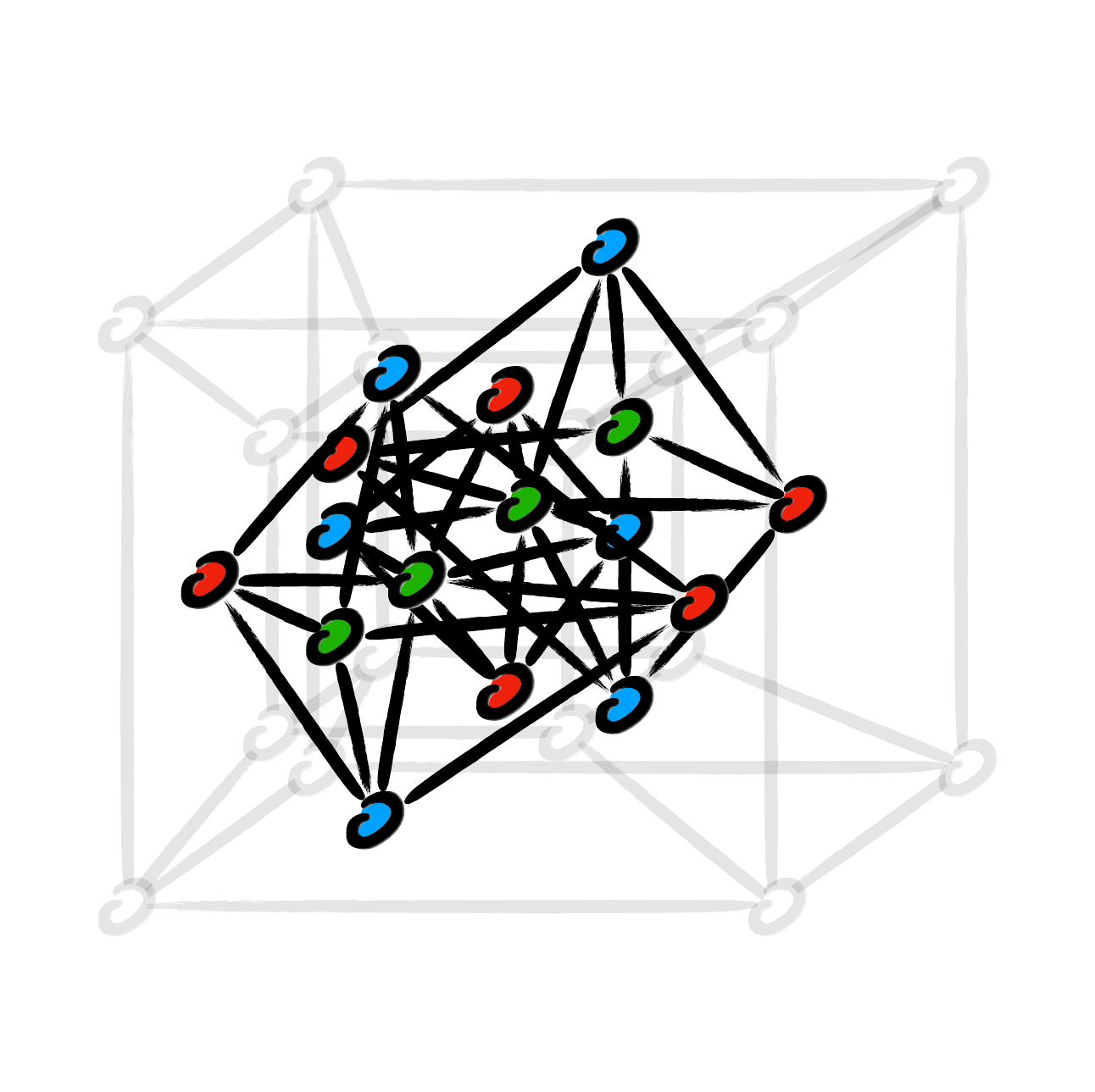}
\label{fig:4DTC_vertices3}}
\hfill
\subfloat[Octaplex!]{
\includegraphics[trim={11.5cm 7cm 11.5cm 7cm},clip,width = 0.4\linewidth]{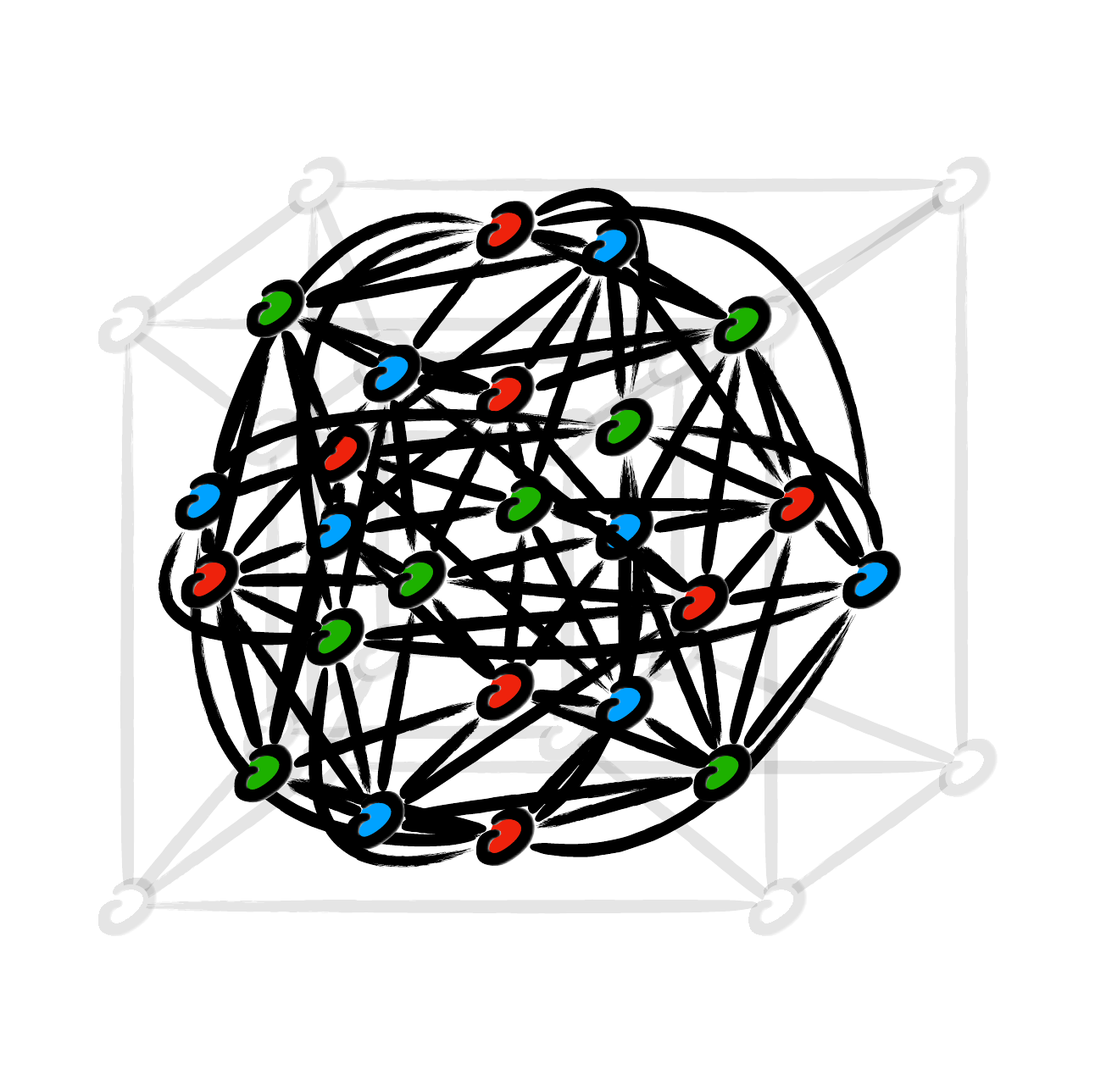}
\label{fig:4DTC_vertices4}}
\caption{Converting between tesseract faces to the orthoplex.}
\label{fig:4DTC_hypercube}
\end{figure}

\subsection{Stabilizer operators of codeblock 0}
\label{sec:CoordCode0}

The $X$~stabilizers of codeblock 0 are associated to 4-cells, which are given by either integer coordinates $(x,y,z,w) \in \mathbb{Z}_2^4$ or half integer coordinates~$(x+\tfrac{1}{2},y+\tfrac{1}{2},z+\tfrac{1}{2},w+\tfrac{1}{2})$. As discussed in the previous subsection, each of the $X$~stabilizers will be weight-24 operators whose support is given by the set of 3-cells that are closest in distance from the given 4-cell.

The $Z$~stabilizers are formed from the intersection of the $X$~stabilizers of the other codes. The $X$~stabilizers represented by different colored vertices only have non-trivial intersection if they form a face in the octaplex tessellation. They correspond to weight-3 operators whose support is given by the 3-cells that contain the given face, see the previous subsection for more details.

\subsection{Logical operators of codeblock 0}
\label{sec:LogicalOp0}

We begin by defining the $Z$~logical operators, which recall are going to be non-contractible loops as per our desiderata. Take the qubit defined by the following 3-cell:~$(0,0,0,\tfrac{1}{2})$, which corresponds to the intersection of the two 24-cells~$(0,0,0,0)$ and $(0,0,0,1)$. Therefore, a $Z$~error on such a qubit would cause a pair of excitations in the $\hat{w}$~direction, indicating that this qubit should belong to a logical~$Z$ operator along that axis. Therefore, the following operator will be a valid logical~$Z$ operator:
\begin{align}
\overline{\mathcal{Z}}_{\hat{w}}^{(0)} = \prod_{w \in \mathbb{Z}_L} Z^{(0)}_{(0,0,0,w+\tfrac{1}{2})},
\label{eq:logZw0}
\end{align}
as it will intersect every $X$~stabilizer~$(0,0,0,w)$ at two locations (for all~$w$, assuming periodic boundary conditions). While this operator commutes with the stabilizers of the code, what remains to be shown is that it is indeed a logical (non-identity) Pauli operator, which implies we must be able to find a logical~$X$ with which it anti-commutes. Before searching for such an operator, note that we can translate the above $Z$ operator by multiplying it by a set of $Z$~stabilizers. As discussed in the last subsection, there is a $Z$-stabilizer supported on qubits in the set:
\begin{align*}
\{&(x,y,z,w+\tfrac{1}{2}),\\
&(x+\tfrac{1}{4},y+\tfrac{1}{4},z+\tfrac{1}{4},w+\tfrac{1}{4}),\\
&(x+\tfrac{1}{4},y+\tfrac{1}{4},z+\tfrac{1}{4},w+\tfrac{3}{4})\}.
\end{align*}
As such, this $Z$~stabilizer will shift any operator supported at~$(x,y,z,w+\tfrac{1}{2})$ by~$\tfrac{1}{4}$ in the $\hat{x}, \ \hat{y}, \ \hat{z}$ directions while also shifting its support~$\pm\tfrac{1}{4}$ in $\hat{w}$. Note that by choosing different sets of $v_r, v_g, v_b$ we could have also shifted by~$-\tfrac{1}{4}$ in any of the $\hat{x}, \ \hat{y}, \ \hat{z}$ directions. We can then continue this shift by multiplying by the $Z$~stabilizer generated by the colored vertices $v_r = (x+\tfrac{1}{2},y+\tfrac{1}{2},z,w), \ v_g = (x+\tfrac{1}{2},y,z+\tfrac{1}{2},w), \ v_b = (x,y+\tfrac{1}{2},z+\tfrac{1}{2},w)$ resulting in the weight-3 operator:
\begin{align*}
\{&(x+\tfrac{1}{2},y+\tfrac{1}{2},z+\tfrac{1}{2},w),\\
&(x+\tfrac{1}{4},y+\tfrac{1}{4},z+\tfrac{1}{4},w-\tfrac{1}{4}),\\
&(x+\tfrac{1}{4},y+\tfrac{1}{4},z+\tfrac{1}{4},w+\tfrac{1}{4})\}.
\end{align*}
Taking the product of these two weight-3 operators thus results in a weight-4 $Z$ stabilizer with support:
\begin{align*}
\{&(x,y,z,w+\tfrac{1}{2}),\\
&(x+\tfrac{1}{4},y+\tfrac{1}{4},z+\tfrac{1}{4},w+\tfrac{3}{4}), \\
&(x+\tfrac{1}{4},y+\tfrac{1}{4},z+\tfrac{1}{4},w-\tfrac{1}{4}), \\
&(x+\tfrac{1}{2},y+\tfrac{1}{2},z+\tfrac{1}{2},w)\},
\end{align*}
which shifts the logical operator from Eq.~\ref{eq:logZw0} as follows:
\begin{align*}
\overline{\mathcal{Z}}_{\hat{w}}^{(0)} &= \prod_{w \in \mathbb{Z}_L} Z^{(0)}_{(0,0,0,w+\tfrac{1}{2})} \\
&\simeq \prod_{w \in \mathbb{Z}_L} Z^{(0)}_{(\tfrac{1}{4},\tfrac{1}{4},\tfrac{1}{4},w+\tfrac{1}{4})} Z^{(0)}_{(\tfrac{1}{4},\tfrac{1}{4},\tfrac{1}{4},w+\tfrac{3}{4})} \\
&\simeq \prod_{w \in \mathbb{Z}_L} Z^{(0)}_{(\tfrac{1}{4},\tfrac{1}{4},\tfrac{1}{4},w-\tfrac{1}{4})}
Z^{(0)}_{(\tfrac{1}{2},\tfrac{1}{2},\tfrac{1}{2},w)}
Z^{(0)}_{(\tfrac{1}{4},\tfrac{1}{4},\tfrac{1}{4},w+\tfrac{3}{4})} \\
&= \prod_{w \in \mathbb{Z}_L} Z^{(0)}_{(\tfrac{1}{2},\tfrac{1}{2},\tfrac{1}{2},w)},
\end{align*}
where the terms given by coordinates~$(\tfrac{1}{4},\tfrac{1}{4},\tfrac{1}{4},w-\tfrac{1}{4})$ and $(\tfrac{1}{4},\tfrac{1}{4},\tfrac{1}{4},w+\tfrac{3}{4})$ cancel each other out given periodic boundary conditions. It is worth pointing out that by choosing a different set of $Z$~stabilizers we could have shifted the above logical operator $\pm\tfrac{1}{2}$ in any $\hat{x}, \ \hat{y}, \ \hat{z}$ direction and as such by iterating this process we can show that all of the following representations are equivalent:
\begin{align*}
\overline{\mathcal{Z}}_{\hat{w}}^{(0)} &= \prod_{w \in \mathbb{Z}_L} Z^{(0)}_{(0,0,0,w+\tfrac{1}{2})} \\
&\simeq \prod_{w \in \mathbb{Z}_L} Z^{(0)}_{(x,y,z,w+\tfrac{1}{2})} \\
&\simeq \prod_{w \in \mathbb{Z}_L} \Big[ Z^{(0)}_{(x+\tfrac{(-1)^{\alpha}}{4},y+\tfrac{(-1)^{\beta}}{4},z+\tfrac{(-1)^\gamma}{4},w+\tfrac{1}{4})} \\ &\hspace{6em} \cdot Z^{(0)}_{(x+\tfrac{(-1)^\alpha}{4},y+\tfrac{(-1)^\beta}{4},z+\tfrac{(-1)^\gamma}{4},w+\tfrac{3}{4})} \Big] \\
&\simeq \prod_{w \in \mathbb{Z}_L} Z^{(0)}_{(x+\tfrac{1}{2},y+\tfrac{1}{2},z+\tfrac{1}{2},w)}, 
\end{align*}
for all $x,y,z \in \mathbb{Z}_L$ and $\alpha, \beta,\gamma \in \mathbb{Z}_2$. Therefore, given that the $Z$~logical operator can be shifted in any $\hat{x}, \ \hat{y}, \ \hat{z}$ direction the corresponding $X$ logical operator will have to span these three axes, thus forming a hyperplane. This is analogous to the $X$~logical operator spanning a plane orthogonal to the $Z$~loop operator in the 3D~toric code. In fact, by the above observation that we can find three disjoint representatives according to the~$\hat{w}$ coordinate being either an integer, half-integer, or quarter-integer, the cooresponding $X$~logical operator will be composed of three hyperplanes with a fixed $\hat{w}$ coordinate of each of these three types. We propose the following $X$~logical operator to be that which is orthogonal to~$\overline{\mathcal{Z}}_{\hat{w}}^{(0)}$:
\begin{align}
\overline{\mathcal{X}}_{\hat{w}}^{(0)} 
&= \prod_{\substack{x,y,z \\ \in \mathbb{Z}_L}} \Big[ \prod_{\substack{\alpha,\beta,\gamma \\ \in \{0,1\}}} X^{(0)}_{(x,y,z,\tfrac{1}{2})} X^{(0)}_{(x+\tfrac{1}{2},y+\tfrac{1}{2},z+\tfrac{1}{2},0)} \nonumber \\
& \hspace{8em} \cdot X^{(0)}_{(x+\tfrac{(-1)^\alpha}{4},y+\tfrac{(-1)^\beta}{4},z+\tfrac{(-1)^\gamma}{4},\tfrac{1}{4})} \Big]. 
\end{align}
We leave the proof that this operator commutes with the stabilizer group, and is thus a logical operator, to Appendix~\ref{app:XlogOps}. It is straightforward to verify that it intersects the logical~$Z$ representative from Eq.~\ref{eq:logZw0} at a single qubit given by coordinate~$(0,0,0,\tfrac{1}{2})$, and as such the two operator anti-commute. Finally, given codeblock~0 is symmetric with respect to all four spatial directions, we can define the other three pairs of logical operators as follows:
\begin{align*}
\overline{\mathcal{Z}}_{\hat{x}}^{(0)} &= \prod_{x \in \mathbb{Z}_L} Z^{(0)}_{(x+\tfrac{1}{2},0,0,0)}, \\
\overline{\mathcal{X}}_{\hat{x}}^{(0)} 
&= \prod_{\substack{y,z,w \\ \in \mathbb{Z}_L}} \Big[ \prod_{\substack{\beta,\gamma,\zeta \\ \in \{0,1\}}} X^{(0)}_{(\tfrac{1}{2},y,z,w)} X^{(0)}_{(0,y+\tfrac{1}{2},z+\tfrac{1}{2},w+\tfrac{1}{2})}  \\
& \hspace{8em} \cdot X^{(0)}_{(\tfrac{1}{4},y+\tfrac{(-1)^\beta}{4},z+\tfrac{(-1)^\gamma}{4},w+\tfrac{(-1)^{\zeta}}{4})} \Big],\\
\overline{\mathcal{Z}}_{\hat{y}}^{(0)} &= \prod_{y\in \mathbb{Z}_L} Z^{(0)}_{(0,y+\tfrac{1}{2},0,0)}, \\
\overline{\mathcal{X}}_{\hat{y}}^{(0)} 
&= \prod_{\substack{x,z,w \\ \in \mathbb{Z}_L}} \Big[ \prod_{\substack{\alpha,\gamma,\zeta \\ \in \{0,1\}}} X^{(0)}_{(x,\tfrac{1}{2},z,w)} X^{(0)}_{(x+\tfrac{1}{2},0,z+\tfrac{1}{2},w+\tfrac{1}{2})}  \\
& \hspace{8em} \cdot X^{(0)}_{(x+\tfrac{(-1)^\alpha}{4},\tfrac{1}{4},z+\tfrac{(-1)^\gamma}{4},w+\tfrac{(-1)^\zeta}{4})} \Big],\\
\overline{\mathcal{Z}}_{\hat{z}}^{(0)} &= \prod_{z\in \mathbb{Z}_L} Z^{(0)}_{(0,0,z+\tfrac{1}{2},0)}, \\
\overline{\mathcal{X}}_{\hat{z}}^{(0)} 
&= \prod_{\substack{x,y,w \\ \in \mathbb{Z}_L}} \Big[ \prod_{\substack{\alpha,\beta,\zeta \\ \in \{0,1\}}} X^{(0)}_{(x,y,\tfrac{1}{2},w)} X^{(0)}_{(x+\tfrac{1}{2},y+\tfrac{1}{2},0,w+\tfrac{1}{2})}  \\
& \hspace{8em} \cdot X^{(0)}_{(x+\tfrac{(-1)^\alpha}{4},y+\tfrac{(-1)^\beta}{4},\tfrac{1}{4},w+\tfrac{(-1)^\zeta}{4})} \Big].
\end{align*}

\subsection{Equivalence of all four codeblocks}
\label{sec:Equivalent}

Recall the other codeblocks are defined by $X$~stabilizers that are supported on the vertices of a given color. In order to show equivalence between codeblocks 0 and 1, suppose we introduce new vertices where all of the 4-cells are centered, while eliminating all of the red vertices and introducing edges following the same rules as in Section~\ref{sec:Coordinates}. The new set of vertices will be:
\begin{align*}
\mathcal{V}'_r &= \{ (x,y,z,w) | \ x,y,z,w \in \mathbb{Z}_L^4\} \nonumber  \\
&\hspace{1em} \cup \{ (x+\tfrac{1}{2},y+\tfrac{1}{2},z+\tfrac{1}{2},w+\tfrac{1}{2}) | \ x,y, z,w \in \mathbb{Z}_L^4 \} \\
\mathcal{V}'_g &= \{ (x,y+\tfrac{1}{2},z,w+\tfrac{1}{2}) | \ x,y,z,w \in \mathbb{Z}_L^4\} \nonumber \\
&\hspace{1em} \cup \{ (x+\tfrac{1}{2},y,z+\tfrac{1}{2},w) | \ x,y, z,w \in \mathbb{Z}_L^4 \} \\
\mathcal{V}'_b&= \{ (x,y+\tfrac{1}{2},z+\tfrac{1}{2},w) | \ x,y,z,w \in \mathbb{Z}_L^4\}  \nonumber \\
&\hspace{1em} \cup \{ (x+\tfrac{1}{2},y,z,w+\tfrac{1}{2}) | \ x,y, z,w \in \mathbb{Z}_L^4 \},
\end{align*}
where vertices share an edge if they are distance~$\tfrac{1}{\sqrt{2}}$ in 2-norm and we define 4-cells centered where all of the old red vertices were located~$\mathcal{V}_r$, where each 4-cell contains all vertices again at distance~$\tfrac{1}{\sqrt{2}}$ in 2-norm. If we then introduce the change of coordinates:~$(x,y,z,w) \rightarrow (x+\tfrac{1}{2}, y+\tfrac{1}{2},z,w)$, we note we have the exact same lattice as previously where the rolls of different sets of vertices and 4-cells have been exhanged:
\begin{align*}
\mathcal{V}_r &\rightarrow \mathcal{O} \\
\mathcal{V}_g &\rightarrow \mathcal{V}_b \\
\mathcal{V}_b &\rightarrow \mathcal{V}_g \\
\mathcal{O} & \rightarrow \mathcal{V}_r \\
\mathcal{Q} & \rightarrow \mathcal{Q}.
\end{align*}
Thus, under this new labelling, codeblock~1 has $X$~stabilizers that were previously labelled by the red vertices and are now defined by 4-cells, with $Z$~stabilizers defined by the intersection of different colored vertices in the new labelling. As such, the properties of codeblock~1 mirror those of codeblock~0. Given our choice of which colored vertices to eliminate was arbitrary all codeblocks are indeed symmetric.

Given the symmetry of the codeblocks, we can use the above change of basis to determine the logical operators for the other codeblocks as well. For example, the logical~$\overline{\mathcal{Z}}_{\hat{w}}^{(1)}$ (red codeblock) will be a shifted version of that from codeblock~0:
\begin{align*}
\overline{\mathcal{Z}}_{\hat{w}}^{(0)} &= \prod_w Z^{(0)}_{(0,0,0,w+\tfrac{1}{2})} & \longrightarrow  \hspace{1em}  &  \overline{\mathcal{Z}}_{\hat{w}}^{(1)} \hspace{-1em} &=&  \prod_w Z^{(1)}_{(\tfrac{1}{2},\tfrac{1}{2},0,w+\tfrac{1}{2})}   \\
&\simeq \prod_w Z^{(0)}_{(-\tfrac{1}{2},-\tfrac{1}{2},\tfrac{1}{2},w)} & \longrightarrow   \hspace{1em}  & & \simeq&  \prod_w  Z^{(1)}_{(0,0,\tfrac{1}{2},w)} 
\end{align*}
We can find the Pauli logical oeprators for all other codeblocks in a systematic way by modifing the change of coordinates, we list them all in Appendix~\ref{app:logPaulis}.

In a similar manner, we can establish the logical~$X$ operator of the other codeblocks by imposing the appropriate shift of coordinates. For example, for codeblock~1:
\begin{align*}
\overline{\mathcal{X}}_{\hat{w}}^{(0)} &= \prod_{\substack{x,y,z \\ \in \mathbb{Z}_L}} \Big[ \prod_{\substack{\alpha,\beta,\gamma \\ \in \{0,1\}}} X^{(0)}_{(x,y,z,\tfrac{1}{2})} X^{(0)}_{(x+\tfrac{1}{2},y+\tfrac{1}{2},z+\tfrac{1}{2},0)}  \\
& \hspace{8em} \cdot X^{(0)}_{(x+\tfrac{(-1)^\alpha}{4},y+\tfrac{(-1)^\beta}{4},z+\tfrac{(-1)^\gamma}{4},\tfrac{1}{4})} \Big]\\
&\longrightarrow \prod_{\substack{x,y,z \\ \in \mathbb{Z}_L}} \Big[ \prod_{\substack{\alpha,\beta,\gamma \\ \in \{0,1\}}} X^{(1)}_{(x+\tfrac{1}{2},y+\tfrac{1}{2},z,\tfrac{1}{2})} X^{(1)}_{(x+1,y+1,z+\tfrac{1}{2},0)}  \\
& \hspace{6em} \cdot X^{(1)}_{(x+\tfrac{1}{2}+\tfrac{(-1)^\alpha}{4},y+\tfrac{1}{2}+\tfrac{(-1)^\beta}{4},z+\tfrac{(-1)^\gamma}{4},\tfrac{1}{4})} \Big] \\
&\simeq \prod_{\substack{x,y,z \\ \in \mathbb{Z}_L}} \Big[ \prod_{\substack{\alpha,\beta,\gamma \\ \in \{0,1\}}} X^{(1)}_{(x+\tfrac{1}{2},y+\tfrac{1}{2},z,\tfrac{1}{2})} X^{(1)}_{(x,y,z+\tfrac{1}{2},0)}  \\
& \hspace{6em} \cdot X^{(1)}_{(x+\tfrac{(-1)^\alpha}{4},y+\tfrac{(-1)^\beta}{4},z+\tfrac{(-1)^\gamma}{4},\tfrac{1}{4})} \Big] = \overline{\mathcal{X}}_{\hat{w}}^{(1)}.
\end{align*}

\subsection{Transversal $\mathsf{CCCZ}$ gate}

Given the code construction, transversal~$\mathsf{CCCZ}$ results in a logical operator. We now verify that it indeed implements the logical~$\mathsf{CCCZ}$ across the four logical qubits. A detailed proof that the given logical operators satisfy the criteria established in Section~\ref{sec:TransversalConditions} is given in Appendix~\ref{app:logPaulis}, yet we summarize the result here.

As presented in the Sec.~\ref{sec:LogicalOp0}, each of the logical~$Z$ operators can be represented by non-contractible strings in each of the Cartesian directions. Each corresponding logical~$X$ is a hyperplane orthogonal to the direction of the logical~$Z$. As such, when considering the overlap of different logical~$X$ operators from different codeblocks, those that have non-trivial overlap span different directions. The intersection of two hyperplanes that are not parallel is a two-dimensional plane. Taking the intersection with yet another orthogonal hyperplane gives rise to a 1D~string. Therefore, the intersection of three non-parallel logical~$X$ operators is a 1D~non-contractible string in the fourth codeblock, which corresponds exactly to the support of the fourth logical~$Z$ operator on that code, as required for the transversality of the logical~$\mathsf{CCCZ}$.

Transversal~$\mathsf{CCCZ}$ thus results in a logical~$\mathsf{CCCZ}$ that couples the following quartets of labelling of the logical operators:
\begin{align*}
&\{\overline{\mathcal{X}}_{\hat{w}}^{(0)},\overline{\mathcal{X}}_{\hat{z}}^{(1)},\overline{\mathcal{X}}_{\hat{y}}^{(2)},\overline{\mathcal{X}}_{\hat{x}}^{(3)}\}, \\
&\{\overline{\mathcal{X}}_{\hat{z}}^{(0)},\overline{\mathcal{X}}_{\hat{w}}^{(1)},\overline{\mathcal{X}}_{\hat{x}}^{(2)},\overline{\mathcal{X}}_{\hat{y}}^{(3)}\}, \\
&\{\overline{\mathcal{X}}_{\hat{y}}^{(0)},\overline{\mathcal{X}}_{\hat{x}}^{(1)},\overline{\mathcal{X}}_{\hat{w}}^{(2)},\overline{\mathcal{X}}_{\hat{z}}^{(3)}\}, \\
&\{\overline{\mathcal{X}}_{\hat{x}}^{(0)},\overline{\mathcal{X}}_{\hat{y}}^{(1)},\overline{\mathcal{X}}_{\hat{z}}^{(2)},\overline{\mathcal{X}}_{\hat{w}}^{(3)}\}.
\end{align*}
Note, that rather than logical operators with the same index being coupled it is those with different indices that are coupled. This reflects that the logical operators must span directions that are orthogonal from one another. In terms of the criteria in Eq.~\ref{eq:Cond5}, the right side of the equality would be 1 if and only if the indices on the left come from one of the sets above.

\section{Boundary conditions}
\label{sec:Boundaries}

Thus far, in order to simply the discussion, we have presented a code construction with periodic boundary conditions, encoding 4 logical qubits across the 4~codeblocks. Moreover, as in the 2D and 3D~toric codes, we can also introduce boundary conditions such that the code does not have to be defined on a periodic lattice at the cost of now only encoding a single logical qubit\footnote{In this work, we omit the cases of introducing holes or twists in the lattice to increase the number of logical qubits.}. 

\begin{figure}[t]
\centering
\subfloat[2D~toric code with boundaries.]{
\includegraphics[trim={8cm 5cm 8cm 5cm},clip,width = 0.45\linewidth]{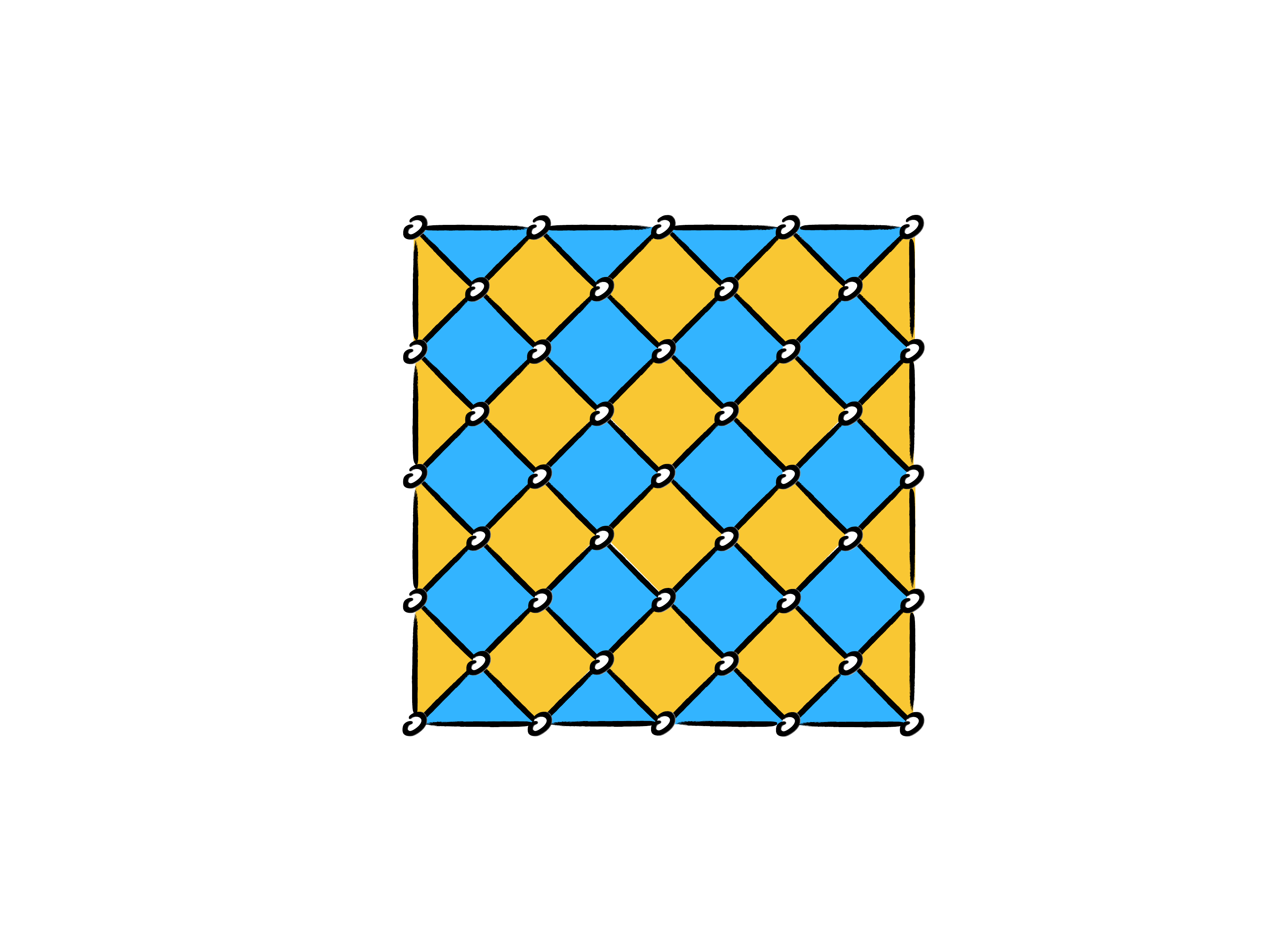}
\label{fig:2DTC_boundaries}}
\hfill
\subfloat[3D~toric code with boundaries.]{
\includegraphics[trim={11cm 9.5cm 11cm 6.5cm},clip,width = 0.45\linewidth]{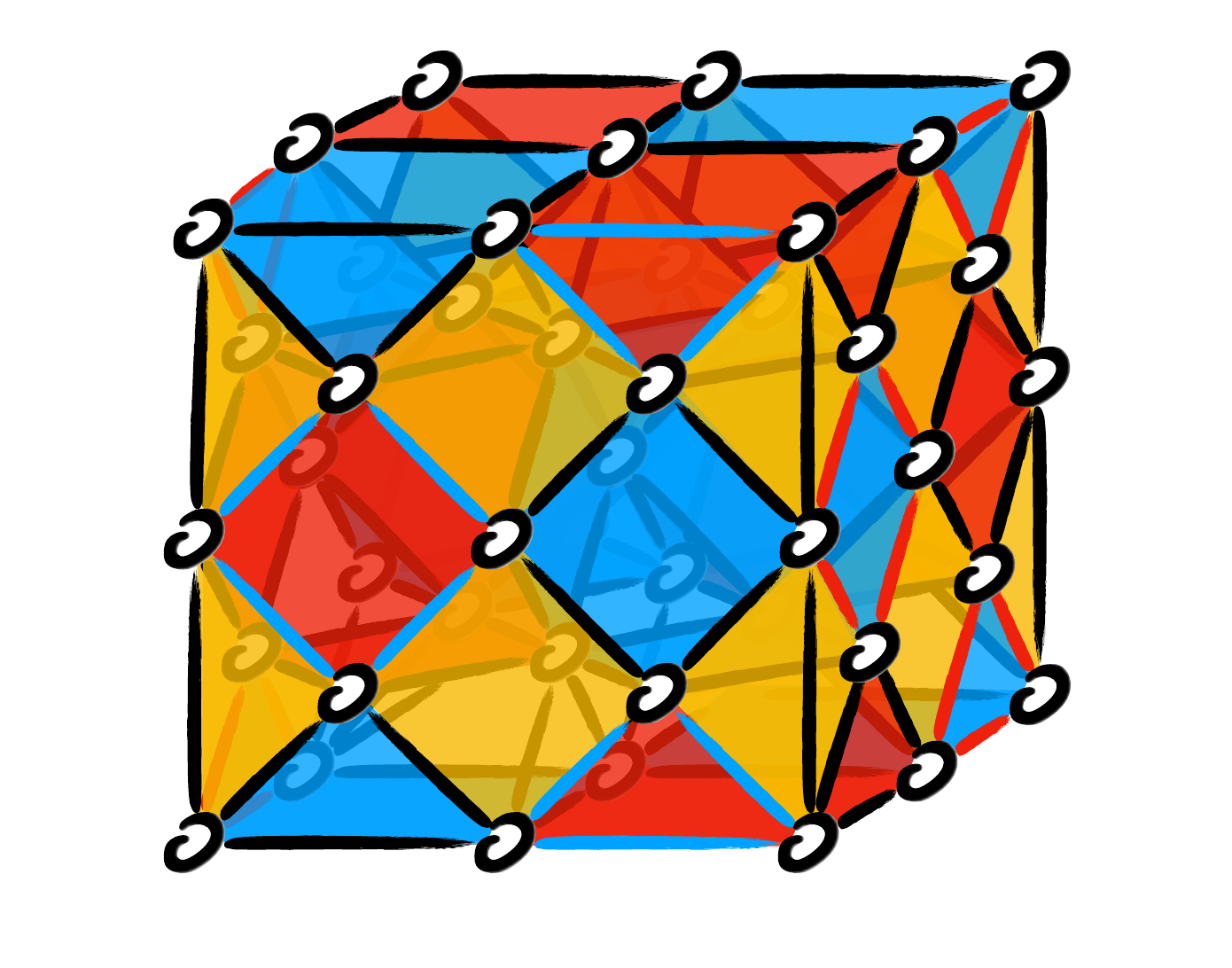}
\label{fig:3DTC_boundaries}}
\caption{Toric codes with boundaries.}
\label{fig:TC_boundaries}
\end{figure}

As first introduced in the 2D~toric code~\cite{BK98}, we can introduce two types of boundaries, rough and smooth, that can serve as endpoints for the different types of logical Pauli operators $X$ and $Z$, respectively. As discussed in the previous section, in order to have a transversal logical $\mathsf{CCCZ}$~gate on the single logical qubit, we require that the logical~$Z$ operators of the four codeblocks span orthogonal axes. As such, the associated rough boundaries should be hyperplanes that are each orthogonal to the respective logical~$Z$ operators in each code. The smooth boundaries in each of the codeblocks should be hyperplanes orthogonal to the other three axes. This is analogous to the case of the 2D and 3D~toric codes with boundaries presented in Fig.~\ref{fig:TC_boundaries}.

The main idea for constructing the code with boundaries is to begin with the periodic case and remove hyperplanes of qubits, thereby cutting the lattice along each of the Cartesian coordinates. As such, we remove a set of X~stabilizers along the hyperplane cut and modify any $X$~stabilizers at the boundary to have slightly smaller support (analogous to both the 2D and 3D~cases). Given a CSS~code can be defined just in terms of its $X$~stabilizers and logical operators (with the $Z$-stabilizers derivable from these alone), we define these sets for the four codeblocks and determine the associated $Z$~stabilizers and logical operators, where there is a single logical operator per codeblock.

We begin with codeblock~0: that is the codeblock where $X$~stabilizers are defined by 4~integer $(x,y,z,w)$ or 4~half-integer $(x+\tfrac{1}{2},y+\tfrac{1}{2},z+\tfrac{1}{2},w+\tfrac{1}{2})$ coordinates, where previously we were working with periodic coordinates in~$\mathbb{Z}_L$. We want to choose boundary conditions such that the logical~$Z$ operator defined in Eq.~\ref{eq:logZw0} has a~$\hat{w}$ coordinate that runs between~$[\tfrac{1}{2},L]$. Therefore, we must choose $X$~stabilizers $(x,y,z,w)$, $(x+\tfrac{1}{2},y+\tfrac{1}{2},z+\tfrac{1}{2},w+\tfrac{1}{2})$ such that $w \in [1,L-1]$. Therefore, we remove all qubits whose support in the $\hat{w}$~coordinate falls outside the interval~$[\tfrac{1}{2},L]$.

By the symmetry arguments of the subsection~\ref{sec:Equivalent}, similar qubits are removed along the other axes as well, thus removing any qubits with a coordinate outside the aforementioned set~$[1,L]$. The associated $Z$~logical operator has the following support:
\begin{align*}
\overline{\mathcal{Z}}_{\hat{w}}^{(0)} &= \prod_{w = 1}^{L} Z_{(x,y,z,w-\tfrac{1}{2})}^{(0)} \\
&\simeq \prod_{w = 1}^{L} Z_{(x-\tfrac{1}{2},y-\tfrac{1}{2},z-\tfrac{1}{2},w)}^{(0)},
\end{align*}
where $x,y,z$ are any integers in the set~$[1,L]$. The corresponding logical~$X$ operator is:
\begin{align*}
\overline{\mathcal{X}}_{\hat{w}}^{(0)}
&= \prod_{x,y,z = 1}^{L} \Big[ \prod_{\substack{\alpha,\beta,\gamma \\ \in \{0,1\}}} X^{(0)}_{(x,y,z,w-\tfrac{1}{2})} X^{(0)}_{(x-\tfrac{1}{2},y-\tfrac{1}{2},z-\tfrac{1}{2},w)}  \\
& \hspace{7em} \cdot X^{(0)}_{(x+\tfrac{(-1)^\alpha}{4},y+\tfrac{(-1)^\beta}{4},z+\tfrac{(-1)^\gamma}{4},w-\tfrac{1}{4})} \Big].
\end{align*}

Now, it should be noted that while we only preserve the $X$~stabilizers whose $\hat{w}$ coordinate is in the interval~$[1,L-\tfrac{1}{2}]$, in the other coordinate directions we will also include those stabilizers whose coordinates are in the set~$\{\tfrac{1}{2},L\}$ and such stabilizers will have smaller support as some of their original qubits have been removed. This is analogous to the boundary stabilizers in the 2D and 3D~toric codes, whose support is also smaller than in the original periodic lattice. Moreover, it is the addition of these stabilizers that prevent the boundaries along these axes from being smooth, and thus preventing a logical~$Z$ operator from terminating there. For example, consider the stabilizer whose coordinates are:~$(\tfrac{1}{2},y+\tfrac{1}{2},z+\tfrac{1}{2},w+\tfrac{1}{2})$, then the analogous logical operator~$\overline{\mathcal{Z}}_{\hat{x}}^{(0)}$, in the $\hat{x}$ direction, would not commute with the example stabilizer as they would intersect at only one qubit:~$(1,y+\tfrac{1}{2},z+\tfrac{1}{2},w+\tfrac{1}{2})$. See Appendix~\ref{app:boundaries} for a summary of the stabilizers at the boundary and their support. 

The $Z$~stabilizers of a given codeblock are formed by taking the intersection of the $X$~stabilizers of the other three codeblocks, thus preserving the requirements for transversal~$\mathsf{CCCZ}$.

\section{Counting stabilizers: Metachecks and single-shot $Z$ stabilizer measurement}
\label{sec:Metachecks}

\begin{figure*}[htbp]
\begin{center}
\includegraphics[trim={1.7cm 3cm 2cm 1.8cm},clip,width = 0.7\linewidth]{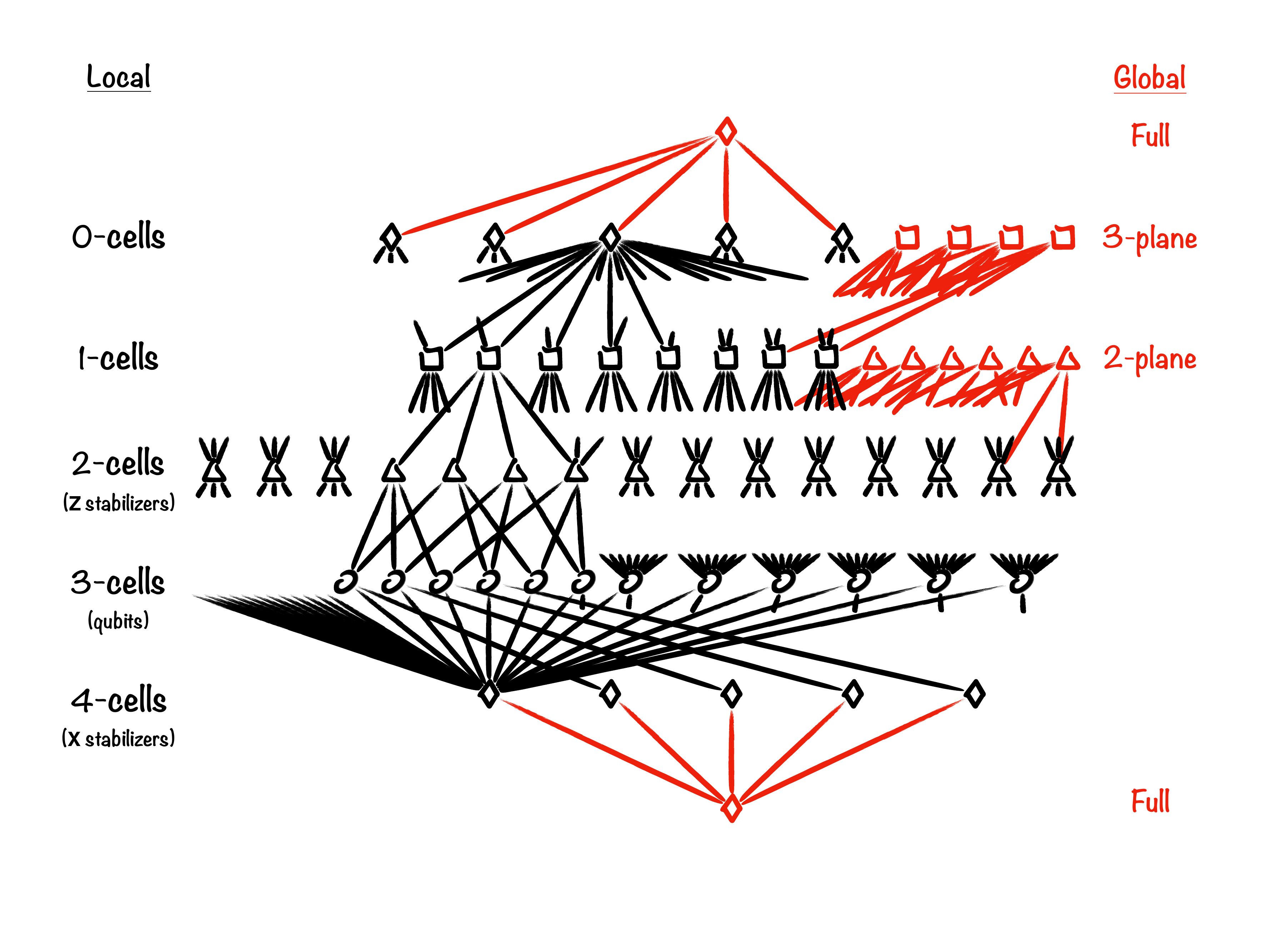}
\caption{Extended Tanner graph description of the octaplex tessellation. Physical qubits are associated with 3-cells. In codeblock 0, $X$~stabilizers are associated with 4-cells where the 3-cells present in a given 4-cell indicate the support of a given $X$~stabilizer. $Z$~stabilizers are associated with 2-cells, which are triangles, whose supports are given by all 3-cells sharing a given face. Not all $X$ and $Z$~stabilizers are independent and are related through metachecks given by the 0- and 1-cells as well as red global symmetry checks.}
\label{fig:TannerExt}
\end{center}
\end{figure*}

In this section we count the degrees of freedom in the code and show that there are indeed only 4~logical qubits in the 4D~octaplex tessellation. This is due to the high amount of redundancy in the $Z$~stabilizer checks. We refer to these redundancies as metachecks following the language of Ref.~\cite{Campbell18}.

Given we have established a symmetry between all of the codeblocks (see Sec.~\ref{sec:Equivalent}), we focus on codeblock~0 as on a periodic lattice. Physical qubits are given by the following coordinates found in Sec.~\ref{sec:Coordinates}: (3i)~one integer and three half-integer, (3ii)~three integer and one half-integer, and (3iii)~four quarter-integer. Therefore, the total number of physical qubits is:~$8L^4 + (2L)^4 = 24 L^4$.

In codeblock~0, the $X$~stabilizers are associated with the 4-cells of the tessellation, which have coordinates that are either all integer or half-integer. As such, the number of such stabilizers is~$2L^4$ and their product is identity, yielding a independent generator set of size~$2L^4 - 1$. We call this redundancy in the stabilizer generators a \emph{global metacheck} as it corresponds to a global symmetry of the lattice. The redundancy of one of the $X$~stabilizers is represented by the bottom metacheck in Fig.~\ref{fig:TannerExt} relating all 4-cells.

Recall the $Z$~stabilizers of codeblock~0 are identified with by the triangular faces (with three different colored vertices) in the octaplex tessellation. Therefore, in order to count the number of such faces it is sufficient to count the associated faces to which a single vertex belongs to, which amounts to determining the number of edges in the vertex operator (see Section~\ref{sec:Schlafli} for a review of vertex operators). The vertex operator of the octaplex tessellation has \schlafli symbol~$\{4,3,3\}$, a tesseract, and as such has 32~edges. The number of vertices of a given color is~$2 L^4$ and as such the number of~$Z$ stabilizers corresponding to triangular faces is~$64 L^4$. Of course, these stabilizers are not all independent and we shall review their dependencies here.

Each 2-cell is a face belonging to three adjacent 3-cells, which corresponds to the qubit support of the Pauli operator defined on the 2-cell. Therefore, a set of 2-cells is said to be dependent if their overlap is even across all 3-cells. There are two forms of such symmetries for the 2-cells: \emph{local} symmetries which are the result of a product of local faces that result in the identity operator, or \emph{global} symmetries which result from a product of faces that span the lattice that result in identity.

We begin with the local symmetry: consider an edge in the octaplex tessellation and the associated triangular faces that contain the given edge. There are four such faces as the associated edge operator\footnote{Vertex operator of the vertex operator.} is a tetrahedron~$\{3,3\}$, where the vertices of the tetrahedron represent neighboring faces and the edges represent neighboring 3-cells. Each neighboring 3-cell is adjacent to two such faces and as such the product over all such faces yields the identity\footnote{An equivalent description of the same statement is that if one were to take the dual of the octaplex tessellation, the 16-cell honeycomb~$\{3,3,4,3\}$, then the underlying 3-cells of the dual (corresponding to edges in the octaplex tessellation) must be tetrahedra. Given qubits are associated with edges in the dual, taking the product over faces of a tetrahedron yields identity.}. Therefore, each edge in the octaplex tessellation provides a redundancy in the $Z$~stabilizers. Each vertex belongs to 16~edges (number of vertices in the vertex operator) and as such given the total number of vertices is~$6L^4$ then the total number of edges is:~$\tfrac{1}{2}\cdot 16 \cdot 6L^4 = 48 L^4$. We represent the local constraints by black squares in Fig.~\ref{fig:TannerExt}. As for the global symmetries with regards to the 2-cells, they are formed by taking the product of faces that form two-dimensional planes of which there are six, as shown in Appendix~\ref{app:2constraints}. They are labelled by red triangles in Fig.~\ref{fig:TannerExt}.

However, not all edges, leading to the aforementioned redundancies in the stabilizers, are themselves independent. As above, a set of 1-cells (edges) is deemed dependent if they overlap an even number of times across all neighboring 2-cells. Again, the symmetries giving rise to these dependencies come in two forms: local and global. The local symmetry arises by considering the set of edges emerging from each of the vertices in the lattice. If we then consider all faces that these edges share, each face must have two of its edges in this set as by definition the chosen vertex is a vertex belonging to the face. Therefore, each vertex provides an additional constraint on the relationship between the 1-cells and there are~$6L^4$ such constraints, represented by black diamonds at the top of Fig.~\ref{fig:TannerExt}. The global symmetries on the 1-cells emerge from taking the product of all 1-cells (edges) forming a hyperplane, totaling 4 such constraints as presented in detail in Appendix~\ref{app:1constraints}, represented by the four red squares in Fig.~\ref{fig:TannerExt}. Finally, there is one additional global symmetry for the vertices themselves each edge in the lattice is in the support of two vertices, therefore taking the product over all vertices in the lattice results in a global symmetry as their support on the edges is all even.

Therefore, the number of independent 0-cell constraints is:~$6L^4-1$. The number of independent 1-cell constraints is:~$48L^4 - (6L^4-1) - 4 = 42L^4 - 3$. Finally, the number of independent 2-cell~$Z$~stabilizers is:~$64L^4 - (42L^4 - 3) - 6 = 22L^4 - 3$. Adding the $X$~stabilizers, the total number of stabilizer generators is:~$2L^4 -1 + 22L^4 - 3 = 24L^4 - 4$, and thus the code has 4 logical qubits as hypothesized.

Finally, we would like to make a comment on the single-shot ability to perform the $Z$~stabilizer measurements. The metachecks of the 0- and 1-cells in Fig.~\ref{fig:TannerExt} form the checks for a classical code where the bits of the code are the measurement outcomes of the $Z$~stabilizer measurements. As such, when a measurement error occurs, it causes a violation in the metachecks, allowing for its correction without having to repeat measurements, unlike the 2D~toric code for example. Given a measurement error will lead to a violation of a metacheck on its three neighboring edges, we can think of the analogy to excitations in the 2D~color code where excitations come in triples where the color of an edge can be the complementary color to the color of its two neighboring endpoints. This alone should be sufficient to guarantee single-shot correction of the $Z$~stabilizer measurements, however we have the additional layer of redundancy that is given by the 0-cells that can identify when violations of 1-cells are misidentified, which further increases the protection against measurement errors.

\section{Discussion}
\label{sec:Discussion}

We have established the existence of a 4D~topological code with a transversal $\mathsf{CCCZ}$ gate. It is then natural to ask whether the presented techniques could be generalized to arbitrary dimension. The natural candidate in 5D would be to consider lattices with \schlafli symbol:~$\{3,3,4,a,b\}$, where $a, b$ are integer degrees of freedom. This choice is motivated by the fact that underlying 4-cells composing the lattice are hyperoctahedra, which are the 4D analog of the octahedron. This would allow for the potential 4-coloring of the vertices as required by the conditions from Sec.~\ref{sec:DualCond}. However, the only regular tessellation of 5D~Euclidean space is from the hypercubic family:~$\{4,3,3,3,4\}$, and thus no regular tessellation with the required conditions exist. Yet, there does exist a regular tessellation in hyperbolic space ($\{3,3,4,3,3\}$) which may have the desired properties and could be of independent interest as it may encode a macroscopic number of logical qubits if the correct boundary conditions can be established. It is worth pointing out again that we only consider regular lattices in this work, where all vertices, edges, etc. are equivalent. It would be natural to conjecture that in higher dimensions breaking full symmetry within the lattice may allow for the implementation of a transversal multi-controlled-$Z$ operation by analogy to what is possible in higher-dimensional color codes~\cite{Bombin15a}.

It may also be of independent interest to consider hyperbolic 3D~space for the existence of regular lattices that may allow for the implementation of $\mathsf{CCZ}$ on a growing number of logical qubits. For example, the lattice given by \schlafli symbol~$\{4,3,5\}$ could be a candidate, yet establishing the correct boundary conditions, whether open or closed, remains open. One direction could be to adapt the approach take in other hyperbolic constructions, which were studied with regards to logical memory and decoding~\cite{BL20}, and adapt them to 3D~hyperbolic space.

\acknowledgements{The authors would like to thank Andrew~Cross, Robert~K\"onig, Aleksander~Kubica, Michael~Vasmer, and Guanyu~Zhu for valuable discussions in the development of this work. T.J.Y.~is partially supported by the IBM Research Frontiers Institute.}

\bibliographystyle{unsrtnat}
\bibliography{bibtex_jochym}

\appendix

\section{Proof of logical operator commutativity with stabilizers}
\label{app:XlogOps}
\normalfont

\lem{The following pair of Pauli operators form a pair of anti-commuting logical operators in codeblock~0:
\begin{align*}
\overline{\mathcal{Z}}_{\hat{w}}^{(0)} &= \prod_{z\in \mathbb{Z}_L} Z^{(0)}_{(0,0,0,w+\tfrac{1}{2})}, \\
\overline{\mathcal{X}}_{\hat{w}}^{(0)} &= \prod_{\substack{x,y,z \\ \in \mathbb{Z}_L}} \Big[ \prod_{\substack{\alpha,\beta,\gamma \\ \in \{0,1\}}} X^{(0)}_{(x,y,z,\tfrac{1}{2})} X^{(0)}_{(x+\tfrac{1}{2},y+\tfrac{1}{2},z+\tfrac{1}{2},0)}  \\
& \hspace{8em} \cdot X^{(0)}_{(x+\tfrac{(-1)^\alpha}{4},y+\tfrac{(-1)^\beta}{4},z+\tfrac{(-1)^\gamma}{4},\tfrac{1}{4})} \Big]
\end{align*}
}

\begin{proof}
It is clear that the pair of Pauli operators anti-commute as the only qubits they have in common is labelled by~$(0,0,0,\tfrac{1}{2})$. What therefore remains to be shown is that these operators do indeed commute with the stabilizers of codeblock~0. 

As discussed in the main text, the $Z$~logical operator will only intersect $X$~stabilizers of with integer coordinates~$(0,0,0,w)$ which will therefore intersect the operator at two qubits:~$(0,0,0,w\pm\tfrac{1}{2})$, again assuming the periodicity of the lattice. As such, the presented $Z$ operator commutes with the stabilizers and is a logical operator.

We now show that the proposed $X$~logical operator also commutes with the $Z$~stabilizers. Suppose a $Z$~stabilizer is supported on a type~(3ii) qubit~$(x,y,z, \tfrac{1}{2})$ (see Sec.~\ref{sec:Coordinates}), we must show that is supported on another qubit that belong to the proposed logical~$X$ operator. The only red vertices that belong to the 3-cell labelled by~$(x,y,z, \tfrac{1}{2})$ are: $v_{r,\gamma} = (x,y,z +(-1)^{\gamma} \tfrac{1}{2}, \tfrac{1}{2})$. In a similar manner, the only two green and blue vertices that belong to the corresponding 3-cell are: $v_{g,\beta} = (x,y +(-1)^{\beta} \tfrac{1}{2}, z, \tfrac{1}{2}) , \ v_{b,\alpha} = (x +(-1)^{\alpha} \tfrac{1}{2},y,z, \tfrac{1}{2})$. Given a triple of different color vertices, the corresponding weight-3 $Z$~stabilizer is supported on the type (3ii) qubit and two type (3iii) qubits:
\begin{align*}
\{ &(x,y,z, \tfrac{1}{2}), \\
&(x + (-1)^{\alpha} \tfrac{1}{4}, y + (-1)^{\beta} \tfrac{1}{4}, z + (-1)^{\gamma} \tfrac{1}{4},  \tfrac{1}{4}), \\
&(x + (-1)^{\alpha} \tfrac{1}{4}, y + (-1)^{\beta} \tfrac{1}{4}, z + (-1)^{\gamma} \tfrac{1}{4},  \tfrac{3}{4}) \}.
\end{align*}
As such, we have shown that any $Z$~stabilizer supported on a qubit (3ii) of the form~$(x,y,z,\tfrac{1}{2})$ must commute with the proposed logical~$X$ operator as it has even overlap. We will argue in a similar manner for a $Z$~stabilizer supported on qubit of type (3i), $(x+\tfrac{1}{2},y+\tfrac{1}{2},z+\tfrac{1}{2}, 0)$. The vertices belonging to that 3-cell are: $v_{r,\gamma} = \{(x +\tfrac{1}{2}, y+ \tfrac{1}{2}, z + \tfrac{1}{2} + (-1)^{\gamma} \tfrac{1}{2},0) \}, v_{g,\beta} = \{ (x + \tfrac{1}{2}, y+\tfrac{1}{2}+(-1)^{\beta} \tfrac{1}{2} , z+ \tfrac{1}{2},0)\} , \ v_{b,\alpha} = \{ (x + \tfrac{1}{2} + (-1)^{\alpha} \tfrac{1}{2}, y + \tfrac{1}{2},z+ \tfrac{1}{2},0)\}$. The corresponding $Z$~stabilizer at the intersection of a given choice of three different colored vertices is:
\begin{align*}
\{
&(x +  \tfrac{1}{2}, y + \tfrac{1}{2}, z + \tfrac{1}{2},  0), \\
&(x + \tfrac{1}{2} + (-1)^{\alpha} \tfrac{1}{4}, y + \tfrac{1}{2} + (-1)^{\beta} \tfrac{1}{4}, z + \tfrac{1}{2} + (-1)^{\gamma} \tfrac{1}{4},  -\tfrac{1}{4}), \\
&(x + \tfrac{1}{2} + (-1)^{\alpha} \tfrac{1}{4}, y + \tfrac{1}{2} + (-1)^{\beta} \tfrac{1}{4}, z + \tfrac{1}{2} + (-1)^{\gamma} \tfrac{1}{4},  \tfrac{1}{4}) \},
\end{align*}
which again has even overlap with the proposed logical~$X$ operator. Thus we have shown that any $Z$~stabilizer supported on qubits of type (3i) and (3ii) in the support of the proposed logical~$X$ operator must commute with the logical operator.

The only case that remains is a $Z$~stabilizer that overlaps with a qubit of type (3iii) in the proposed logical operator, but not necessarily one of type (3i) or (3ii). Suppose a $Z$~stabilizer is supported on qubit~$(x+(-1)^{\alpha} \tfrac{1}{4},y+(-1)^{\beta} \tfrac{1}{4},z+(-1)^{\gamma} \tfrac{1}{4},\tfrac{1}{4})$, then the vertices that are in the support of the associated 3-cell are: $v_{r} = \{(x,y,z +(-1)^{\gamma} \tfrac{1}{2}, \tfrac{1}{2}), (x+ (-1)^{\alpha} \tfrac{1}{2},y+(-1)^{\beta} \tfrac{1}{2},z,0) \}, v_{g} = \{ (x,y + (-1)^{\beta} \tfrac{1}{2}, z, \tfrac{1}{2}),  (x+ (-1)^{\alpha} \tfrac{1}{2}, y, z+(-1)^{\alpha} \tfrac{1}{2} ,0)\} , \ v_{b} = \{ (x + (-1)^{\alpha} \tfrac{1}{2}, y ,z, \tfrac{1}{2}), (x, y + (-1)^{\beta} \tfrac{1}{2}, z + (-1)^{\gamma} \tfrac{1}{2} , 0)\}$. We already covered the cases when the $Z$~stabilizer is supported on a face with a fixed $\hat{w}$ coordinate in the set $\{0,\tfrac{1}{2} \}$, therefore the chosen vertices cannot all agree in the $\hat{w}$~coordinate. Without loss of generality, choose the following set of vertices: $\{(x,y,z +(-1)^{\gamma} \tfrac{1}{2}, \tfrac{1}{2}), (x,y + (-1)^{\beta} \tfrac{1}{2}, z, \tfrac{1}{2}), (x, y + (-1)^{\beta} \tfrac{1}{2}, z + (-1)^{\gamma} \tfrac{1}{2} , 0) \}$, which will support a $Z$~stabilizer on the following set of coordinates:
\begin{align*}
\{
&(x,y+(-1)^{\beta}\tfrac{1}{2},z+(-1)^{\gamma}\tfrac{1}{2},\tfrac{1}{2}), \\
&(x-\tfrac{1}{4},y+(-1)^{\beta}\tfrac{1}{4},z+(-1)^{\gamma}\tfrac{1}{4},\tfrac{1}{4}), \\
& (x+\tfrac{1}{4},y+(-1)^{\beta}\tfrac{1}{4},z+(-1)^{\gamma}\tfrac{1}{4},\tfrac{1}{4})\}.
\end{align*}
Note that the presented $Z$~stabilizer does indeed commute with the proposed logical operator as it intersects with two type~(3iii) qubits. This will necessarily be the case for any face whose vertex coordinates do not all agree in the $\hat{w}$~coordinate.
\end{proof}

\section{Proof of distance of logical $X$ and $Z$ operators}
\label{app:logOpsDistance}
\normalfont

Given the symmetry between the different logical operators within a codeblock as well as the different codeblocks, we will prove the distance for a single pair of logical operators~$\overline{\mathcal{X}}_{\hat{w}}^{(0)}, \ \overline{\mathcal{Z}}_{\hat{w}}^{(0)}$.

As shown in the main section of the text, we can find many different representations of logical~$\overline{\mathcal{Z}}_{\hat{w}}^{(0)}$:
\begin{align*}
\overline{\mathcal{Z}}_{\hat{w}}^{(0)} &= \prod_{w \in \mathbb{Z}_L} Z_{(x,y,z,w+\tfrac{1}{2})} \\
&\simeq \prod_{\substack{w \in \mathbb{Z}_L\\ \zeta \in \mathbb{Z}_2}} Z_{(x+ \tfrac{(-1)^\alpha}{4},y+\tfrac{(-1)^\beta}{4},z+\tfrac{(-1)^\gamma}{4},w+\tfrac{(-1)^\zeta}{4})} \\
&= \prod_{w \in \mathbb{Z}_L} Z_{(x+\tfrac{(-1)^\alpha}{2},y+\tfrac{(-1)^\beta}{2},z+\tfrac{(-1)^\gamma}{2},w)}, \ \forall \ x, y, z \in \mathbb{Z}.
\end{align*}
There are $2L^3 + (2L)^3 = 8L^3$ different non-overlapping representations of the above logical~$\overline{Z}_w^{(0)}$ operator. Therefore, the corresponding conjugate pair must be be supported on at least $8L^3$~qubits. The following representatives have exactly that weight:
\begin{align*}
\overline{\mathcal{X}}_{\hat{w}}^{(0)} &= \prod_{\substack{x,y,z \\ \in \mathbb{Z}_L}} \Big[ \prod_{\substack{\alpha,\beta,\gamma \\ \in \{0,1\}}} X^{(0)}_{(x,y,z,\tfrac{1}{2})} X^{(0)}_{(x+\tfrac{1}{2},y+\tfrac{1}{2},z+\tfrac{1}{2},0)}  \\
& \hspace{8em} \cdot X^{(0)}_{(x+\tfrac{(-1)^\alpha}{4},y+\tfrac{(-1)^\beta}{4},z+\tfrac{(-1)^\gamma}{4},\tfrac{1}{4})} \Big].
\end{align*}
Moreover, this logical operator can be shifted in the $\hat{w}$ direction by multiplying by the appropriate set of stabilizers:
\begin{align*}
\overline{\mathcal{X}}_{\hat{w}}^{(0)} &= \prod_{\substack{x,y,z \\ \in \mathbb{Z}_L}} \Big[ \prod_{\substack{\alpha,\beta,\gamma \\ \in \{0,1\}}} X^{(0)}_{(x,y,z,w+\tfrac{1}{2})} X^{(0)}_{(x+\tfrac{1}{2},y+\tfrac{1}{2},z+\tfrac{1}{2},w)}  \\
& \hspace{7em} \cdot X^{(0)}_{(x+\tfrac{(-1)^\alpha}{4},y+\tfrac{(-1)^\beta}{4},z+\tfrac{(-1)^\gamma}{4},w+\tfrac{1}{4})} \Big],\\
\end{align*}
which gives us $L$~different disjoint representatives of the logical~$\overline{\mathcal{X}}_{\hat{w}}^{(0)}$~operator, implying the $Z$~distance must be at least~$L$, of which the above logical~$\overline{\mathcal{Z}}_{\hat{w}}^{(0)}$ satisfy this lower bound.

Therefore, the respective $X$ and $Z$~distances of the code are: $d_X = 8L^3, \ d_Z = L$.

\section{List of logical Pauli operators of 4D toric code and associated $\mathsf{CCCZ}$ gate}
\label{app:logPaulis}

\begin{widetext}
Codeblock 0:
\begin{align*}
\overline{\mathcal{Z}}_{\hat{x}}^{(0)} &= \prod_{x \in \mathbb{Z}_L} Z^{(0)}_{(x+\tfrac{1}{2},0,0,0)}, \qquad
\overline{\mathcal{X}}_{\hat{x}}^{(0)} = \prod_{\substack{y,z,w \\ \in \mathbb{Z}_L}} \Big[ \prod_{\substack{\beta,\gamma,\zeta \\ \in \{0,1\}}} X^{(0)}_{(\tfrac{1}{2},y,z,w)} X^{(0)}_{(0,y+\tfrac{1}{2},z+\tfrac{1}{2},w+\tfrac{1}{2})} X^{(0)}_{(\tfrac{1}{4},y+\tfrac{(-1)^\beta}{4},z+\tfrac{(-1)^\gamma}{4},w+\tfrac{(-1)^{\zeta}}{4})} \Big],\\
\overline{\mathcal{Z}}_{\hat{y}}^{(0)} &= \prod_{y\in \mathbb{Z}_L} Z^{(0)}_{(0,y+\tfrac{1}{2},0,0)}, \qquad
\overline{\mathcal{X}}_{\hat{y}}^{(0)} = \prod_{\substack{x,z,w \\ \in \mathbb{Z}_L}} \Big[ \prod_{\substack{\alpha,\gamma,\zeta \\ \in \{0,1\}}} X^{(0)}_{(x,\tfrac{1}{2},z,w)} X^{(0)}_{(x+\tfrac{1}{2},0,z+\tfrac{1}{2},w+\tfrac{1}{2})} X^{(0)}_{(x+\tfrac{(-1)^\alpha}{4},\tfrac{1}{4},z+\tfrac{(-1)^\gamma}{4},w+\tfrac{(-1)^\zeta}{4})} \Big],\\
\overline{\mathcal{Z}}_{\hat{z}}^{(0)} &= \prod_{z\in \mathbb{Z}_L} Z^{(0)}_{(0,0,z+\tfrac{1}{2},0)}, \qquad
\overline{\mathcal{X}}_{\hat{z}}^{(0)} = \prod_{\substack{x,y,w \\ \in \mathbb{Z}_L}} \Big[ \prod_{\substack{\alpha,\beta,\zeta \\ \in \{0,1\}}} X^{(0)}_{(x,y,\tfrac{1}{2},w)} X^{(0)}_{(x+\tfrac{1}{2},y+\tfrac{1}{2},0,w+\tfrac{1}{2})} X^{(0)}_{(x+\tfrac{(-1)^\alpha}{4},y+\tfrac{(-1)^\beta}{4},\tfrac{1}{4},w+\tfrac{(-1)^\zeta}{4})} \Big],\\
\overline{\mathcal{Z}}_{\hat{w}}^{(0)} &= \prod_{z\in \mathbb{Z}_L} Z^{(0)}_{(0,0,0,w+\tfrac{1}{2})}, \qquad
\overline{\mathcal{X}}_{\hat{w}}^{(0)} = \prod_{\substack{x,y,z \\ \in \mathbb{Z}_L}} \Big[ \prod_{\substack{\alpha,\beta,\gamma \\ \in \{0,1\}}} X^{(0)}_{(x,y,z,\tfrac{1}{2})} X^{(0)}_{(x+\tfrac{1}{2},y+\tfrac{1}{2},z+\tfrac{1}{2},0)} X^{(0)}_{(x+\tfrac{(-1)^\alpha}{4},y+\tfrac{(-1)^\beta}{4},z+\tfrac{(-1)^\gamma}{4},\tfrac{1}{4})} \Big].
\end{align*}

Codeblock 1:
\begin{align*}
\overline{\mathcal{Z}}_{\hat{x}}^{(1)} &= \prod_{x \in \mathbb{Z}_L} Z^{(1)}_{(x,\tfrac{1}{2},0,0)}, \qquad
\overline{\mathcal{X}}_{\hat{x}}^{(1)} = \prod_{\substack{y,z,w \\ \in \mathbb{Z}_L}} \Big[ \prod_{\substack{\beta,\gamma,\zeta \\ \in \{0,1\}}} X^{(1)}_{(0,y+\tfrac{1}{2},z,w)} X^{(1)}_{(\tfrac{1}{2},y,z+\tfrac{1}{2},w+\tfrac{1}{2})}  X^{(1)}_{(\tfrac{1}{4},y+\tfrac{(-1)^\beta}{4},z+\tfrac{(-1)^\gamma}{4},w+\tfrac{(-1)^{\zeta}}{4})} \Big],\\
\overline{\mathcal{Z}}_{\hat{y}}^{(1)} &= \prod_{y\in \mathbb{Z}_L} Z^{(1)}_{(\tfrac{1}{2},y,0,0)}, \qquad
\overline{\mathcal{X}}_{\hat{y}}^{(1)} = \prod_{\substack{x,z,w \\ \in \mathbb{Z}_L}} \Big[ \prod_{\substack{\alpha,\gamma,\zeta \\ \in \{0,1\}}} X^{(1)}_{(x+\tfrac{1}{2},0,z,w)} X^{(1)}_{(x,\tfrac{1}{2},z+\tfrac{1}{2},w+\tfrac{1}{2})}  X^{(1)}_{(x+\tfrac{(-1)^\alpha}{4},\tfrac{1}{4},z+\tfrac{(-1)^\gamma}{4},w+\tfrac{(-1)^\zeta}{4})} \Big],\\
\overline{\mathcal{Z}}_{\hat{z}}^{(1)} &= \prod_{z\in \mathbb{Z}_L} Z^{(1)}_{(0,0,z,\tfrac{1}{2})}, \qquad
\overline{\mathcal{X}}_{\hat{z}}^{(1)} = \prod_{\substack{x,y,w \\ \in \mathbb{Z}_L}} \Big[ \prod_{\substack{\alpha,\beta,\zeta \\ \in \{0,1\}}} X^{(1)}_{(x,y,0,w+\tfrac{1}{2})} X^{(1)}_{(x+\tfrac{1}{2},y+\tfrac{1}{2},\tfrac{1}{2},w)} X^{(1)}_{(x+\tfrac{(-1)^\alpha}{4},y+\tfrac{(-1)^\beta}{4},\tfrac{1}{4},w+\tfrac{(-1)^\zeta}{4})} \Big],\\
\overline{\mathcal{Z}}_{\hat{w}}^{(1)} &= \prod_{z\in \mathbb{Z}_L} Z^{(1)}_{(0,0,\tfrac{1}{2},w)}, \qquad
\overline{\mathcal{X}}_{\hat{w}}^{(1)} = \prod_{\substack{x,y,z \\ \in \mathbb{Z}_L}} \Big[ \prod_{\substack{\alpha,\beta,\gamma \\ \in \{0,1\}}} X^{(1)}_{(x,y,z+\tfrac{1}{2},0)} X^{(1)}_{(x+\tfrac{1}{2},y+\tfrac{1}{2},z,\tfrac{1}{2})} X^{(1)}_{(x+\tfrac{(-1)^\alpha}{4},y+\tfrac{(-1)^\beta}{4},z+\tfrac{(-1)^\gamma}{4},\tfrac{1}{4})} \Big].
\end{align*}

Codeblock 2:
\begin{align*}
\overline{\mathcal{Z}}_{\hat{x}}^{(2)} &= \prod_{x \in \mathbb{Z}_L} Z^{(2)}_{(x,0,\tfrac{1}{2},0)}, \qquad
\overline{\mathcal{X}}_{\hat{x}}^{(2)} = \prod_{\substack{y,z,w \\ \in \mathbb{Z}_L}} \Big[ \prod_{\substack{\beta,\gamma,\zeta \\ \in \{0,1\}}} X^{(2)}_{(0,y,z+\tfrac{1}{2},w)} X^{(2)}_{(\tfrac{1}{2},y+\tfrac{1}{2},z,w+\tfrac{1}{2})} X^{(2)}_{(\tfrac{1}{4},y+\tfrac{(-1)^\beta}{4},z+\tfrac{(-1)^\gamma}{4},w+\tfrac{(-1)^{\zeta}}{4})} \Big],\\
\overline{\mathcal{Z}}_{\hat{y}}^{(2)} &= \prod_{y\in \mathbb{Z}_L} Z^{(2)}_{(0,y,0,\tfrac{1}{2})}, \qquad
\overline{\mathcal{X}}_{\hat{y}}^{(2)} = \prod_{\substack{x,z,w \\ \in \mathbb{Z}_L}} \Big[ \prod_{\substack{\alpha,\gamma,\zeta \\ \in \{0,1\}}} X^{(2)}_{(x,0,z,w+\tfrac{1}{2})} X^{(2)}_{(x+\tfrac{1}{2},\tfrac{1}{2},z+\tfrac{1}{2},w)} X^{(2)}_{(x+\tfrac{(-1)^\alpha}{4},\tfrac{1}{4},z+\tfrac{(-1)^\gamma}{4},w+\tfrac{(-1)^\zeta}{4})} \Big],\\
\overline{\mathcal{Z}}_{\hat{z}}^{(2)} &= \prod_{z\in \mathbb{Z}_L} Z^{(2)}_{(\tfrac{1}{2},0,z,0)}, \qquad
\overline{\mathcal{X}}_{\hat{z}}^{(2)} = \prod_{\substack{x,y,w \\ \in \mathbb{Z}_L}} \Big[ \prod_{\substack{\alpha,\beta,\zeta \\ \in \{0,1\}}} X^{(2)}_{(x+\tfrac{1}{2},y,0,w)} X^{(2)}_{(x,y+\tfrac{1}{2},\tfrac{1}{2},w+\tfrac{1}{2})} X^{(2)}_{(x+\tfrac{(-1)^\alpha}{4},y+\tfrac{(-1)^\beta}{4},\tfrac{1}{4},w+\tfrac{(-1)^\zeta}{4})} \Big],\\
\overline{\mathcal{Z}}_{\hat{w}}^{(2)} &= \prod_{z\in \mathbb{Z}_L} Z^{(0)}_{(0,\tfrac{1}{2},0,w)}, \qquad
\overline{\mathcal{X}}_{\hat{w}}^{(2)} = \prod_{\substack{x,y,z \\ \in \mathbb{Z}_L}} \Big[ \prod_{\substack{\alpha,\beta,\gamma \\ \in \{0,1\}}} X^{(2)}_{(x,y+\tfrac{1}{2},z,0)} X^{(2)}_{(x+\tfrac{1}{2},y,z+\tfrac{1}{2},\tfrac{1}{2})} X^{(2)}_{(x+\tfrac{(-1)^\alpha}{4},y+\tfrac{(-1)^\beta}{4},z+\tfrac{(-1)^\gamma}{4},\tfrac{1}{4})} \Big].
\end{align*}

Codeblock 3:
\begin{align*}
\overline{\mathcal{Z}}_{\hat{x}}^{(3)} &= \prod_{x \in \mathbb{Z}_L} Z^{(3)}_{(x,0,0,\tfrac{1}{2})}, \qquad 
\overline{\mathcal{X}}_{\hat{x}}^{(3)} = \prod_{\substack{y,z,w \\ \in \mathbb{Z}_L}} \Big[ \prod_{\substack{\beta,\gamma,\zeta \\ \in \{0,1\}}} X^{(3)}_{(0,y,z,w+\tfrac{1}{2})} X^{(3)}_{(\tfrac{1}{2},y+\tfrac{1}{2},z+\tfrac{1}{2},w)} X^{(3)}_{(\tfrac{1}{4},y+\tfrac{(-1)^\beta}{4},z+\tfrac{(-1)^\gamma}{4},w+\tfrac{(-1)^{\zeta}}{4})} \Big],\\
\overline{\mathcal{Z}}_{\hat{y}}^{(3)} &= \prod_{y\in \mathbb{Z}_L} Z^{(3)}_{(0,y,\tfrac{1}{2},0)}, \qquad
\overline{\mathcal{X}}_{\hat{y}}^{(3)} = \prod_{\substack{x,z,w \\ \in \mathbb{Z}_L}} \Big[ \prod_{\substack{\alpha,\gamma,\zeta \\ \in \{0,1\}}} X^{(3)}_{(x,0,z+\tfrac{1}{2},w)} X^{(3)}_{(x+\tfrac{1}{2},\tfrac{1}{2},z,w+\tfrac{1}{2})}X^{(3)}_{(x+\tfrac{(-1)^\alpha}{4},\tfrac{1}{4},z+\tfrac{(-1)^\gamma}{4},w+\tfrac{(-1)^\zeta}{4})} \Big],\\
\overline{\mathcal{Z}}_{\hat{z}}^{(3)} &= \prod_{z\in \mathbb{Z}_L} Z^{(3)}_{(0,\tfrac{1}{2},z,0)}, \qquad
\overline{\mathcal{X}}_{\hat{z}}^{(3)} = \prod_{\substack{x,y,w \\ \in \mathbb{Z}_L}} \Big[ \prod_{\substack{\alpha,\beta,\zeta \\ \in \{0,1\}}} X^{(3)}_{(x,y+\tfrac{1}{2},0,w)} X^{(3)}_{(x+\tfrac{1}{2},y,\tfrac{1}{2},w+\tfrac{1}{2})} X^{(3)}_{(x+\tfrac{(-1)^\alpha}{4},y+\tfrac{(-1)^\beta}{4},\tfrac{1}{4},w+\tfrac{(-1)^\zeta}{4})} \Big],\\
\overline{\mathcal{Z}}_{\hat{w}}^{(3)} &= \prod_{z\in \mathbb{Z}_L} Z^{(3)}_{(\tfrac{1}{2},0,0,w)}, \qquad
\overline{\mathcal{X}}_{\hat{w}}^{(3)} = \prod_{\substack{x,y,z \\ \in \mathbb{Z}_L}} \Big[ \prod_{\substack{\alpha,\beta,\gamma \\ \in \{0,1\}}} X^{(3)}_{(x+\tfrac{1}{2},y,z,0)} X^{(3)}_{(x,y+\tfrac{1}{2},z+\tfrac{1}{2},\tfrac{1}{2})} X^{(3)}_{(x+\tfrac{(-1)^\alpha}{4},y+\tfrac{(-1)^\beta}{4},z+\tfrac{(-1)^\gamma}{4},\tfrac{1}{4})} \Big].
\end{align*}
\end{widetext}

The intersection of the $X$~logical operators from codeblocks 1--3 (say, $z,y,x$) will result in the following $Z$ operator under $\mathsf{CCCZ}$ on codeblock~0:
\begin{align*}
 \prod_{\substack{w  \in \mathbb{Z}_L}} \prod_{\zeta \in \{0,1\}}  Z_{(0,0,0,w+\tfrac{1}{2})}^{(0)} Z_{(\tfrac{1}{4},\tfrac{1}{4},\tfrac{1}{4},w+\tfrac{(-1)^\zeta}{4})}^{(0)} Z_{(\tfrac{1}{2},\tfrac{1}{2},\tfrac{1}{2},w)}^{(0)}.
\end{align*}
Yet, when one takes the intersection of the following three $X$~stabilizers from codeblocks 1--3, $v_r = (\tfrac{1}{2},\tfrac{1}{2},0,w), \ v_g = (\tfrac{1}{2},0,\tfrac{1}{2},w), \ v_b = (0,\tfrac{1}{2},\tfrac{1}{2},w) \}$, we arrive at the following 3-body $Z$~stabilizer:
\begin{align*}
\{  
&(\tfrac{1}{2},\tfrac{1}{2},\tfrac{1}{2},w), \\
&(\tfrac{1}{4},\tfrac{1}{4},\tfrac{1}{4},w-\tfrac{1}{4}), \\
&(\tfrac{1}{4},\tfrac{1}{4},\tfrac{1}{4},w+\tfrac{1}{4})
\}.
\end{align*}
Therefore the above logical operator is equivalent to logical~$\overline{\mathcal{Z}}_{\hat{w}}^{(0)}$. Given the symmetry of the logical operators across all logical codeblocks, conditions~\ref{eq:Cond4}--\ref{eq:Cond5} are clearly satisfied when logical~$X$ operators with different labels are taken across different codeblocks.

Finally, we need to show that Eq.~\ref{eq:Cond3} is satified when $k \ne l$. Take a pair of logical~$X$ operators with the different indices, say $\overline{\mathcal{X}}_{\hat{z}}^{(2)}, \ \overline{\mathcal{X}}_{\hat{w}}^{(3)}$. Under the action of the transversal~$\mathsf{CCCZ}$, $(\mathsf{CCCZ} \cdot \overline{\mathcal{X}}_{\hat{z}}^{(2)} \cdot \mathsf{CCCZ}^{\dagger})\overline{\mathcal{X}}_{\hat{w}}^{(3)}(\mathsf{CCCZ} \cdot \overline{\mathcal{X}}_{\hat{z}}^{(2)} \cdot \mathsf{CCCZ}^{\dagger})^{\dagger}$, the action on their overlap will yield:
\begin{align*}
 \prod_{\substack{x,y  \in \mathbb{Z}_L \\ \alpha,\beta \in \mathbb{Z}_2}} & \Big[ \mathsf{CZ}^{(0,1)}_{(x+\tfrac{1}{2},y,0,0)} \mathsf{CZ}^{(0,1)}_{(x,y+\tfrac{1}{2},\tfrac{1}{2},\tfrac{1}{2})} \\
&\hspace{1em} \cdot\mathsf{CZ}^{(0,1)}_{(x + \tfrac{(-1)^\alpha}{4},y+\tfrac{(-1)^\beta}{4},\tfrac{1}{4},\tfrac{1}{4})} \Big]. 
\end{align*}

What remains to be shown is that this operator has even overlap with the product of any $X$~stabilizers from codeblocks~0 and~1, and is thus equivalent to logical identity. Without loss of generality we can consider the $X$~stabilizers given by the set of coordinates: $(x_0, y_0, z_0, w_0)$ for codeblock 0 and~$(x_1, y_1, z_1 +\tfrac{1}{2}, w_1 + \tfrac{1}{2})$ for codeblock~1. However, in order to have non-zero overlap we must have~$x_1 = x_0, \ y_1 = y_0$, $z_1 \in \{ z_0, z_0+1\}$,  $w_1 \in \{ w_0, w_0+1\}$. Again, without loss of generality we will choose $z_1 = z_0$ and~$w_1 = w_0$. Therefore, the corresponding $Z$~operator above is equivalent to logical~$\overline{Z}_x^{(3)}$. They have overlap on six~3-cells, given by coordinates:
\begin{align*}
\{
&(x_0,y_0,z_0,w_0+\tfrac{1}{2}), \\
&(x_0,y_0,z_0+\tfrac{1}{2},w_0),\\
&(x_0-\tfrac{1}{4},y_0-\tfrac{1}{4},z_0+\tfrac{1}{4},w_0+\tfrac{1}{4}),\\ 
&(x_0-\tfrac{1}{4},y_0+\tfrac{1}{4},z_0+\tfrac{1}{4},w_0+\tfrac{1}{4}),\\
&(x_0+\tfrac{1}{4},y_0-\tfrac{1}{4},z_0+\tfrac{1}{4},w_0+\tfrac{1}{4}), \\
&(x_0+\tfrac{1}{4},y_0+\tfrac{1}{4},z_0+\tfrac{1}{4},w_0+\tfrac{1}{4}) \},
\end{align*}
of which it is fairly straightforward to check that this operator has even overlap with the tensor product of $\mathsf{CZ}$~operators above. Following a symmetric argument, one can show that Eq.~\ref{eq:Cond3} is always satisfied for the choice of logical~$X$ operators. In conclusion, the transversal~$\mathsf{CCCZ}$ will implement a logical~$\mathsf{CCCZ}$, coupling logical operators in the following quartets:
\begin{align*}
&\{\overline{\mathcal{X}}_{\hat{w}}^{(0)},\overline{\mathcal{X}}_{\hat{z}}^{(1)},\overline{\mathcal{X}}_{\hat{y}}^{(2)},\overline{\mathcal{X}}_{\hat{x}}^{(3)}\}, \\
&\{\overline{\mathcal{X}}_{\hat{z}}^{(0)},\overline{\mathcal{X}}_{\hat{w}}^{(1)},\overline{\mathcal{X}}_{\hat{x}}^{(2)},\overline{\mathcal{X}}_{\hat{y}}^{(3)}\}, \\
&\{\overline{\mathcal{X}}_{\hat{y}}^{(0)},\overline{\mathcal{X}}_{\hat{x}}^{(1)},\overline{\mathcal{X}}_{\hat{w}}^{(2)},\overline{\mathcal{X}}_{\hat{z}}^{(3)}\}, \\
&\{\overline{\mathcal{X}}_{\hat{x}}^{(0)},\overline{\mathcal{X}}_{\hat{y}}^{(1)},\overline{\mathcal{X}}_{\hat{z}}^{(2)},\overline{\mathcal{X}}_{\hat{w}}^{(3)}\}.
\end{align*}

\section{Global constraints on 2-cells}
\label{app:2constraints}
The goal of this Appendix is to explicitly provide a global redundancy among the 2-cells corresponding to the $Z$~stabilizers. We focus here on codeblock~0 as described in the text, yet due to the equivalence between the different codeblocks all arguments will port over to the other codeblocks as well. We will show that by taking a set of faces that form a plane, their corresponding support will cancel out and thus the set of faces will have one fewer degree of freedom.

Consider the support of the face whose vertices are given by:~$\{ (x,0,\tfrac{1}{2}, w + \tfrac{1}{2}), (x,\tfrac{1}{2}, 0, w + \tfrac{1}{2}), (x,\tfrac{1}{2},\tfrac{1}{2}, w )\}$. The support of the operator will thus be given by the following triple of qubits:
\begin{align*}
(x, \tfrac{1}{2}, \tfrac{1}{2}, w+ \tfrac{1}{2}), \ (x-\tfrac{1}{4}, \tfrac{1}{4}, \tfrac{1}{4}, w+ \tfrac{1}{4}), \ (x+\tfrac{1}{4}, \tfrac{1}{4}, \tfrac{1}{4}, w+ \tfrac{1}{4}).
\end{align*}
If we now consider face given by the following triple of vertices:~$\{ (x,0,\tfrac{1}{2}, w + \tfrac{1}{2}), (x,\tfrac{1}{2}, 0, w + \tfrac{1}{2}), (x,\tfrac{1}{2},\tfrac{1}{2}, w + 1 )\}$. Then the corresponding support will be:
\begin{align*}
(x, \tfrac{1}{2}, \tfrac{1}{2}, w+ \tfrac{1}{2}), \ (x-\tfrac{1}{4}, \tfrac{1}{4}, \tfrac{1}{4}, w+ \tfrac{3}{4}), \ (x+\tfrac{1}{4}, \tfrac{1}{4}, \tfrac{1}{4}, w+ \tfrac{3}{4}).
\end{align*}
Taking the product of these two operators, will yield an operator with support on the following quartet: 
\begin{align*}
&(x-\tfrac{1}{4}, \tfrac{1}{4}, \tfrac{1}{4}, w+ \tfrac{1}{4}), \\ 
&(x+\tfrac{1}{4}, \tfrac{1}{4}, \tfrac{1}{4}, w+ \tfrac{1}{4}), \\
&(x-\tfrac{1}{4}, \tfrac{1}{4}, \tfrac{1}{4}, w+ \tfrac{3}{4}), \\ 
&(x+\tfrac{1}{4}, \tfrac{1}{4}, \tfrac{1}{4}, w+ \tfrac{3}{4}).
\end{align*}

Along a similar vein, the face whose vertices are~$\{ (x+\tfrac{1}{2},\tfrac{1}{2},0,w), \ (x+\tfrac{1}{2},0,\tfrac{1}{2},w), (x,\tfrac{1}{2},\tfrac{1}{2},w)\}$ yields support:
\begin{align*}
(x+\tfrac{1}{2}, \tfrac{1}{2}, \tfrac{1}{2}, w), \ (x+\tfrac{1}{4}, \tfrac{1}{4}, \tfrac{1}{4}, w- \tfrac{1}{4}), \ (x+\tfrac{1}{4}, \tfrac{1}{4}, \tfrac{1}{4}, w+ \tfrac{1}{4}).
\end{align*}
Similarly, the face whose vertices are~$\{ (x+\tfrac{1}{2},\tfrac{1}{2},0,w), \ (x+\tfrac{1}{2},0,\tfrac{1}{2},w), (x+1,\tfrac{1}{2},\tfrac{1}{2},w)\}$ yields support:
\begin{align*}
(x+\tfrac{1}{2}, \tfrac{1}{2}, \tfrac{1}{2}, w), \ (x+\tfrac{3}{4}, \tfrac{1}{4}, \tfrac{1}{4}, w- \tfrac{1}{4}), \ (x+\tfrac{3}{4}, \tfrac{1}{4}, \tfrac{1}{4}, w+ \tfrac{1}{4}).
\end{align*}
Therefore, taking the product of the latter two operators will yield an operator with support:
\begin{align*}
&(x+\tfrac{1}{4}, \tfrac{1}{4}, \tfrac{1}{4}, w- \tfrac{1}{4}), \\ 
&(x+\tfrac{1}{4}, \tfrac{1}{4}, \tfrac{1}{4}, w+ \tfrac{1}{4}), \\
&(x+\tfrac{3}{4}, \tfrac{1}{4}, \tfrac{1}{4}, w- \tfrac{1}{4}), \\ 
&(x+\tfrac{3}{4}, \tfrac{1}{4}, \tfrac{1}{4}, w+ \tfrac{1}{4}).
\end{align*}
Therefore, taking the plane of such weight-4 operators by spanning all integer~$x$ and~$w$, the resulting operators will all cancel out and yield a global dependence. Moreover, the same argument will hold for any pair of cartesian directions, thus totally six such planar redundancies. These can be thought of analogously to the planes that form the six logical operators in the traditional 4D~toric code where there is an equivalence between the $X$ and $Z$~stabilizers.

\section{Global constraints on 1-cells}
\label{app:1constraints}
In this Appendix we explore the redundancy between the 1-cell constraints forming a hyperplane in the lattice. The goal of this Appendix is to provide a set of edges that share a global symmetry where all faces that emerge from these edges share have exactly two edges belonging to the set. We will consider the set of all edges between vertices~$v_0$ and~$v_1$ such that the $x$ coordinate of $v_0$ is~$x_0 = 0$ and the $x$~coordinate of $v_1$ is~$x_1 = \tfrac{1}{2}$. Thus there will be three such forms of possible edges: 
\begin{align*}
&\{(0,y,z+\tfrac{1}{2},w+\tfrac{1}{2}),  (\tfrac{1}{2},y, z+\tfrac{1}{2},w) \} \\
&\{ (0,y,z+\tfrac{1}{2},w+\tfrac{1}{2}), (\tfrac{1}{2},y, z, w+\tfrac{1}{2}) \} \\
&\{ (0,y+\tfrac{1}{2},z,w+\tfrac{1}{2}), (\tfrac{1}{2},y, z, w+\tfrac{1}{2}) \}.
\end{align*}
Without loss of generality, we shall focus on an edge between a red and green vertex point, that is: $\{(0,y,z+\tfrac{1}{2},w+\tfrac{1}{2}),  (\tfrac{1}{2},y, z+\tfrac{1}{2},w) \}$. Such an edge can only share faces with the following set of blue vertices:~$(0,y-\tfrac{1}{2},z+\tfrac{1}{2},w), \ (0,y+\tfrac{1}{2},z+\tfrac{1}{2},w) , \ (\tfrac{1}{2}, y, z, w+\tfrac{1}{2}), \ (\tfrac{1}{2}, y, z+1, w+\tfrac{1}{2})$, any choice of which will form an edge from the desired criteria with one of the original pair. Thus, any face sharing an edge from the criteria above will necessarily share two such edges.

Finally, given the above set of edges form a hyperplane with fixed~$x_0 = 0$ and $x_1 = \tfrac{1}{2}$, the choice of the $x$~coordinate here is arbitrary and can be generalized for all four Cartesian directions, thus totaling four such hyperplane constraints.

\section{Boundary stabilizers}
\label{app:boundaries}

In this Appendix, we explore the set of boundary stabilizers and their corresponding support. We will devote this appendix to the case of codeblock 0, however similar arguments will hold for codeblocks 1--3 by symmetry. Recall we are removing all qubits that have a coordinate outside the interval~$[\tfrac{1}{2},L]$.

The $X$~stabilizers of codeblock~0 are given by the following sets of coordinates: $(x,y,z,w), \ (x+\tfrac{1}{2},y+\tfrac{1}{2},z+\tfrac{1}{2},w+\tfrac{1}{2})$, where the $\hat{x}, \hat{y}, \hat{z}$ coordinates are constrained to be in the interval $[\tfrac{1}{2},L]$ while the $\hat{w}$ coordinate is constrained to be in the interval~$[1,L-\tfrac{1}{2}]$. We will call one of the coordinates~$\hat{x}, \hat{y}, \hat{z}$ a \textit{boundary coordinate} if their value is at endpoint of the allowed interval. Normally, the weight of an $X$~stabilizer in the octaplex tessellation is 24, yet when boundary coordinates are present that will be changed as some of the qubits within the support of a given operator have been removed. Given a number of boundary coordinates $b$, the adjusted weight of a boundary operator is: $8-b+16/2^b$. This can be derived as follows, suppose we have the following stabilizer: $(\tfrac{1}{2},y+\tfrac{1}{2},z+\tfrac{1}{2}, w+\tfrac{1}{2})$, where $y,z,w$ are integers in the interval $[1,L-1]$, then the stabilizer will no longer have support on any qubit with $\hat{x}$ coordinate smaller than~$\tfrac{1}{2}$ of which there is one with value 0 and eight with value~$\tfrac{1}{4}$, so the total weight of the operator is $24 - 9 = 15$. Similarly, if there were two boundary coordinates (say $y=0$ above) then there would be the additional qubits that would be absent from the normal support, one $\hat{y}$ coordinate equal to $0$ and the eight with value~$\tfrac{1}{4}$ (four of which we already accounted for when the $\hat{x}$ coordinate is equal to~$\tfrac{1}{4}$, bringing the stabilizer to: $15 - 1 - 4 = 10$.

There will also be modified $Z$~stabilizers on the boundary. Recall the $Z$~stabilizers defined in Sec.~\ref{sec:Coordinates}--\ref{sec:CoordCode0}. They are defined by intersection of $X$~stabilizers from the other codeblocks and have coordinates that are either: (2i) 3 quarter-integer and 1 half-integer, (2ii) 3 quarter-integer, 1 integer. Focusing on case~(2i), the boundary $Z$~stabilizer we have to be concerned with is that where the half-integer coordinate is equal to~$\tfrac{1}{2}$. That is, suppose we have the following $Z$~stabilizer:~$(\tfrac{1}{2}, y + (-1)^\beta\tfrac{1}{4}, z+(-1)^{\alpha}\tfrac{1}{4}, w+\tfrac{1}{2}+(-1)^\zeta\tfrac{1}{4})$, for integers $y,z,w$ in the interval~$[1,L-1]$, then normally the support of such a stabilizer will be on the following set of 3-cells (qubits): 
\begin{align*}
\{ &( \tfrac{1}{4}, y + (-1)^\beta\tfrac{1}{4}, z+(-1)^{\alpha}\tfrac{1}{4}, w+\tfrac{1}{2}+(-1)^\zeta\tfrac{1}{4}),\\
 &(\tfrac{3}{4}, y + (-1)^\beta\tfrac{1}{4}, z+(-1)^{\alpha}\tfrac{1}{4}, w+\tfrac{1}{2}+(-1)^\zeta\tfrac{1}{4}), \\
 &(\tfrac{1}{2}, y + (-1)^\beta\tfrac{1}{2}, z+(-1)^{\alpha}\tfrac{1}{2}, w+\tfrac{1}{2}) \}.
\end{align*}
The first such qubit has been removed, and therefore the true support of such $Z$~stabilizers will be of weight~2. One can similarly argue for all boundary faces, and show that they will be cut to only have support on a pair of qubits, this is analogous to the weight~2 $Z$~stabilizers we obtain in the 3D toric code with boundaries.

\end{document}